\documentclass[11pt]{article}

\usepackage{amsmath}    %For more commands for lay-outing formulae
\usepackage{amssymb}
\usepackage{graphicx}
\usepackage{amsfonts}   %For more commands for lay-outing formulae
\usepackage{latexsym}   %For extra symbols
\usepackage{slashed}
\usepackage{yfonts} 
\usepackage{verbatim}
\usepackage{wrapfig}
\usepackage{mathrsfs}
\usepackage{cancel}

\global\newcount\itemno \global\itemno=0

\def\itemaut#1{\global\advance\itemno by1\noindent\item{\the\itemno.}#1}

\newif{\ifeq}		% defines a new condition @eq tested by the conditional \ifeq
\eqfalse		% if uncommented, declares @eq to be false
\newcommand{\be}{\begin{equation}}
\newcommand{\ee}{\end{equation}}
\newcommand{\bes}{\begin{equation*}}
\newcommand{\ees}{\end{equation*}}
\newcommand{\bea}{\begin{eqnarray}}
\newcommand{\eea}{\end{eqnarray}}
\newcommand{\beas}{\begin{eqnarray*}}
\newcommand{\eeas}{\end{eqnarray*}}

\newcommand{\bb}{\mathbb}
\newcommand{\ph}{\phantom}

\def\({\left(}
\def\){\right)}
\def\[{\left[}
\def\]{\right]}

\def\frac#1#2{{#1 \over #2}}

\renewcommand{\tfrac}[2]{{\textstyle\frac{#1}{#2}}}

\newcommand{\rt}{{\sqrt 2}}

\renewcommand{\a}{\alpha}
\renewcommand{\b}{\beta}

\newcommand{\vt}{\vartheta}

\newcommand{\G}{\Gamma}

\newcommand{\U}{\Upsilon}

\newcommand{\GB}{\overline{\Gamma}}

\newcommand{\CD}{{\cal D}}

\newcommand{\CG}{{\cal G}}

\newcommand{\CN}{{\cal N}}

\newcommand{\MB}{{\overline{M}}}
\newcommand{\bbar}{\overline}

\newcommand{\tG}{\tilde{\Gamma}}

\newcommand{\ma}{~\!{\textrm{\dn m}}}

\newcommand{\tr}{{\rm tr \,}}

\def\ie{{\it i.e.}}

\newcommand{\lsim}{\,\raise.3ex\hbox{$<$\kern-.75em\lower1ex\hbox{$\sim$}}\,}
\newcommand{\gsim}{\,\raise.3ex\hbox{$>$\kern-.75em\lower1ex\hbox{$\sim$}}\,}
\def\p{\partial}
\newcommand{\tp}{{\theta^+}}
\newcommand{\tbp}{{\bar{\theta}^+}}

\newcommand{\DB}{\overline{\rm D}}
\newcommand{\CDB}{\overline{\cal D}}

\newcommand{\QB}{{\overline{Q}}}

\def\susy{supersymmetry}
\def\susic{supersymmetric}

\def\Ka{K\"{a}hler}

\def\nK{non-K\"{a}hler}

\def\ZT{$(0,2)$}

\def\II{\relax{I\kern-.10em I}}

\newcommand{\V}{\mathcal{V}}
\renewcommand{\ma}{{m_{a'}}}
\newcommand{\Ups}{\Upsilon}
\newcommand{\THB}{{\overline{\Theta}}}
\newcommand{\thp}{\tp}
\newcommand{\thpb}{\tbp}

\numberwithin{equation}{section}

\begin{document}
\begin{titlepage}
\begin{flushright}
February 9, 2011 \\
MIT-CTP 4066 \\
NSF-KITP-09-170
\end{flushright}
\vskip 1in

\begin{center}
{\Large{Computing the Spectrum of a Heterotic Flux Vacuum}} 
\vskip 0.5in Allan Adams$^{1}$ and Joshua M. Lapan$^{2}$
\vskip 0.4in {\it $^{1}$ Center for Theoretical Physics\\  Massachusetts Institute of Technology\\ Cambridge, MA  02139}
\vskip 0.2in {\it $^{2}$ Kavli Institute for Theoretical Physics\\  University of California\\  Santa Barbara, CA 93106}
\end{center}
\vskip 0.5in

\begin{abstract}\noindent
We compute the massless spectra of a set of flux vacua of the heterotic string.  The vacua we study include well-known \nK\ $T^{2}$-fibrations over $K3$ with $SU(3)$ structure and intrinsic torsion.
%\footnote{Thse are the so-called Dasgupta-Rajesh-Sethi-Goldstein-Prokuskin-Fu-Strominger-Yau-Witten manifolds, or, for brevity, DRSGPFSYW manifolds.}.
Following gauged linear sigma models of these vacua into phases governed by asymmetric Landau-Ginzburg orbifolds allows us to compute the spectrum using generalizations of familiar LG-orbifold techniques.  We study several four- and six-dimensional examples with spacetime $\CN=2$ supersymmetry in detail.  
%The same techniques can be applied to more general WZW-fibred LG models and to examples with reduced spacetime supersymmetry.
	
\end{abstract}

\end{titlepage}

\section{Introduction}
Since the discovery of the heterotic string \cite{Gross:1984dd}\ and its landscape of vacua \cite{Strominger:1986uh}, the search has been on for heterotic compactifications with realistic phenomenology (see  {\it e.g.} \cite{Anderson:2007nc,Andreas:2007ei,Bouchard:2005ag,Braun:2005nv} for some recent examples).   Traditionally, this involves wading through the space of stable vector bundles over Calabi-Yau manifolds in search of a suitable GUT.  The problem then becomes lifting the massless moduli which parameterize this space.

In type II, it is well understood that these troublesome moduli may be lifted at the level of supergravity by turning on background RR and NS-NS fluxes and decorating the result with D-branes and orientifolds \cite{Giddings:2001yu,Kachru:2003aw}.  Meanwhile, interesting phenomenology may be orchestrated by arranging suitable singularities or brane intersections in the compactification manifold.  The result is a teaming ecology of increasingly sophisticated models (for some recent F-theory examples, see {\it e.g.} \cite{Beasley:2008dc,Donagi:2008ca,Marsano:2009ym}).  Unfortunately, due to the difficulty of introducing RR fluxes into the worldsheet theory of the type II string, it has proven surprisingly difficult to go beyond the supergravity approximation to a microscopic worldsheet description, leaving potentially important tracts of the stringy landscape largely unexplored.

In heterotic string theory, the situation is much the reverse.  At the level of supergravity, the conditions for unbroken $\CN=1$ supersymmetry, phrased as conditions on the compactification geometry and background fluxes in \cite{Strominger:1986uh}\ more than two decades ago, turn out to be surprisingly difficult to solve.  For example, setting $\langle H \rangle \neq 0$ forces the compactification geometry to be \nK, so that many of the basic tools used to study Calabi-Yau examples --- such as Hodge theory and special geometry --- do not apply.  As a result, torsional compactifications of the heterotic string have received relatively little attention, and until very recently only a single class of solutions were known.  First constructed via dualities by Dasgupta, Rajesh, and Sethi \cite{Dasgupta:1999ss}, and later studied geometrically by Goldstein and Prokushkin \cite{Goldstein:2002pg}, these solutions involve non-K\"ahler complex $T^2$ fibrations over $K3$ bases with $H$-flux mixing fiber and base.  The existence of non-trivial vector bundles over these geometries which, together with the dilaton and $H$-flux, satisfy the modified Bianchi identity was subsequently proven by Fu and Yau \cite{Fu:2006vj} and elaborated upon by Becker, Becker, Fu, Tseng, and Yau \cite{Becker:2006et}.\footnote{A judicious choice of connection, which boils down to a choice of worldsheet renormalization scheme, can simplify this analysis; see \cite{Strominger:1986uh,Bergshoeff:1989de,Becker:2009df} for discussion of this choice and its consequences.}  
A computation of the massless spectrum in these torsional compactifications, however, has remained elusive despite considerable effort \cite{Cyrier:2006pp,Becker:2006xp,Becker:2007ea,Fu:2008ga,Becker:2008rc}, largely due to the complexity of even the linearized supergravity equations of motion.  
Of course, since these compactifications involve cancelations between terms at different orders in the $\alpha'$ expansion, the supergravity analysis of these torsional compactifications is not guaranteed to be reliable.  To be sure, in the equations of motion, all terms at a given order in $\alpha'$ scale homogeneously under constant rescalings $H\rightarrow \lambda H$, $G\rightarrow \lambda G$, implying that, for each solution, there is a family of solutions.  
However, the heterotic Bianchi identity,
\be
\label{eqn:bianchi}
dH = \frac{\alpha'}{4} \Big[ \tr\big( R \wedge R \big) - \textrm{tr} \big( F \wedge F \big) \Big] \, ,
\ee
scales inhomogeneously, so if it is satisfied nontrivially (neither side separately vanishing), this suggests that some radii may be fixed in units of $\a'$ with the precise value determined by the flux quantum numbers (see appendix \ref{app:no-large-radius}).
If the flux quanta could be taken large so as to ensure that all curvature invariants are parametrically small, as occurs in many type II flux vacua, we could trust the supergravity description.  In fact, the flux quanta 
are bounded by $c_{2}(K3)=24$, so no large-radius limit is available (see \cite{Adams:2009tt} for a more detailed discussion of this point), leaving the supergravity description as a potentially uncontrolled approximation.    

On the worldsheet, however, the situation is much better.  With no RR fields to complicate the analysis and no need to assume the existence of any small-curvature limit, there is no {\em a priori} obstruction to constructing exact worldsheet descriptions of heterotic flux vacua.   A powerful tool for constructing such worldsheet descriptions is the torsion linear sigma model (TLSM) developed in \cite{Adams:2006kb}\ and generalized in \cite{Adams:2009av}, in which a one loop gauge anomaly is cancelled by the classical anomaly of an asymmetrically gauged WZW model.  The semi-classical Higgs branch of the resulting gauge theory is a \nK\ complex manifold supporting gauge- and $H$-fluxes satisfying the heterotic Bianchi identity.  The power of the TLSM is that it allows us to quantize these vacua without assuming the validity of the semi-classical approximation.  More precisely, by studying phase transitions in the gauge theory \cite{Adams:2009tt}, these vacua may be related to chiral Landau-Ginzburg orbifolds in which a discrete orbifold anomaly in the LG model is cancelled by an orbifold Green-Schwarz mechanism. Suitable modifications of familiar orbifold tools should then allow us to compute the exact massless spectra of these heterotic flux vacua.

The goal of this paper is to develop the technology needed to compute the exact spectrum of massless fermions in these chiral LG orbifolds, and to then apply these techniques to a set of concrete examples, focusing on abelian linear models for simplicity.  When the compactification preserves spacetime \susy, this is enough to determine the full massless spectrum.  The logic of the calculation will follow that of existing GLSM lore \cite{Kachru:1993pg,Distler:1993mk,Silverstein:1994ih}, with appropriate modifications to account for the intricate interplay between tree-level and one-loop effects in our models.  It will turn out that the presence of the free fermion in the torsion multiplet will force us to have either $\mathcal{N}=2$ or $\CN=0$ 4d supersymmetry, but never $\CN=1$; this is in agreement with \cite{Becker:2006et} when the curvatures of the circle bundles are anti-self-dual, $(1,1)$-forms.  Indeed, in abelian TLSMs we can only realize circle bundles whose curvature forms are purely $(1,1)$, as they must be linear combinations of hyperplane classes.  This unfortunately implies that the net number of generations in these abelian models is zero, a result that is also expected from \cite{Becker:2006et} since $c_3(\V_K)=0$ even when the circle-bundle curvatures have $(0,2)$ components.  It would be interesting to extend our analysis to non-abelian models, where the fermions in the torsion multiplet are interacting and the number of generations may be non-zero.  For now, we will focus on the more simple abelian models.

The remainder of this paper is organized as follows.
Section 2 presents a quick review of the structure of non-linear sigma models and torsion gauged linear sigma models of heterotic flux vacua.
Section 3 constructs the technology needed to quantize TLSMs at chiral LG-orbifold points, working in considerable generality.
Section 4 applies this technology to two specific examples.
Section 5 concludes with a discussion of some remaining puzzles and directions for future research.
Three appendices review our \ZT\ conventions, the geometry of heterotic flux manifolds, and a quantization of the heterotic string on an Iwasawa-type compactification.

\section{Linear Models for Heterotic Flux Vacua: Review}

The torsional $SU(3)$-manifolds discussed in this paper are \nK\ complex 3-folds constructed as a $T^{2}$-fibration over a \Ka\ base, $T^{2}\to K\to S$.  For simplicity, the gauge bundles, $\V_K$, living over our manifolds will be pullbacks to $K$ of stable bundles, $\V_S$, over the base $S$.  In particular, for the explicit compact examples whose spectra we compute, we'll take $S$ to be either $T^4$ or $K3$. (While $T^4$ was ruled out in \cite{Becker:2006et} using constraints from spacetime supersymmetry, we won't constrain ourselves to \susic\ examples {\it a priori}.)

In \cite{Adams:2006kb}, we showed how to construct a non-linear sigma model (NLSM) describing these compactifications as the low-energy limit of a novel gauged linear sigma model (GLSM) which we dubbed a ``torsion linear sigma model'' (TLSM).  The idea is to begin with a standard (0,2) GLSM describing the \Ka\ base, $S$, and its bundle, $\V_{S}$, then add a chiral multiplet (the ``torsion multiplet'') whose lowest bosonic components are a pair of periodic scalars that shift under worldsheet gauge transformations --- essentially a pair of dynamical axions.  These real scalar components parameterize two circles that twist as they move around the base, producing a $T^2$-fibration over $S$.  For later reference, note that in Wess-Zumino gauge, the right-handed fermion in the torsion multiplet is uncharged and, in fact, entirely free.

As it turns out, it is impossible to write a supersymmetric and gauge-invariant kinetic term for the torsion multiplet.  We \emph{can}, however, write a supersymmetric action for the torsion multiplet whose gauge variation, while non-vanishing,  is of exactly the same form as the anomalous variation of the measure for a gauge-charged fermion.  We can thus cancel the classical gauge-variation of the torsion multiplet action off a one-loop anomaly of some non-standard set of gauge-charged fermions in the base GLSM.  

The remainder of this section will review the salient features of the TLSM  construction as needed for the computation to follow.  We begin by recalling how NS-NS flux appears in heterotic NLSMs, then review the GLSM description of a \Ka\ compactification, and finish with the construction of the action of the TLSM.

\subsection{Non-Linear Sigma Models with NS-NS Flux}
\label{sec:review}

The worldsheet action of a heterotic nonlinear sigma model includes the terms
\bea
S_{\Sigma} &=& \ldots + \frac{1}{\pi\alpha'} \int_{\Sigma} X^* (B) + \frac{1}{2\pi}\int_\Sigma \!\bigg\{ g_{MN}(X) \, \psi_+^M \Big( \delta^N_Q \p_- + \p_- X^P \, \Gamma_{(-)PQ}^{\ph{(+}N} \Big) \psi_+^Q  \nonumber \\
&& \qquad\qquad\qquad + \lambda_-^I \Big( \delta_{IJ} \p_+ + i\p_+ X^M A_{M\, IJ} \Big) \lambda_-^J \bigg\} d^2\sigma \, ,
\eea
where $\Gamma_{(-)}$ are the Christoffel symbols minus $dB$.  Since the chiral worldsheet fermions $\psi_+^M$ and $\lambda_-^I$ are charged under spacetime Lorentz ($\Lambda$) and gauge ($\alpha$) transformations, we have to worry about an anomalous transformation of the fermion measure, and indeed it contributes
\be
\delta S_{\mathit{eff}} \propto \frac{1}{4\pi} \int_\Sigma  X^* \Big\{ \tr\big( \Lambda d\Omega_{(+)} \big) - \textrm{tr} \big( \alpha dA \big) \Big\} \, ,
\ee
where $\Omega_{(+)}$ is the torsion-free spin connection plus $dB$.\footnote{Note that this is now the opposite connection of what appears in the action.  This seems the natural convention to use, following \cite{Hull:1986xn}.}  So the fermion measure by itself is not a gauge invariant quantity, but we can make the path integral a gauge invariant quantity by defining
\be
B \rightarrow B - \frac{\alpha'}{4} \Big\{ \tr\big( \Lambda d\Omega_{(+)} \big) - \textrm{tr} \big( \alpha dA \big) \Big\}
\ee
so that a variation of the classical action cancels a variation of the measure.  This is the Green-Schwarz mechanism from the NLSM perspective, and shortly we will see the TLSM avatar of this.  This leads one to define the gauge-invariant three form
\be
H \equiv dB + \frac{\alpha'}{4} \Big\{ \Omega_{\mathit{CS}} \big( \Omega_{(+)} \big) - \Omega_{\mathit{CS}} \big( A \big) \Big\} \, ,
\ee
where $\Omega_{\mathit{CS}}$ refers to the Chern-Simons three form of the stated connection, which in turn leads to the modified Bianchi identity\footnote{In fact, $\Omega_{(+)}$ also appears more naturally in the supergravity analysis.  See \cite{Becker:2009df} for a recent discussion of possible connections to use in computing $R$.}
\be
dH = \frac{\alpha'}{4} \Big\{ \tr \Big( R\big(\Omega_{(+)}\big) \wedge R\big(\Omega_{(+)}\big) \Big) - \textrm{tr} \big( F \wedge F \big) \Big\} \, .
\ee
We will reproduce a cohomological statement of this Bianchi identity in the TLSM analysis.

\subsection{The Gauged-Linear Sigma Model for the \Ka\ Base}
\label{sec:base}

We now briefly review the structure of the standard \ZT\ GLSM that we will use to describe the \Ka\ base of our \nK\ compactifications, emphasizing the symmetries and potential anomalies of the GLSM and of the NLSM to which the GLSM flows.  For a considerably more detailed review of how an NLSM with a geometric interpretation emerges from a GLSM, see \cite{Distler:1995mi,Witten:1993yc,Morrison:1994fr}.  Our conventions for $(0,2)$ supersymmetry appear in appendix \ref{app:conventions}.

For generality, we consider an $N$-dimensional, complete intersection $S$ in an $(N+\iota)$-dimensional toric variety $T$ sitting inside $\mathbb{C}^{N+s+\iota}$, and $s$ monad gauge bundles over $S$ of ranks $r^{a'}$, $a'=1,\ldots, s$.  Thus, we start with $s$ vector superfields $V_a,~V_{-a}$, labeled by $a=1,\ldots,s$,~\,  $\iota$ fermi superfields $\tG^\mu$, $\mu=1,\ldots,\iota$, with charges $-d_\mu^a$, and $s$ chiral superfields $P^{a'}$ with charges $-n_{a'}^a$.  To this we add $N+s+\iota$ chiral superfields $\Phi^i$ labeled by $i=1,\ldots,N+s+\iota$ with charges $Q^a_i$, and $s$ sets of $r^{a'}+1$ fermi superfields $\Gamma^\ma$ labeled by $\ma=1,\ldots,r^{a'}+1$ with charges $q^a_{m_{a'}}$.  To summarize this, we write the superpotential with $U(1)^s$ charges written beneath
\be
\label{eqn:superpot}
W = \frac{1}{\sqrt{2}}\int\! \! d\tp \Big\{   \sum_\mu \underbrace{\tG^\mu}_{-d^a_\mu} \underbrace{G^\mu(\Phi)}_{d^a_\mu} + \sum_{a',m_{a'}}\underbrace{P^{a'}}_{-n_{a'}^a} \underbrace{\G^\ma}_{q_\ma^a} \underbrace{J^\ma(\Phi)}_{n^a_{a'}-q_\ma^a}   \Big\} \bigg|_{\bar{\theta}^+=0} \, .
\ee
Gauge invariance of $W$ forces the $G^\mu(\Phi)$ and $J^\ma(\Phi)$ to be quasi-homogeneous functions of the $\Phi^i$ of degree shown above.\footnote{We could be a bit more general but for simplicity, we'll work with models where $\CDB_+ \G^\ma = \CDB_+ \tG^\mu = 0$ and that lack $\Sigma$ fields (see, for example, \cite{Distler:1995mi}).}

In the ultraviolet, the $\Phi^i$ and $P^{a'}$ parameterize $\mathbb{C}^{N+2s+\iota}$.  In the geometric phase of the theory (defined when the FI parameters $r_{\mathit{FI}}^a$ are in a suitable range), the D-terms, $s$ of the F-terms, and the $U(1)^s$ symmetry combine to leave $N+\iota$ gauge-invariant, massless scalars that parameterize our toric variety $T$, while the left-handed fermions are restricted to transform as sections of the pullback of a vector bundle (or sheaf) $\V_T \rightarrow T$ of rank $\sum_{a'} r^{a'}$, defined by the $J^\ma(\phi)$.  The rest of the F-terms further restrict our massless scalars to those parameterizing the complete intersection $S=\{\vec{\phi}\, |\, G^1(\phi) =0\} \cap \ldots \cap \{ \vec{\phi}\, | \, G^\iota(\phi) =0 \} \subset T$, restricting the sheaf $\V_T$ to the sheaf $\V_S$.  In fact, we want $\V_S$ to be a bundle, so we must restrict the sets $\big\{ G^1 , \ldots , G^\iota , J^{1_{a'}} , \ldots , J^{r^{a'} + 1} \big\}$, for each $a'$, to be non-degenerate (they only all vanish when $\vec{\phi} = 0$) --- this ensures that the dimension of $\V_S$ does not jump over $S$.  We are thus left with the geometry of a (0,2) NLSM, with scalar fields parameterizing a manifold $S$, right-handed fermions transforming as sections of $T_S$, and left-handed fermions transforming as sections of a vector bundle $\V_S = \oplus_{a'} \V^{a'}_S$.

For each generator of the $U(1)^s$ gauge group, there is a corresponding generator $\eta_a$ of $H^2(T)$.  In terms of these generators, we can write the Chern characters for $T_T$, $T_S$, and $\V^{a'}_S$, as
\bea
ch(T_T) &=& \sum_i e^{\sum_a Q_i^a \eta_a} - s   \\
ch(T_S) &=& \Big[ ch(T_T) - \sum_\mu e^{\sum_a d_\mu^a \eta_a}  \Big] \Big|_S   \\
ch(\V^{a'}_S) &=& \Big[ \sum_\ma e^{\sum_a q_\ma^a \eta_a} - e^{\sum_a n_{a'}^a \eta_a} \Big] \Big|_S \, .
\eea
In particular, we see that
\bea
\label{eqn:chern}
ch_1(T_S) &=& \sum_a \Big[ \sum_i Q_i^a - \sum_\mu d_\mu^a \Big] \eta_a \big|_S    \nonumber  \\
ch_2(T_S) &=& \sum_{a,b}\Big[ \sum_i Q_i^a Q_i^b - \sum_\mu d_\mu^a d_\mu^b \Big] \eta_a \wedge \eta_b \big|_S   \nonumber  \\
ch_1(\V^{a'}_S) &=& \sum_a \Big[ \sum_\ma q_\ma^a - n_{a'}^a \Big] \eta_a \big|_S     \nonumber  \\
ch_2(\V^{a'}_S) &=& \sum_{a,b} \Big[ \sum_\ma q_\ma^a q_\ma^b - n_{a'}^a n_{a'}^b \Big] \eta_a \wedge \eta_b \big|_S \, .
\eea

Since we are working with a $(0,2)$ model, we have to worry about the anomaly under a gauge transformation which, for super gauge parameter $\Lambda_a$, is
\bea
\label{eqn:anomaly}
\ln( \delta_\Lambda {\rm Measure} ) &=& -\frac{1}{16\pi}\sum_{a,b}\left( \sum_i Q_i^a Q_i^b + \sum_{a'} n_{a'}^a n_{a'}^b - \sum_{a',\ma} q_\ma^a q_\ma^b - \sum_\mu d_\mu^a d_\mu^b \right)     \nonumber \\
&& \times \int d^2y \left[ \int d\tp \Lambda_a\Ups_b + {\rm h.c.} \right] \, ,
\eea
where $\Ups_{a}$ is the gauge fieldstrength superfield.
Comparing to (\ref{eqn:chern}), we see that the gauge anomaly is proportional to $ch_2(T_S) - ch_2(\V_S)$, a point to which we'll return later.

%%%%%%%%%%%%%%%%%%%%%  TORSION MULTIPLET  %%%%%%%%%%%%%%%%%%%%%%

\subsection{Adding the $T^{2}$ Fibration and Canceling the Anomaly}
\label{sec:torsion}

To build a $T^{2}$ fibration over the \Ka\ base constructed above, we add to the model a chiral superfield, $\Theta$, whose lowest component is a $T^{2}$-valued boson, and which shifts under gauge transformations as $\delta_{\Lambda} \Theta = - \sum_a M^a \Lambda_a$, where $M^a = M_1^a + iM_2^a$.  While $\Theta$ is gauge variant, we can create an invariant derivative of $\Theta$,
\be
\CD_- \Theta \equiv \p_- \Theta - \frac{i}{2} \sum_a M^a ( 2\p_- V_a + i V_{-a} )
\ee
which satisfies the useful relation $\DB_+ \CD_- \Theta = -\frac{i}{2}\sum_a M^a \Ups_a$.  We can thus construct a manifestly supersymmetric action for $\Theta$, the torsion multiplet, as
\bea
(4\pi) S_{\mathit{tor}}  &=&   \int \! d^2 y \, \bigg\{  - \frac{i}{2} \int \! d^2 \theta ( \THB + 2i\sum_a\MB{}^a V_a ) \CD_-\Theta     -     \frac{1}{4}  \sum_a \int \! d\thp \Theta \MB{}^a \U_a            \nonumber \\
&& \qquad\qquad  +  \p_+ \Big[ (\THB + 2i\sum_a \MB{}^aV_a) \CD_- \Theta \Big]  \bigg\} \bigg|_{\thp=\thpb=0}   +    \mathrm{h.c.} 
\eea
where we have included the total derivative to avoid ambiguities over integration by parts.  Now, using the fact that $\delta_\Lambda ( \THB + 2i\sum_a \MB{}^a V_a ) = - \sum_a \MB{}^a \Lambda_a$, we see that this supersymmetric action is {\em not} gauge-invariant, but rather transforms as,
\be
\delta_\Lambda S_{\mathit{tor}} =  \frac{1}{16\pi} \sum_{a,b} (M^a\MB{}^b + \MB{}^a M^b ) \int \! d^2 y \, d\thp \Lambda_a \U_b   +   \mathrm{h.c.}
\ee
This classical anomaly combines with the quantum anomaly of the measure (\ref{eqn:anomaly}) to give
\be
\label{eqn:full-anomaly}
\ln(\delta_\Lambda \mathrm{Measure}) + \delta_\Lambda S_{\mathit{tor}}  =  -\frac{1}{16\pi}\sum_{a,b} \mathcal{A}^{ab} \int \! d^2y \, d\thp \Lambda_a \Ups_b    +    \mathrm{h.c.}
\ee
where
\be
\label{eqn:anom} 
\mathcal{A}^{ab} = \sum_i Q_i^a Q_i^b + \sum_{a'} n_{a'}^a n_{a'}^b - \sum_{a',\ma} q_\ma^a q_\ma^b - \sum_\mu d_\mu^a d_\mu^b   -   2 M^{(a}\MB{}^{b)} \, .
\ee

It is somewhat illuminating to write the component action in Wess-Zumino gauge,\footnote{Actually, we can only fix to WZ gauge while preserving \susy\ if the total gauge anomaly (\ref{eqn:anom}) vanishes since the supersymmetry algebra in WZ gauge only closes up to a gauge transformation.  At the classical level, then, we shouldn't work in WZ gauge because we'll miss some equations of motion.  It is thus extremely useful to work in superspace without fixing any gauge. See the TeX source for more details.}
\be
S_{\mathit{tor}} = \frac{1}{4\pi} \int \! d^2 y  \bigg\{  2 \nabla_+ \vt \nabla_- \bar{\vt}  +  2 \nabla_+ \bar{\vt} \nabla_- \vt    +    2i\bar{\chi}_+ \p_- \chi_+   +   2\sum_a(M^a \bar{\vt} + \MB{}^a \vt ) v_{+-a} \bigg\}
\ee
where the only integration by parts we performed involved gauge invariant quantities: the fermion $\chi_+$ (hereafter $\chi$) and $\nabla_\pm \vt \equiv \p_\pm \vt + \sum_a M^a v_{\pm a}$.  Having fixed Wess-Zumino gauge, the only unbroken gauge symmetry is the bosonic $U(1)^s$ acting on $\vt$ as $\delta_\alpha \vt = - \sum_a M^a \alpha_a$.  Thus, $\vt$ is a dynamical axion whose superpartner $\chi$ is a free, right-handed fermion.

Since $\vt$ naturally parameterizes a $T^2$, the kinetic term for $\vt$ is the metric for a $T^2$ fibration with connection $\sum_a M^a v_{\pm a}$.  In the NLSM, this should translate into a pair of circle bundles with curvature 2-forms, $\sum_a M_l^a \eta_a$.  However for the circle bundles to be well-defined, the curvature must live in integer cohomology, so the $M_l^a$ should be suitably quantized in terms of the moduli of the $T^{2}$.  This arises as follows in the geometric (Higgs) phase of the gauge theory.  For any given gauge field configuration, the instanton number is quantized, $\frac{1}{\pi} \int \! d^2 y \, v_{+-a}  \in \mathbb{Z}$.\footnote{In our conventions, $d^2y = dy^0 dy^1 = \frac{1}{2} dy^+ dy^-$.}  For the action to be single-valued, we thus need
\be
\vt \cong \vt + 2\pi R  \cong  \vt + 2\pi (\tau_1 + i\tau_2) \, ,
\ee
which tells us that 
\be
k_1^a \equiv M_1^a R  \in \mathbb{Z}\, , \qquad   k_2^a \equiv M_1^a \tau_1 + M_2^a \tau_2 \in \mathbb{Z} \, .
\ee
So the quantization of $M^a$ depends on the moduli of the $T^2$.  Meanwhile, we also want our $U(1)^s$ to be compact, which means that we can normalize our charges so that $\alpha_a = -2\pi$ is the identity operator.  So we must also have
\be
M_1^a = l_1^a R + l_2^a \tau_1 \, ,   \qquad   M_2^a = l_2^a \tau_2
\ee
where $l_1^a, l_2^a \in \mathbb{Z}$.  From these, we learn that $\sum_l M_l^a M_l^b = k_1^a l_1^b + k_2^a l_2^b$, which is certainly compatible with integer choices for the charges determining the base $S$ and gauge bundle $\V_S$.  The $T^2$ fibration then has only one continuous parameter and a few integer parameters that are constrained by the above equations as well as anomaly cancelation.

If we define $\xi \equiv \tau_1 R$, then we can rewrite the constraints as
\be
R^2 = \frac{k_1^a - l_2^a \xi}{l_1^a}  \qquad\quad  \textrm{and}  \qquad\quad  |\tau|^2 = \frac{k_2^a - l_1^a \xi}{l_2^a}  \, .
\ee
When $\xi \neq 0$, for this to be true for all $a=1,\ldots,s$, we must have that $k_l^a = \mathscr{M}^a k_l$ and $l_l^a = \mathscr{M}^a l_l$, where $l_1,~l_2,~k_1,$ and $k_2$, are integers with greatest common divisor $1$, and $\mathscr{M}^a \in \mathbb{Z}$.  When $\xi=0$, we can instead have $l_l^a = \mathscr{M}_l^a l_l$ and $k_l^a = \mathscr{M}_l^a k_l$, where $l_1$ and $k_1$ are relatively prime, $l_2$ and $k_2$ are relatively prime, and $\mathscr{M}_l^a \in \mathbb{Z}$.  For the examples in this paper, we will restrict ourselves to the rectangular $T^2$ case where $\xi = 0$.  In this case, we have
\be
\label{eqn:rect-t2-radii}
M_1^a = \mathscr{M}_1^a \frac{k_1}{R}  =  \mathscr{M}_1^a l_1 R    \qquad\quad \textrm{and}  \qquad\quad   M_2^a = \mathscr{M}_2^a \frac{k_2}{\tau_2} = \mathscr{M}_2^a l_2 \tau_2 \, .   
\ee

Thus, as explained more thoroughly in \cite{Adams:2006kb}, we have reproduced the structure of the torsional solutions \cite{Fu:2006vj}.  One further check comes from noticing that our anomaly cancelation condition
\be
\label{eqn:anomaly-t2-bundle}
\mathcal{A}^{ab} \eta_a \wedge \eta_b \Big|_S = ch_2(T_S) - ch_2(\V_S) - 2\sum_l \omega_l \wedge \omega_l  =  0
\ee
implies the integrated modified Bianchi identity of \cite{Becker:2006et}, where $\omega_l \equiv \sum_a M_l^a \eta_a \big|_S$ is the curvature of the circle bundle and $\alpha'=1$ in our conventions.

\section{Computing the Low-Energy Spectrum (a TLSM Toolkit)}

So far, we have constructed a TLSM that, in the geometric phase, flows in the IR to a NLSM describing a class of non-K\"ahler $T^2$ bundles.  Since this provides a microscopic description of such a compactification, the existence of this TLSM demonstrates the existence these non-K\"ahler compactifications perturbatively in $\alpha'$, and we have reason to believe non-perturbatively as well (see our comments in section 5 of \cite{Adams:2006kb}).  This is already exciting news, but we'd like to go further and use this microscopic description to get a handle on the physics of the compactification; in particular, we will focus on the massless spectrum.

To understand the massless spectrum of the theory, we need to understand which states in the TLSM descend to massless states in the IR NLSM, but to do this we have to understand what their $L_0$ and $\tilde{L}_0$ eigenvalues will be.  Fortunately, \cite{Kachru:1993pg,Distler:1993mk,Silverstein:1994ih} have laid the groundwork for ordinary GLSMs that we can modify for our TLSMs.

The basic idea for finding $\tilde{L}_0=0$ states in the UV is to conjecture that the supercharges $Q_+,~ \QB_+ $, become the zero-modes of the right-moving supercurrents in the (0,2) superconformal algebra in the IR.  In that case, $\tilde{L}_0=0$ states in the right-moving Ramond sector are in one-to-one correspondence with representatives of $\QB_+$-cohomology in the UV theory.

For $L_0=0$, we don't have left-handed supercharges to come to the rescue.  Instead, as in  \cite{Silverstein:1994ih} for GLSMs, we'll identify a chiral operator in the UV theory that has the OPE of a stress-tensor.  Its existence will depend on the existence of a non-anomalous $U(1)_R$ symmetry, but nothing more, and the central charge in the $T_{--}T_{--}$ OPE will depend on the R-charges.  The fact that $T_{--}$ is chiral suggests that it survives to the IR theory, though it could pair up with another chiral operator to become massive.  In the grand tradition of (0,2) GLSMs, however, we will conjecture that this operator $T_{--}$ survives to the IR theory to generate the left-moving conformal algebra and we'll use it to compute $L_0$ eigenvalues of our $\QB_+$-cohomology representatives.  The $U(1)_R$ symmetry becomes part of the right-moving superconformal algebra and is used in the right-moving GSO projection.

Similarly, we can find operators $J^a_L$ with the OPEs (with each other and with $T$) of $U(1)_L$ currents that we can use in the IR theory to implement GSO projections.  In section \ref{sec:IR-algebra}, we'll identify the chiral operators in the UV theory corresponding to $T$ and $J_L^a$ and use them to find the central charges and vector-bundle ranks of the conjectured IR SCFT.  In section \ref{sec:charges}, we'll analyze the implications of anomaly cancelation (both gauge and conformal anomalies) on consistent charge assignments in our theories and will find some restrictions.  In particular,  in many cases we will be forced to choose $c_1(T_S)=0$.  In section \ref{sec:LG}, we will describe how to extract the massless spectrum from the Landau-Ginzburg phase.

%%%%%%%%%%%%%%%%%%%%%%%  IR ALGEBRA  %%%%%%%%%%%%%%%%%%%

\subsection{Infrared Algebra}
\label{sec:IR-algebra}

\emph{For ease of exposition, in this subsection all chiral superfields will be labeled $\Phi^i$ with charges $Q_i^a$, fermi superfields $\G^m$ with charges $q_m^a$, and the torsion multiplet is still $\Theta$ with shift-charge $M^a$.  Additionally, we will call the superpotential}
\be
\label{eqn:superpot-temp}
W = \frac{1}{\sqrt{2}}\int d\thp \sum_m \G^m F^m(\Phi)
\ee
\emph{and will relax the chirality constraint on $\G^m$ to $\CDB_+ \G^m = \sqrt{2} E^m(\Phi)$, where \newline $\sum_m E^m(\Phi) F^m(\Phi)=0$.}

\newcommand{\J}{{\mathscr{J}}}
\newcommand{\Jb}{{\bar{\mathscr{J}}}}
\newcommand{\Lb}{{\overline{\Lambda}}}
\newcommand{\Phib}{{\overline{\Phi}}}
\newcommand{\Gb}{{\overline{\Gamma}}}

As a first step toward computing the exact massless spectrum, we would like to identify right-chiral operators in the massive theory whose OPEs flow to those of a left-moving $U(1)_{L}$ current, $J_{(\b)}$,  and a left-moving stress tensor, $T_{--}$, in the deep IR.  To this end, consider the gauge-invariant operator,
\be
\label{eqn:U1L-operator}
J_{(\beta)}    =       -{\textstyle \sum_i} \beta_i \Phi^i \CD_- \overline{\Phi}{}^i   +   \tfrac{i}{2} {\textstyle \sum_m} \beta_m \G^m \GB{}^m    +   i\overline{\beta} \CD_- \Theta  +   i\beta \CD_- \THB   \,. % \nonumber \\
%&& ~ -  \tfrac{1}{4} ( \overline{\beta} M^a + \beta \MB{}^a ) \big( 2 \p_- V_a + i V_{-a} \big) \textrm{ \Large :}  
\ee
We will first check whether it's chiral at tree level by using the classical equations of motion
\bea
\CDB_+ \bbar{\Gamma}^m & = & \sqrt{2} F^m(\Phi)   \\
\CDB_+ \CD_- \bbar{\Phi}{}^i & = & \frac{i}{\rt}\sum_m \left( \frac{\p E^m(\Phi)}{\p\Phi^i} \bbar{\Gamma}{}^m + \frac{\p F^m(\Phi)}{\p \Phi^i}\Gamma^m \right) \\
\DB_+ \CD_- \THB  & = &  - \tfrac{i}{2} \sum_a \MB{}^a \Ups_a \\
\label{eqn:gauge-eom}
-\frac{1}{2e_a^2}\p_- \DB_+ \bbar{\Ups}_a & = & 2i\sum_i Q_i^a\Phi^i\CD_-\bbar{\Phi}{}^i + \sum_m q_m^a\Gamma^m\bbar{\Gamma}{}^m + 2\CD_- ( M^a \bbar{\Theta} + \MB{}^a \Theta )   \nonumber \\
&& + i\sum_b M^{(a}\bbar{M}{}^{b)}(2\p_- V_{b} + iV_{-b}) \, ,
\eea
as well as the identity $\DB_+ \CD_- \Theta = - \frac{i}{2} \sum_a M^a \Ups_a$.  Applying $\DB_+$ to the gauge field equation of motion, using the matter equations of motion, and recalling the quasi-homogeneity properties of $E^m(\Phi)$ and $F^m(\Phi)$
\be
\sum_i Q_i^a \Phi^i \p_i E^m  =  q_m^a E^m  \, ,   \qquad   \sum_i Q_i^a \Phi^i \p_i F^m = - q_m^a F^m \, ,
\ee
we learn that the equations of motion also imply that $\sum_b M^{(a}\MB{}^{b)}\Ups_b = 0$, which holds if and only if $\sum_a M^a \Ups_a = \sum_a \MB{}^a\Ups_a = 0$.  This is just a reflection of the fact that gauge transformations satisfying $\sum_a M^a\Lambda_a = \sum_a \MB{}^a\Lambda_a=0$ are symmetries of the classical action whereas the others cancel against the variation of the measure.  Said another way, the combinations $M_l^a v_{\pm a}$ are classically massive vector fields.  We see, then, that for our current
\be
\DB_+ J_{(\beta)} \big|_{EOMs}  =  -\frac{i}{2\sqrt{2}} \sum_{m} \bigg[ \Big( {\textstyle \sum_i} \, \beta_i \Phi^i \p_i E^m - \beta_m E^m \Big) \GB{}^m  + \Big( {\textstyle \sum_i} \, \beta_i \Phi^i \p_i F^m + \beta_m F^m \Big) \G^m \bigg]   \, .
\ee
For the classical action to be invariant under a global $U(1)_L$ action with charges $\beta_i,\,\beta_m\,,\beta$, the functions $E^m(\Phi)$ and $F^m(\Phi)$ must be quasi-homogeneous with weights $\beta_m$ and $-\beta_m$, respectively.  It's not surprising, then, that this is the same condition we need in order for our current to be classically chiral $\DB_+ J_{(\beta)} \big|_{EOMs} = 0$.

Next, we should check whether our current remains chiral at the quantum level.  As in \cite{Silverstein:1994ih}, we can check the chirality by computing $\DB_+ J_{(\beta)}$ within a correlation function and making use of the fact that we'll be concerned only with supersymmetric vacua.  In this case, we know that $\langle [ \hat{\DB}_+, J_{(\beta)}(x) ] \mathcal{O}(y) \rangle = - \langle J_{(\beta)}(x) [ \hat{\DB}_+, \mathcal{O}(y) ]_\pm \rangle$, so we can choose an operator $\mathcal{O}$ whose transformation we know exactly, {\it e.g.} one of the fundamental fields (by $\hat{\DB}_+$, we mean the charge generating the action of $\DB_+$).  Again, we follow \cite{Silverstein:1994ih} and choose $\mathcal{O} = \p_- \overline{\Ups}_a$.  Making use of the gauge field equation of motion, as well as the free field OPEs (which we'll define through Wick rotation from the Euclidean version)
\bea
\phi^i (y) \bar{\phi}{}^j(0)  &\sim& -\delta^{ij}\ln(y^+ y^-)      \nonumber  \\
\gamma^m(y) \bar{\gamma}{}^n(0) &\sim& -\frac{2i}{y^-}\delta^{mn}    \nonumber  \\
\vt (y) \bar{\vt} (0) &\sim & -\ln(y^+y^-)      \nonumber  \\
\lambda_a (y) \bar{\lambda}_b(0) &\sim& -\frac{2ie^2}{y^-}\delta_{ab} 
\eea
we find that
\bea
\big\langle \DB_+ J_{(\beta)}(x) \p_- \overline{\Ups}_a(y) \big\rangle &\approx& - \big\langle J_{(\beta)}(x) \DB_+ \p_-\overline{\Ups}_a(y) \big|_{EOMs} \big\rangle   \nonumber \\
& \sim & (4ie^2) \frac{\sum_i \beta_i Q^a_i - \sum_m \beta_m q^a_m - \beta \MB{}^a - \bar{\beta} M^a }{(x^- - y^-)^2} + \ldots
\eea
where $\ldots$ are terms either less singular in $(x-y)$, or terms containing higher powers of the coupling constants $e$ or $\mu$.\footnote{$\mu$ is a parameter with dimensions of mass that appears in front of the superpotential --- we have implicitly absorbed it into the definition of $F^m(\Phi)$, but it is there and flows to zero in the ultraviolet.}  Again appealing to the free field OPEs, this leads us to the identification
\be
\DB_+ J_{(\beta)} \sim -\frac{1}{2} \Big( \sum_i \beta_i Q^a_i - \sum_m \beta_m q^a_m - \beta \MB{}^a - \bar{\beta}M^a \Big) \Ups_a\, .
\ee
This has precisely the same form as the variation of the effective action under this global $U(1)_L$ (similar to the gauge anomaly (\ref{eqn:full-anomaly})), so we find that when this $U(1)_L$ is non-anomalous, the operator $J_{(\beta)}$ is part of the chiral algebra.

The $U(1)$ currents that will concern us in our models will be related to the gauge symmetries and, as such, will encode information about the spacetime vector bundles.  In particular, the most singular term in the $J_{(\beta)} J_{(\beta)}$ OPE, $J_{(\beta)}(y) J_{(\beta)}(0) \sim \frac{r_L^{(\beta)}}{(y^-)^2}+\ldots$, determines the rank of the associated vector bundle; a simple computation gives,
\be
r_L^{(\beta)} =  \sum_m \beta_m \beta_m  +  2|\beta|^2 - \sum_i \beta_i \beta_i \, .
\ee

In a similar vein, we can define an operator $T_{--}$ as
\bea
\label{eqn:stress-tensor}
T_{--} &=& - \frac{i}{8e^2}\sum_a \Ups_a \p_- \overline{\Ups}_a - \sum_i \CD_- \Phi^i \CD_- \Phib{}^i - \frac{i}{4} \sum_m \G^m \CD_- \Gb{}^m + \frac{i}{4}\sum_m \CD_- \G^m \Gb{}^m   -  \CD_- \Theta \CD_- \overline{\Theta} \nonumber \\
&& + \p_- \bigg( \sum_i \frac{\alpha_i}{2}\Phi^i \CD_- \Phib{}^i - \frac{i}{4}\sum_m  \alpha_m \G^m \Gb{}^m - \frac{i}{2} \CD_- \big( \bar{\alpha} \Theta + \alpha \overline{\Theta} \big) \bigg)  \, .
\eea
When the $\alpha_i$ and $\alpha_m$ correspond to the charges of $\Phi^i$ and $\G^m$ under a classical $U(1)$ R-symmetry (so that $E^m(\Phi)$ and $F^m(\Phi)$ are quasi-homogeneous of degrees $1+\alpha_m$ and $1-\alpha_m$, respectively), then the classical equations of motion yield
\be
\DB_+ T_{--} \big|_{EOMs} = 0  \, .
\ee
Computing $\langle \DB_+ T_{--}(x) \p_- \overline{\Ups}_a (y) \rangle$, as we did for $J_{(\beta)}$, suggests from the most singular terms the identification
\be
\DB_+ T_{--} \sim \frac{1}{4}\sum_a \Big( \sum_i ( \alpha_i - 1)Q_i^a - \sum_m \alpha_m q_m^a - \alpha \MB{}^a - \bar{\alpha}M^a \Big) \p_- \Ups_a \, ,
\ee
which vanishes whenever our $U(1)_R$ symmetry is non-anomalous.

We can evaluate the $T_{--}T_{--}$ OPE in the ultraviolet to check whether it is warranted to conjecture that $T_{--}$ corresponds to a left-moving stress tensor in the infrared.  Doing so, we find that it has the OPE of a stress tensor
\be
T_{--}(y) T_{--}(0) \sim \frac{c_L}{2(y^-)^4} + \frac{2}{(y^-)^2} T_{--}(0) + \frac{1}{y^-} \p_- T_{--}(0) + \ldots .
\ee
with central charge
\be
c_L =  \sum_i (3(\alpha_i-1)^2 - 1) + \sum_m (1 - 3 \alpha_m^2) + (2 - 6 |\alpha|^2) - \sum_a 2 \, .
\ee
Next, we can check whether our currents $J_{(\beta)}$ are left-moving, dimension $1$ currents under this stress tensor,
\be
T_{--}(y) J_{(\beta)}(0) \sim \frac{\sum_i \beta_i(\alpha_i - 1) - \sum_m \beta_m \alpha_m - \alpha\bar{\beta} - \bar{\alpha}\beta }{(y^-)^3}  +  \frac{1}{(y^-)^2} J_{(\beta)}(0) + \frac{1}{y^-} \p_- J_{(\beta)}(0) + \ldots .
\ee
This has the expected form when the $\frac{1}{(y^-)^3}$ term vanishes, which occurs exactly when we can non-anomalously gauge the $U(1)_L$ symmetry while maintaining the $U(1)_R$ symmetry at the quantum level.  This means that in our infrared theory, we will have $U(1)_L$ and $U(1)_R$ currents that have no OPE with each other, supporting the identification of $U(1)_R$ as right-moving and $U(1)_L$ as left-moving in the infrared CFT.  Finally, assuming that the $U(1)_R$ current becomes part of the $(0,2)$ superconformal algebra in the infrared, we can compute $\hat{c}_R$ using the fact that $\bar{J}_R (y) \bar{J}_R(0) \sim \frac{\hat{c}_R}{2(y^+)^2}+\ldots$ in a $(0,2)$ SCFT, yielding
\be
\frac{\hat{c}_R}{2} = \sum_i (\alpha_i-1)^2 - \sum_m \alpha_m^2  + \underbrace{1}_{\chi_+} - 2|\alpha|^2  - \sum_a \underbrace{1}_{\lambda_{-a}}   \, .
\ee

%%%%%%%%%%%%%%%%%% RELATION TO NOETHER %%%%%%%%%%%%%%%%%

\subsubsection{Relation to Na\"ive Algebra}

As in \cite{Kachru:1993pg}, we would like to relate these chiral operators to the $U(1)_L$ charge and stress tensor that one would derive from a Noether procedure.  We will find that they differ by $\overline{Q}_+$-exact terms and total derivatives so, on $\overline{Q}_+$-cohomology, our chiral operators will implement the same actions as the Noether charges.  This is an important observation because the massless spectrum of the compactified theory is in one-to-one correspondence with $\overline{Q}_+$-cohomology, so these are precisely the states that will interest us in this paper.  First, the $U(1)_L$ charge that one would derive from a Noether procedure is
\be
\label{eqn:noether-u1}
j'_{(\beta)}  =  \frac{1}{2}\int \! dy^1  \Big[ \sum_i \beta_i \big( \bar{\phi}{}^i \!\stackrel{\leftrightarrow}{\nabla}_0\! \phi^i   - i  \bar{\psi}^i \psi^i \big)  - i \sum_m \beta_m \bar{\gamma}{}^m \gamma^m   +   i \nabla_0 \big( \beta \bar{\vt} + \bar{\beta} \vt \big) \Big] \, ,
\ee
which is related to the charge we would derive from (\ref{eqn:U1L-operator}), $j_{(\beta)} = \int dy^1 J_{(\beta)}$, by
\be
j_{(\beta)} = j'_{(\beta)}  - \frac{i}{2} \int \! dy^1 \, \nabla_1 \big( \beta \bar{\vt} + \bar{\beta} \vt \big)   +  \Big\{ \overline{Q}_+ \,  , \,  \frac{i}{2\sqrt{2}}\int\! dy^1 \sum_i \beta_i \bar{\phi}{}^i  \psi^i \Big\}   \, .
\ee
So we see that $j_{(\beta)}$ and $j'_{(\beta)}$ differ by total derivatives and $\overline{Q}_+$-exact terms.

Similarly, the left-moving $L'_0 = H-P$ that one derives from the Noether procedure is
\bea
\label{eqn:noether-stress-tensor}
L'_0 &=& \int \! dy^1 \bigg[ - \sum_i \nabla_- \phi^i \nabla_- \bar{\phi}{}^i  + \frac{1}{4}\sum_m \Big( i \bar{\gamma}{}^m \!\stackrel{\leftrightarrow}{\nabla}_1 \!\gamma^m  +  |G^m|^2 - |E^m|^2 - G^m F^m - \bar{G}{}^m \bar{F}{}^m \Big)    \nonumber \\
&&  \quad\qquad  - \nabla_- \vt \nabla_- \bar{\vt} + \frac{1}{8e^2} \sum_a \Big( 2i \bar{\lambda}_a \!\stackrel{\leftrightarrow}{\p}_1\!\lambda_a +  D_a^2 - 4(v_{+-a})^2 - 2e^2 r_{\textit FI}^a D_a \Big)       \nonumber \\
&& \quad\qquad +    \frac{1}{4} \sum_{i,a} \Big(  i\sqrt{2} Q_i^a \bar{\phi}{}^i \psi^i \lambda_a  + i \sqrt{2} Q_i^a \phi^i \bar{\psi}{}^i \bar{\lambda}_a  + Q_i^a D_a |\phi^i|^2 \Big)      \nonumber \\
&&  \quad\qquad  +  \frac{1}{4}\sum_{i,m} \Big(  \psi^i \bar{\gamma}{}^m \p_i E^m - \bar{\psi}{}^i \gamma^m \bar{\p}_i \bar{E}{}^m  +  \psi^i \gamma^m \p_i F^m  -  \bar{\gamma}{}^m \bar{\psi}{}^i \bar{\p}_i \bar{F}{}^m \Big)  \bigg]  \,  .
\eea
We would like to relate this to (\ref{eqn:stress-tensor}).  To that end, note that in the second line of (\ref{eqn:stress-tensor}) we have a term of the form $-\frac{1}{2}\p_- J_{(\alpha)}$, where $J_{(\alpha)}$ is not chiral since the $\alpha_i$ correspond to R-charges.  Up to a boundary term, 
\bea
-\frac{1}{2} \int dy^1 \p_- J_{(\alpha)} &=& -\frac{1}{2} \int dy^1 \p_+ J_{(\alpha)} = -\frac{1}{8i} \int dy^1 \big[ \{ Q_+, \overline{Q}_+ \} , J_{(\alpha)} \big]        \nonumber \\
&= &    -\frac{1}{8i} \int dy^1 \Big( \big\{ Q_+ ,   \big[\overline{Q}_+, J_{(\alpha)} \big]   \big\}   +  \big\{ \overline{Q}_+, \big[ Q_+, J_{(\alpha)} \big] \big\}  \Big)      \, .
\eea
Again, we ignore $\overline{Q}_+$-exact terms since we will work in cohomology.  Using the equations of motion, we compute that
\be
-\frac{1}{2} \int dy^1 \p_- J_{(\alpha)} \Big|_{\textit EOMs}  \cong  -\frac{1}{8} \int dy^1 \sum_m \Big[  |F^m|^2 + |E^m|^2 -\sum_i \big( \psi^i \gamma^m \p_i F^m + \psi^i \bar{\gamma}{}^m \p_i E^m \big)  \Big] \, .
\ee
Next, we rewrite $L_0'$ using the following: $\int \!dy^1 \bar{\lambda}_a \!\!\stackrel{\leftrightarrow}{\p}_1\! \!\lambda_a  = \int \!dy^1 \big( \bar{\lambda}_a \!\!\stackrel{\leftrightarrow}{\p}_1\! \!\lambda_a  -  \p_1 (\bar{\lambda}_a \lambda_a)  \big)   = \int\! dy^1 \big( 2 \lambda_a \p_+ \bar{\lambda}_a - 2 \lambda_a \p_- \bar{\lambda}_a \big)$, $\int \!dy^1 \bar{\gamma}{}^m \!\stackrel{\leftrightarrow}{\nabla}_1\! \gamma^m  =  \int \! dy^1 \bar{\gamma}{}^m \big( \!\stackrel{\leftrightarrow}{\nabla}_+ \! - \! \stackrel{\leftrightarrow}{\nabla}_- \! \big) \gamma^m$, and the equations of motion for $\gamma^m,~G^m,~\lambda_a,~D_a$, and complex conjugates.  Doing so, we find that
\bea
L'_0 \big|_{\textit EOMs}  &=&  L_0 + \bigg\{  \overline{Q}_+ \, ,  \,  \int\! \frac{dy^1}{8}\bigg[ 2i\sum_{i,a} Q_i^a |\phi^i|^2 \bar{\lambda}_a - \frac{1}{\sqrt{2}} \sum_{m} \big( \gamma^m \bar{E}{}^m + \bar{\gamma}{}^m \bar{F}{}^m  \big)       \nonumber \\
&&  \qquad\qquad\qquad\qquad~~  + \frac{i}{e^2}\sum_a \bar{\lambda}_a \big( 2iv_{+-a} - 2e^2 r_{\mathit{FI}}^a - D_a  \big)    \bigg] \bigg\}
\eea
so that, up to equations of motion and boundary terms, $L'_0 \cong L_0$ on ${\cal H}_{\overline{Q}_+}$.

The purpose in all of this is that, on the one hand, we have chiral operators $T_{--}$ and $J_{(\beta)}$ that have the operator algebra of a left-moving stress tensor and $U(1)$ current in a conformal theory while, on the other hand, we can relate the charges generated by these currents acting on $\QB_+$-cohomology to the charges generated by the full-fledged Noether charges of the theory.  When we are interested in the charges of states in the theory, it will be more convenient to use the Noether charges (\ref{eqn:noether-u1}) and (\ref{eqn:noether-stress-tensor}).  In fact, it is worth noting that rescaling the superpotential can be undone by an appropriate rescaling of the fields which, in turn, rescales the kinetic terms and therefore corresponds to a $\QB_+$-exact deformation.  That means that when working within $\QB_+$-cohomology, the terms in $L'_0$ containing $E^m$ and $F^m$ will not contribute and so we can work with a simplified version of $L'_0$ by dropping these terms (this can also be seen from the fact that neither appears in $L_0$).

\subsubsection{Supercharge}

\newcommand{\mathcalQB}{\overline{\mathcal{Q}}}

We can also derive a Noether supercurrent and associated supercharge from our action, 
\bea
J^{\mathcalQB_+}_-  &=&  \sum_a \lambda_a \Big(  \frac{2i}{e^2} v_{+-a} -  r_{\textit FI}^a + \sum_i Q_i^a |\phi^i|^2 \Big)  - \sqrt{2}i \sum_m \big( \bar{\gamma}{}^m E^m + \gamma^m F^m \big)    \nonumber \\
&=& -\frac{1}{e^2}\sum_a \lambda_a \big(  D_a - 2i v_{+-a}  \big)  - \sqrt{2}i \sum_m \big( \bar{\gamma}{}^m E^m + \gamma^m F^m \big)     \\
J^{\mathcalQB_+}_+ &=&  2\sqrt{2} \Big(  \sum_i \bar{\psi}{}^i \nabla_+ \phi^i   +   \bar{\chi} \nabla_+ \vt \Big)    \\
\mathcalQB_+  &=&   \int \! dy^1 \Big(      J^{\mathcalQB_+}_+  +  J^{\mathcalQB_+}_-    \Big) \, .
\eea
We will use this supercharge in computing the massless spectrum.

%%%%%%%%%%%%%%%%%%%%%%  CONSISTENT CHARGES  %%%%%%%%%%%%%%%%%

\subsection{Consistent $U(1)_L$ and $U(1)_R$ Charges}
\label{sec:charges}

\emph{Note: We now return to the notation of subsection \ref{sec:base}, with superpotential given by}
\be
\label{eqn:superpot2}
W = \frac{1}{\sqrt{2}}\int\! \! d\thp \Big\{   \sum_\mu \underbrace{\tG^\mu}_{-d^a_\mu} \underbrace{G^\mu(\Phi)}_{d^a_\mu} + \sum_{a',\ma} \underbrace{P^{a'}}_{-n_{a'}^a} \underbrace{\G^\ma}_{q_\ma^a} \underbrace{J^\ma(\Phi)}_{n^a_{a'}-q_\ma^a}   \Big\} \bigg|_{\bar{\theta}^+=0} \, .
\ee

The most general $U(1)_L$ and $U(1)_R$ charges of the scalar superfields $\Phi^{i}$ consistent with a completely generic superpotential (\ref{eqn:superpot2}) are linear combination of their gauge charges, and similarly for the other fields.\footnote{Of course, specific choices of polynomials $G^\mu(\phi)$ and $J^m_{a'}(\phi)$ may be covariant under other $\Phi^i$ charge assignments --- we make this assumption for simplicity only.  It is straightforward to relax this assumption.}  Let us summarize the charge assignments in a table
\be
{\setlength\arraycolsep{8pt}
\begin{array}{|c||c|c|c|c|c|} \hline    & \Phi^i & P^{a'} & \tG^\mu & \G^\ma & \Theta       \\
\hline\hline U(1)_a  &  Q_i^a  &  -n_{a'}^a  &  -d_\mu^a  &  q_\ma^a  &  (M_l^a)_s    \\
\hline U(1)_{L_a}  & \beta^a_b Q_i^b &  \rho_{a'}^a - \beta^a_b n_{a'}^b  &  - \beta^a_b d^b_\mu  &  \beta^a_b q^b_\ma - \rho_{a'}^a  &  (\beta^a_b M_l^b + m_l^a)_s  \\
\hline U(1)_{R}  &   \alpha_a Q_i^a &  1 + \sigma_{a'} -  \alpha_a n_{a'}^a  & 1 -  \alpha_a d^a_\mu  &   \alpha_a q^a_\ma  - \sigma_{a'}  &  ( \alpha_a M_l^a + \tilde{m}_l)_s   \\
\hline
\end{array}}  \nonumber
\ee
where sums over repeated indices are implicit.  Now we have to check the various anomalies, as well as the factorization of $U(1)_{L_a}$ and $U(1)_R$ into left-moving and right-moving currents in the infrared:
\bea
\label{eqn:GG}
G_a G_b : & \quad &  Q_i^a Q_i^b + n_{a'}^a n_{a'}^b - d_\mu^a d_\mu^b - \sum_{a'} q_\ma^a q_\ma^b - 2 M_l^a M_l^b  = 0   \\
\label{eqn:GL}
G_a L_b :  &  \quad  & \sum_{a'} \left( - n^a_{a'}  + \sum_\ma q_\ma^a \right) \rho_{a'}^b - 2 M_l^a m_l^b  =  0  \\
\label{eqn:GR}
G_a R :  &  \quad &  \left( \sum_\mu d_\mu^a -  \sum_i Q_i^a  \right) -  \sum_{a'} \sigma_{a'} \left(  n_{a'}^a - \sum_\ma q_\ma^a \right) - 2 M_l^a \tilde{m}_l  =  0   \\
\label{eqn:RL}
R L_a :  & \quad &  \sum_{a'} \sigma_{a'} \rho_{a'}^a \left( 1 - \sum_{\ma} 1 \right)  -  2 \tilde{m}_l m_l^a =  \sigma_{a'} \rho_{a'}^a (-r^{a'}) - 2\tilde{m}_l m_l^a  = 0
 \eea
where in (\ref{eqn:GL}) and (\ref{eqn:GR}) we made use of (\ref{eqn:GG}), and in (\ref{eqn:RL}) we used (\ref{eqn:GG})--(\ref{eqn:GR}).

We also wish to demand that our TLSM yield a NLSM with our desired central charges $\hat{c}_R = 2N+2$, $c_L = 2N+2+\sum_{a'} r^{a'}$, and vector bundles of rank $r^{a'}_L=r^{a'}$.  Demanding $\hat{c}_R = 2N+2$
\bea
\frac{\hat{c}_R}{2} &=& \sum_i 1 + \sum_{a'} \sigma_{a'}^2 - \sum_\mu 1 - \sum_{a',m} \sigma^2 - 2 \tilde{m}_l \tilde{m}_l \underbrace{+1}_{\textrm{from }\chi\textrm{ in }\Theta} - \sum_a 1    \nonumber \\
&=& N + \sum_{a'} \sigma_{a'}^2 (-r^{a'}) - 2\tilde{m}_l\tilde{m}_l + 1 = N + 1
\eea
tells us that $\sigma_{a'} = \tilde{m}_l = 0$.  This forces the base $S$ to be Calabi-Yau
\be
G_a R :   \quad  \sum_\mu d_\mu^a - \sum_i Q_i^a = 0  ~~~ \Longrightarrow ~~~ c_1(T_S) = 0 ~.
\ee
So, unless there exists an R-charge assignment that is compatible with $G^\mu(\phi)$ and $J^\ma(\phi)$ and is \emph{not} a linear combination of the gauge charges, the base must be Calabi-Yau; we hope to explore such examples in future work.  The value $c_L = 2N+2+\sum_{a'} r^{a'}$ follows from the $\hat{c}_R$ computation since $c_L$ is computed from the R-charges.

In the examples below, to have more confidence that there exists an infrared fixed point, we will restrict our attention to models with a pure Landau-Ginzburg phase where all the $p^{a'}$ get VEVs.  We should thus assign $U(1)_{L_a}$ and $U(1)_R$ charges such that the $p^{a'}$ are invariant (otherwise, their VEVs would break these symmetries), which restricts us to
\be
\label{eqn:u1r-constraint}
\sum_a \alpha_a n^a_{a'} = 1  \, ,   \qquad   \quad \rho^a_{a'}  =  \sum_b \beta^a_b n^b_{a'} \, .
\ee
The simplest way to satisfy the first constraint is by setting $\alpha_a = \sum_{a'} (n^{-1})^{a'}_a$, which we shall do for the remainder of the paper (we must choose $n^a_{a'}$ to be invertible, otherwise some linear combination of the $p^{a'}$ would be uncharged which would violate the assumption of a pure LG phase).  In the anomaly constraints, there is an invariance under rescaling our $U(1)_{L_a}$ charges which corresponds to rescaling $\beta^a_b$ and $m_l^a$.  We can fix this rescaling symmetry by demanding $r_L^{a'} = r^{a'}$, where $r_L^{a'}$ is the coefficient of the most singular term in the $J_{L_{a}} J_{L_{a'}}$ OPE:
\be
r^a_L \equiv \sum_{a',b,c} r^{a'} \beta^a_{b}\beta^a_c n^b_{a'}n^c_{a'} + 2m_l^a m_l^a = r^a \, .
\ee
The reason for choosing this scaling is so that we can identify $J_{L_a}$ with the fermion number current of a certain subset of the $r^a$ left-moving fermions in the infrared CFT.  The most obvious way to satisfy this constraint is by choosing $m_l^a = 0$ and $\beta^a_b = \big(n^{-1}\big)^a_b$, in which case we are forced to set $\sum_{m_{a'}} q^a_{m_{a'}} - n^a_{a'} = 0$, or $c_1(\V^a_S) = 0$, but this is not necessary.  In the examples in this paper we will make this choice, so the charge assignments will be
\be
\label{eqn:our-charges}
{\setlength\arraycolsep{0pt}
\begin{array}{|c||c|c|c|c|c|} \hline    & \Phi^i & P^{a'} & \tG^\mu & \G^\ma & \Theta       \\
\hline\hline U(1)_a  &  Q_i^a  & ~ -n_{a'}^a ~ &  -d_\mu^a  &  q_\ma^a  &  (M_l^a)_s    \\
\hline ~ U(1)_{L_a} ~ &  (n^{-1})^a_b Q_i^b &  0  &  - (n^{-1})^a_b d^b_\mu  &  (n^{-1}) q^b_\ma - \delta_{a'}^a  &  ((n^{-1})^a_b M_l^b )_s  \\
\hline U(1)_{R}  &  ~ \sum_{a'}(n^{-1})^{a'}_a Q_i^a ~ &  0  & ~1 - \sum_{a'}(n^{-1})^{a'}_a d^a_\mu  ~&~  \sum_{b'}(n^{-1})^{b'}_a q^a_\ma  ~ &~  (\sum_{a'}(n^{-1})^{a'}_a M_l^a )_s ~  \\
\hline
\end{array}} 
\ee
where sums over repeated indices are implicit.

To summarize, unless the $G^\mu(\phi)$ and $J^\ma(\phi)$ admit R-charge assignments for the $\Phi^i$ that are \emph{not} a linear combination of the gauge charges, our models will only be consistent and describe the theory we desire when $c_1(T_S) =  0$.  Additionally, we will constrain ourselves to models where $c_1(\V^a_S)=0$.  It is important to remember that it may be possible to relax both of these conditions, and we hope to find and analyze such examples in the future.

%%%%%%%%%%%%%%%%%%%%%  LG-Orbifold  %%%%%%%%%%%%%%%%%

\subsection{Landau-Ginzburg Orbifold and Massless Spectrum}
\label{sec:LG}

We are interested in vacua annihilated by $\overline{Q}_+$, so we must demand that our vacua satisfy
\be
{\setlength\arraycolsep{8pt}
\begin{array}{lclcl}
D_a - 2i v_{+-a} = 0 \, ,     &   \qquad   &     \nabla_+ \phi^i = 0  \, ,    &   \qquad   &   \nabla_+ p^{a'} = 0  \, ,   \\
G^\mu(\phi) = 0  \, ,   &   \qquad   &   \sum_{a'} p^{a'} J^{m_{a'}}(\phi)   =  0  \, ,   & &   \nabla_+ \vt = 0  \, .
\end{array}}
\ee
In Lorentzian signature, the real and imaginary parts of the first constraint tell us that $D_a = v_{+-a} = 0$.  In Euclidean signature, we can have small instantons supported by winding of the phases of the scalar fields $\phi^i$ and $p^{a'}$, as well as by winding of $\vt$, around the instantons.  The conditions $\nabla_+ \phi^i = \nabla_+ p^{a'} = 0$ ensure that $D_a = e^2 \big(  r_{\textit FI}^a + \sum_{a'} n^a_{a'} |p^{a'}|^2 - \sum_i Q_i^a |\phi^i|^2 \big)$ has support precisely where $v_{12a}$ does, allowing the more general condition $D_a - v_{12a} = 0$.

Away from the core of the instantons, we will still have that $D_a \rightarrow 0$.  In the Landau-Ginzburg phase, we have $\sum_a (n^{-1})_a^{a'} r^a_{\textit FI} \ll -1 $ so that $|p^{a'}|>0$ for all $a'$ (we also choose charges such that $\sum_a (n^{-1})_a^{a'} Q_i^a \geq 0$).  The non-degeneracy of the sets $\big\{ G^1 , \ldots , G^\iota,\linebreak J^{1_{a'}} , \ldots , J^{r^{a'}+1} \big\}$, for each $a'$, now guarantees that the only way $\sum_{a'} p^{a'} J^{m_{a'}} = G^\mu = 0$ for each $m_{a'}$ and $\mu$, is if $\phi^i = 0$ for all $i$.  Now we can use our $U(1)^s$ gauge symmetry to set
\be
\langle p^{a'} \rangle = \sqrt{- {\textstyle\sum_a} \big(n^{-1})^{a'}_a r^a_{\textit FI}} \, ,
\ee
but the discrete subgroup $\Gamma \subset U(1)^s$ that leaves $p^{a'}$ invariant will remain unbroken ($\Gamma = \big\{ k_a \in [0,1)^s \big| \sum_a k_a n^a_{a'} \in \mathbb{Z},~\forall~ a'\big\}$).  In the large $|r_{\textit FI}^a|$ limit, we can simply replace $p^{a'}$ by its VEV since corrections will be proportional to $1/|r_{\textit FI}^a|$.  After rescaling the coefficients of polynomials $J^{m_{a'}}$ to absorb the VEV of $p^{a'}$, we have a Landau-Ginzburg orbifold with superpotential
\be
\label{eqn:LG-superpot}
W = \frac{1}{\sqrt{2}}\int\! \! d\thp \Big\{   \sum_\mu \tG^\mu G^\mu(\Phi) + \sum_{a',m_{a'}}\G^\ma J^\ma(\Phi)   \Big\} \bigg|_{\bar{\theta}^+=0} \, .
\ee
We also have the charges
\bea
j_{(\beta^a)} &=& \int\! dy^1 \bigg(  \tfrac{1}{2}\sum_{i} \beta_i^a \big( \bar{\phi}{}^i \!\stackrel{\leftrightarrow}{\p}_0\! \phi^i  - i \bar{\psi}{}^i \psi^i \big)   -  \tfrac{i}{2} \sum_{\mu} \beta_\mu^a \bar{\gamma}{}^\mu \gamma^\mu   -  \tfrac{i}{2} \sum_{a',\ma}  \beta_\ma^a \bar{\gamma}{}^\ma \gamma^\ma   \nonumber \\
&& \qquad\qquad +\, i \sum_{l} \beta_l^a \p_0 \vt_l               \bigg)     \\
L_0 &=& \int\! dy^1 \bigg( \!\!     - \sum_i \p_- \phi^i \p_- \bar{\phi}{}^i   - \p_-\vt\p_-\bar{\vt}  +  \tfrac{i}{4} \sum_\mu  \bar{\gamma}{}^\mu \!\stackrel{\leftrightarrow}{\p}_1\! \gamma^\mu  +   \tfrac{i}{4} \sum_{a',\ma}  \bar{\gamma}{}^\ma \!\stackrel{\leftrightarrow}{\p}_1\! \gamma^\ma    \!\! \bigg) ~~   \\
\mathcalQB_+& = &\mathcalQB_0 + \mathcalQB_1 \equiv 2\sqrt{2} \int\! dy^1 \bigg(  \sum_i \bar{\psi}{}^i \p_+ \phi^i  +  \bar{\chi}\p_+ \vt \bigg)     \nonumber \\
&& \qquad\qquad\quad  - \sqrt{2}i\int\! dy^1  \bigg( \sum_\mu \gamma^\mu G^\mu(\phi)  +  \sum_{a',\ma} \gamma^\ma J^\ma(\phi) \bigg) \, ,
\eea
where
\be
\beta_i^a \equiv {\textstyle \sum_b} \beta_b^a Q_i^b \, ,   \quad   \beta_\mu^a \equiv -{\textstyle\sum_b} \beta_b^a d_\mu^b \, ,   \quad   \beta_{\ma}^a \equiv {\textstyle\sum_b} \beta_b^a \big( q_\ma^b - n^b_{a'} \big) \, ,  \quad  \beta_l^a \equiv {\textstyle\sum_b} \beta_b^a M_l^b + m_l^a \, .
\ee
As in \cite{Kachru:1993pg}, we have that $\{ \mathcalQB_0 , \mathcalQB_1 \} = \mathcalQB_0^2 = \mathcalQB_1^2 = 0$, and we have an operator $U$ satisfying $[U,\mathcalQB_0]=\mathcalQB_0$ and $[U,\mathcalQB_1]=0$ that assigns charge $1$ to $\bar{\psi}{}^i$ and $\bar{\chi}$, $-1$ to $\psi^i$ and $\chi$, and $0$ to all other fields.  As Kachru and Witten explained, since the only nontrivial states in $\mathcalQB_0$-cohomology have $U=0$, the cohomology of $\mathcalQB_+$ is isomorphic to the cohomology of $\mathcalQB_1$ computed within $\mathcalQB_0$-cohomology.

The right-moving component of elements of the cohomology of $\mathcalQB_0$ are in one-to-one correspondence with the right-moving ground state acted on only by holomorphic functions of the zero modes of $\phi^i$, the ``right-moving'' zero modes of $\vt$, and the zero modes of $\chi$ ({\it i.e.} these representatives have no $\psi^i$, $\bar{\psi}{}^i$, $\chi$, $\bar{\chi}$, or right-moving $\phi^i$, $\bar{\phi}{}^i$, $\vt$, or $\bar{\vt}$, raising modes, and are annihilated by $\psi^i_{0}$), call it $\mathcal{F}(\phi^i_{0},\tilde{\vt}_0,\bar{\chi}_0)$, where we define the ground state to be annihilated by $\chi_0$.  However, since $\vt$ takes values on a torus, this severely restricts the dependence of this holomorphic function to be invariant under the defining identifications of the torus and, in fact, implies that $\mathcal{F}$ cannot depend at all on the right-moving zero modes of $\vt$.  Thus, elements of $\mathcalQB_0$-cohomology are in one-to-one correspondence with states that depend only on left-moving oscillators, $\chi_0$, $\bar{\chi}_0$, holomorphic functions of $\phi_0^i$, and arbitrary functions of the ``left-moving'' zero modes of $\vt$ that are invariant under the torus identifications.

%%%%%%%%%%%%%%%%%%%%%%  TWISTED SECTORS  %%%%%%%%%%%%%%%%%%%

\subsubsection{Twisted Sectors --- GLSM}

Since we must still orbifold by the discrete group $\Gamma \subset U(1)^s$, we will have to include twisted sectors where $\phi^i$, $\psi^i$, $\tilde{\gamma}^\mu$, and $\gamma^\ma$, return to their original values when circling the origin, up to multiplication by an appropriate element of $\Gamma$.  This has the usual effect of changing the modings of their respective oscillators as well as the energy and $U(1)_{L_a}$ and $U(1)_R$ charges of the ground state.

If we consider the sector twisted by $k_a \in \Gamma \subset U(1)^s$, then left-moving fermions will satisfy $\tilde{\gamma}^\mu(y^- - 2\pi) = e^{-2\pi id_\mu\cdot k} \tilde{\gamma}^\mu (y^-)$ ($\gamma^\ma$ is similar) and right-moving fermions will satisfy $\psi^i (y^+ + 2\pi) = e^{2\pi iQ_i \cdot k} \psi^i (y^+)$.  Bosonizing $\tilde{\gamma}^\mu = e^{iH_\mu}$ and $\psi^i = e^{iH_i}$, we see that for the ground states $\mathcal{O}_{\tilde{\gamma}^\mu} = e^{i a_\mu H_\mu}$ and $\mathcal{O}_{\psi^i} = e^{i a_i H_i}$ to have the right branch cut in their OPEs with $\tilde{\gamma}^\mu$ and $\psi^i$, respectively, then $a_\mu \in -(-d_\mu \cdot k+\frac{1}{2}) + \mathbb{Z}$ and $a_i \in  (Q_i \cdot k - \frac{1}{2}) + \mathbb{Z}$ (the $\frac{1}{2}$ shift comes from the conformal transformation between the cylinder and the plane).  If we further require that $\tilde{\gamma}^\mu_{0}$ and $\psi^i_{0}$ annihilate the ground state in twisted sectors that have fermion zero modes, then the energy minimizing ground state must have
\bea
\label{eqn:fermion-grnd-state}
a_\mu &=& d_\mu \cdot k  +  \big\lfloor -d_\mu \cdot k  \big\rfloor + \frac{1}{2}  \, ,   \\
a_i & =& Q_i \cdot k + \big\lfloor -Q_i \cdot k \big\rfloor + \frac{1}{2} \, .
\eea
Similarly, the bosons satisfy $\phi^i (y^++2\pi,y^- - 2\pi) = e^{2\pi iQ_i\cdot k }\phi^i(y^+,y^-)$, so the energy, $U(1)_{L_a}$ and $U(1)_R$ charges of the ground states in the $k_a \in \Gamma$ twsited sector are
\bea
%L_0 &=& \frac{1}{24} \Big(  \sum_i 1 -  \sum_\mu 1  - \sum_{a',\ma} 1 \Big)  + \frac{1}{2} \sum_\mu  \Big( - \frac{\beta_\mu \cdot k}{2\pi}  +  \big\lfloor \frac{\beta_\mu \cdot k}{2\pi}   \big\rfloor  + \frac{1}{2}  \Big)^2    \nonumber \\
%&& + \frac{1}{2} \sum_{a',\ma} \Big( - \frac{\beta_\ma \cdot k}{2\pi}  +  \big\lfloor \frac{\beta_\ma \cdot k}{2\pi}  \big\rfloor  +  \frac{1}{2}  \Big)^2     -  \frac{1}{2} \sum_i  \Big( \frac{\beta_i\cdot k}{2\pi} + \big\lfloor -\frac{\beta_i \cdot k}{2\pi} \big\rfloor  + \frac{1}{2} \Big)^2     \qquad  \\
L_0  &=& \frac{(N-r)}{24}  + \frac{1}{2} \sum_\mu  \Big( d_\mu \cdot k  +  \big\lfloor -d_\mu \cdot k   \big\rfloor  + \frac{1}{2}  \Big)^2   + \frac{1}{2} \sum_{a',\ma} \Big( - q_\ma \cdot k  +  \big\lfloor q_\ma \cdot k  \big\rfloor  +  \frac{1}{2}  \Big)^2          \nonumber \\
&&   -  \frac{1}{2} \sum_i  \Big( Q_i\cdot k + \big\lfloor - Q_i \cdot k \big\rfloor  + \frac{1}{2} \Big)^2  \, ,            \\
q_{L_a} &=& \sum_\mu \beta_\mu^a \Big( d_\mu \cdot k  +  \big\lfloor -d_\mu \cdot k   \big\rfloor  +  \frac{1}{2} \Big) + \sum_{a',\ma} \beta_\ma^a \Big( - q_\ma \cdot k  +  \big\lfloor q_\ma \cdot k   \big\rfloor  +  \frac{1}{2} \Big)      \nonumber \\
&& +  \sum_i  \beta_i^a \Big(   Q_i \cdot k + \big\lfloor -Q_i \cdot k  \big\rfloor  +  \frac{1}{2}  \Big) \, ,     \\
q_R &=&  \sum_\mu \alpha_\mu \Big( d_\mu \cdot k  +  \big\lfloor -d_\mu \cdot k \big\rfloor + \frac{1}{2}\Big) + \sum_{a',\ma} \alpha_\ma \Big( - q_\ma \cdot k  +  \big\lfloor q_\ma \cdot k  \big\rfloor + \frac{1}{2} \Big)      \nonumber \\
&& +  \sum_i  \big(\alpha_i - 1\big) \Big(  Q_i \cdot k + \big\lfloor -Q_i \cdot k  \big\rfloor  + \frac{1}{2}  \Big) \, .      
\eea
The twist sector also determines the modings of the left-moving oscillators to be
\be
\phi^i_{n+Q_i\cdot k}\, , ~~~~ \bar{\phi}{}^i_{n - Q_i\cdot k}\, ,~~~~ \tilde{\gamma}^\mu_{n - d_\mu \cdot k } \, ,~~~~ \bar{\tilde{\gamma}}{}^\mu_{n + d_\mu \cdot k }  \, , ~~~~ \gamma^{\ma}_{n + q_\ma \cdot k } \, ,  ~~~~  \bar{\gamma}{}^{\ma}_{n - q_\ma \cdot k } \, ,
\ee
for $n\in\mathbb{Z}$.

%%%%%%%%%%%%%%%%%%  TORSION TWIST SECTORS  %%%%%%%%%%%%%%%%

\subsubsection{Twisted Sectors --- Torsion Multiplet}

States that involve the zero modes of $\vt$ must transform correctly under the torus identifications (section \ref{sec:torsion})
\be
\vt \cong \vt + 2\pi R \cong \vt + 2\pi \tau \, ,
\ee
where $\tau = \tau_1 + i \tau_2$.  Thus, they can only depend on the zero modes via terms of the form
\be
\label{eqn:twisted-states}
\exp\bigg\{i\sum_l ( a_l \vt_{0,l} + \tilde{a}_l \tilde{\vt}_{0,l} )\bigg\} \, ,
\ee
where $\vt_0$ and $\tilde{\vt}_0$ are ``left-moving'' and ``right-moving'' zero modes, respectively.  One might think that these states should be invariant under the identifications, but a subtlety arises here: in the original GLSM, the twisted sectors of the Landau-Ginzburg phase have the interpretation of configurations with fractional flux $\frac{1}{\pi} \int d^2y\, v_{+-a} = -k_a \in \Gamma$ supported by the VEV of $p^{a'}$.  The partition function in the GLSM is only invariant under the torus identifications when the gauge flux is integral, so the path integral picks up a phase that must be reproduced by the twisted sector states:
\bea
\exp\!\big\{ i\pi R (a_1 + \tilde{a}_1 ) \big\}  &=& \exp\!\big\{ -2\pi i R M_1\cdot k \big\}    \\
\exp\!\big\{i\pi {\textstyle\sum_l} \tau_l (a_l + \tilde{a}_l ) \big\} &=& \exp\!\big\{ - 2\pi i {\textstyle\sum_l} \tau_l M_l \cdot k \big\} \, .
\eea
We can write this as
\be
(a_1 + \tilde{a}_1)  =  \frac{2 z_1}{R} - 2 M_1 \cdot k  \, ,   \qquad\quad   (a_2 + \tilde{a}_2) = \frac{2 z_2}{\tau_2} - \frac{\tau_1}{\tau_2}\frac{2 z_1}{ R}  -  2 M_2 \cdot k \, ,
\ee
where $z_1,z_2\in\mathbb{Z}$.

In addition to this projection, we must also allow twisted states from both the orbifold by $\Gamma$ and from winding states allowed by the torus identifications.  The twisted sectors depend on the integers defining the winding sector, $w_1$, $w_2$, and on the orbifold twist by $k_a \in \Gamma$:
\be
\vt( y^+ + 2\pi, y^- - 2\pi) = \vt( y^+, y^- )  +  2\pi M \cdot k + 2\pi w_1 R + 2\pi w_2 \tau \, .
\ee
The OPE of $\vt$ with the state (\ref{eqn:twisted-states}) implies that
\be
(\tilde{a}_1 - a_1) = 2 M_1 \cdot k + 2 w_1 R + 2 w_2 \tau_1 \, ,   \qquad\quad  (\tilde{a}_2 - a_2) = 2 M_2 \cdot k  + 2 w_2 \tau_2 \, ,
\ee
so we have
\bea
\tilde{a}_1 &=& \frac{z_1}{R} + \big( w_1 R + w_2 \tau_1 \big)   \, ,    \qquad\qquad\qquad\quad~   \tilde{a}_2 = \frac{z_2}{\tau_2} - \frac{\tau_1}{\tau_2}\frac{z_1}{R}  +  w_2\tau_2  \, ,   \\
a_1 &=& \frac{z_1}{R} - \big(  w_1 R + w_2 \tau_1 \big) - 2M_1 \cdot k   \, ,    \qquad\quad   a_2 = \frac{z_2}{\tau_2} - \frac{\tau_1}{\tau_2}\frac{z_1}{R}  -  w_2\tau_2  -  2M_2\cdot k  \, .
\eea

States in $\mathcalQB_+$-cohomology are independent of $\tilde{\vt}_{0,l}$, so there only exist states in theories that have twisted sectors in which there exist choices of $z_1,~z_2,~w_1$, and $w_2$, for which $\tilde{a}_1 = \tilde{a}_2 = 0$.  In any given twisted sector $k_a$, the ground state is the one in which $a_1^2+a_2^2$ is minimized.  When the torus is not rectangular, $\xi \neq 0$, then $\tilde{a}_l = 0$ implies that
\be
l_1 z_1 = \xi \big( l_2 w_1 - l_1 w_2 \big)  - w_1 k_1    \qquad\quad  \textrm{and}  \qquad \quad   l_2 z_2 =  \xi \big( l_1 w_2 - l_2 w_1 \big)  - w_2 k_2 \, .
\ee
When $\xi$ is irrational, we must have $w_1 = l_1 n',~w_2 = l_2 n',~z_1 = -k_1 n',~z_2 = -k_2 n'$, yielding
\be
a_1 = -\frac{2 k_1}{R} \big( \mathscr{M} \cdot k + n' \big)  \qquad\quad  \textrm{and}  \qquad\quad  a_2 = - 2 l_2 \tau_2 \big( \mathscr{M} \cdot k + n' \big) \, .
\ee
The ground state has $n' = - \big\lfloor \mathscr{M}\cdot k + \frac{1}{2} \big\rfloor$ or $n' = \big\lfloor -\mathscr{M}\cdot k + \frac{1}{2} \big\rfloor$.  When $\xi \neq 0$ is rational, the analysis is similar but more involved.

For the models in this paper, we'll restrict our attention to rectangular torii, $\xi_1=0$.  In this case,  the charges are $M_1^a = \mathscr{M}_1^a \sqrt{k_1 l_1}$ and $M_2^a = \mathscr{M}_2^a \sqrt{k_2 l_2}$, where $R^2 = \frac{k_1}{l_1}$ for relatively prime integers $k_1$ and $l_1$, and similarly for $\tau_2^2 = \frac{k_2}{l_2}$ (see equation (\ref{eqn:rect-t2-radii})).  The conditions that $\tilde{a}_1 = \tilde{a}_2 = 0$ become
\be
\label{eqn:state-constraints}
z_l +  w_l \frac{k_l}{l_l}  =  0   \, .  
\ee
Thus, we see that we must choose $z_l = -k_l n_l$ and $w_l = l_l n_l$, for any $n_l \in \mathbb{Z}$, so
\be
a_l = - 2\sqrt{k_l l_l} \big( \mathscr{M}_l \cdot k +  n_l \big) \,  .
\ee
Then the ground state will have 
\be
n_l = - \Big\lfloor \mathscr{M}_l \cdot k + \frac{1}{2} \Big\rfloor \, ,   \qquad  \textrm{or}   \qquad   n_l = \Big\lfloor -  \mathscr{M}_l \cdot k + \frac{1}{2} \Big\rfloor \, ,
\ee
which only differ when $\mathscr{M}_l\cdot k \in \mathbb{Z}$; note that these look similar to the fermion ground states (\ref{eqn:fermion-grnd-state}), which is not surprising since this discussion contains the case where the circles are at the free fermion radius.  Let's define
\be
n_l \equiv - \Big\lfloor \mathscr{M}_l \cdot k + \frac{1}{2} \Big\rfloor + \hat{n}_l \, ,
\ee
so that $\hat{n}_l \neq 0$ corresponds to excited states.    

The energy, $U(1)_{L_a}$ and $U(1)_R$ charges of the states involving zero modes of the torsion multiplet are, thus,
\bea
L_0 &=& - \frac{1}{12} + \frac{1}{4} \sum_l a_l^2  ~~\stackrel{\xi=0}{\longrightarrow}~ - \frac{1}{12} + \sum_l k_l l_l \Big(  \mathscr{M}_l\cdot k - \Big\lfloor \mathscr{M}_l \cdot k + \frac{1}{2}\Big\rfloor + \hat{n}_l  \Big)^2   \, ,  \\
q_{L_a} &=& \sum_l \beta_l^a a_l   ~~  \stackrel{\xi=0}{\longrightarrow}~ - 2\sum_l \beta_l^a \sqrt{k_l l_l}  \,\Big(  \mathscr{M}_l\cdot k - \Big\lfloor \mathscr{M}_l \cdot k + \frac{1}{2} \Big\rfloor + \hat{n}_l  \Big) \, ,   \\
q_{R} &=& -\frac{1}{2} + \sum_l \alpha_l a_l   ~~  \stackrel{\xi=0}{\longrightarrow}~ -\frac{1}{2} - 2\sum_l \alpha_l \sqrt{k_l l_l} \,\Big(  \mathscr{M}_l\cdot k - \Big\lfloor \mathscr{M}_l \cdot k + \frac{1}{2} \Big\rfloor + \hat{n}_l \Big) \, ,
\eea
where the $-\frac{1}{2}$ term in $q_R$ comes from $\chi$; remember that when $\xi \neq 0$, we have to remember to satisfy the Diophantine constraints $\tilde{a}_l = 0$ which we already accounted for when $\xi = 0$.  The left-moving oscillator modes of $\vt$ are unaffected by the twist: $\vt_{n}$ and $\bar{\vt}_{n}$ for $n\in\mathbb{Z} \backslash \{0\}$.

Finally, a crucial point in the analysis of the model is the fact that the states \emph{do} depend on the zero modes of the superpartner of $\vt$, $\chi_0$ and $\bar{\chi}_0$.  The reason this differs from the chiral multiplets $\Phi^i$ is because, for them, the state annihilated by $\bar{\psi}{}^i_{0}$ is exact in $\mathcalQB_0$ cohomology ($\mathcalQB_0 (\bar{\phi}{}^i_{0} |0\rangle )$) but for the torsion multiplet, the state is \emph{not} exact because $\bar{\tilde{\vt}}_0 |0\rangle$ does not respect the torus identifications and, therefore, is not a state in the Hilbert space of the theory.  This will lead us to find pairs of states in the massless spectrum that differ only by a shift in R-charges which, in turn, leads to a spectrum with non-chiral spacetime fermions.  When $d=4$, $\CN=1$ supersymmetry is not broken, it is thus automatically enhanced to a manifest $\CN=2$.

%%%%%%%%%%%%%%%%%%%%%%  GSO PROJECTION  %%%%%%%%%%%%%%%%%%

\subsubsection{GSO Projection}

To combine these components into a full heterotic model, we must add three more ingredients: first, we have to add $(3-N)$ uncharged chiral multiplets corresponding to the non-compact, $(10-2N-2)$-dimensional degrees of freedom in light-cone gauge; second, we have to add $(32-2r)$ left-handed, Majorana-Weyl fermions to fill out the $32$ gauge bundle fermions needed in our infrared CFT; and finally, we must implement left and right GSO orbifolds.

For the $Spin(32)/\mathbb{Z}_2$ heterotic string, we introduce $(32-2r)$ left-handed, Majorana-Weyl fermions $\lambda^I$.  It is, then, natural to identify the generator of the left GSO twist with
\be
\exp \!\bigg\{  -i\pi \sum_a J_{L_a}  -  i\pi \sum_I F_{\lambda^I} \bigg\} \, ,
\ee
where $F_{\lambda^I}$ is the fermion number operator for $\lambda^I$.  For the right GSO twist, we take
\be
\label{eqn:right-GSO}
\exp\!\bigg\{ -i\pi J_R  -  i\pi \sum_{M=2}^{7-2N} F_{\psi^M}  \bigg\} \, ,
\ee
where $F_{\psi^M}$ is fermion number operator for the $(8-2N)$ right-handed, Majorana-Weyl fermions that are the worldsheet superpartners of the spacetime coordinates $X^M$.

For the $E_8 \times E_8$ heterotic string, we must divide our gauge bundle fermions $\gamma^\ma$ into two sets, so we write $r = r_{(1)} + r_{(2)}$; similarly, we divide the $J_{L_a}$ into two sets that we will label by $a_{(1)}$ and $a_{(2)}$.  Next, we introduce $(16-2r_{(1)})$ left-handed, Majorana-Weyl fermions $\lambda^{I_{(1)}}$ and $(16-2r_{(2)})$ left-handed, Majorana-Weyl fermions $\lambda^{I_{(2)}}$.  Now, our left GSO twist is generated by the two elements
\be
\exp\!\bigg\{ - i\pi \sum_{a_{(1)}} J_{L_{a_{(1)}}} - i \pi \sum_{I_{(1)}} F_{\lambda^{I_{(1)}}}  \bigg\} \, ,   \quad\qquad  \exp\!\bigg\{ - i\pi \sum_{a_{(2)}} J_{L_{a_{(2)}}} - i \pi \sum_{I_{(2)}} F_{\lambda^{I_{(2)}}}  \bigg\} \, ,
\ee
where the $F_\lambda$'s are, again, fermion number operators.  The right GSO twist is still generated by (\ref{eqn:right-GSO}).

In our models, the $U(1)_R$ assignments are given in (\ref{eqn:our-charges}).  Since we are looking for states with right-moving $\tilde{L}_0=0$ by examining $\mathcalQB_+$-cohomology, we are necessarily restricting our attention to the Ramond sector where the supercurrent in the CFT has integral moding.\footnote{The sector in which the supercurrent has half-integral moding does not allow us to turn the problem of finding $\tilde{L}_0=0$ states into a cohomology computation.}  Thus, we are only interested in the even sectors of the right GSO twist, under which the fields will have periodicities given by $\exp\!\big\{ - 2\pi i k' J_R \big\}$, for $k'\in\mathbb{Z}$.  Recall that the unbroken subgroup of the original $U(1)^s$ is given by $\Gamma = \big\{ k_a \in [0,1)^s \, \big| \, k\cdot n_{a'} \in \mathbb{Z}\big\}$.  If we pick an element $k_a = -k' \sum_{a'}(n^{-1})^{a'}_a \in \Gamma$, then we see that the fields in the sector of our theory twisted by this $k_a\in\Gamma$ have the same periodicities as those in the sector twisted by $\exp\!\big\{ -2\pi i k' J_R \big\}$, so the even twisted sectors of the right GSO twist are already contained in the $\Gamma$ twist; the only additional work that we must do to compute the right GSO orbifold is to project onto states where the right GSO operator evaluates to 1.  %(by a similar logic, it is already guaranteed to evaluate to $\pm 1$).  
In the examples we'll consider in this paper, the left GSO twisted sectors will actually contain the $\Gamma$ twisted sectors, so we will only have to worry about implementing the left GSO orbifold and the right GSO projection, as well as the projection onto states with $L_0 = 0$.

For completeness, for the $E_8 \times E_8$ heterotic string, we list the contribution to the energy of the ground state from the non-compact chiral multiplets and the free left-handed fermions $\lambda^{I_{(1)} = 1,\ldots, 16-2r_{(1)}}$ and $\lambda^{I_{(2)} = 1,\ldots,16-2r_{(2)}}$, in the sector twisted by $\exp\!\Big\{ -i\pi k^{(1)} J^{(1)}_{GSO} - i \pi k^{(2)} J^{(2)}_{GSO} \Big\}$:
\bea
L_0 &=& -\frac{(6-2N)}{24} - \frac{\big(16-2r_{(1)}\big)}{48} - \frac{\big(16-2r_{(2)}\big)}{48} +\frac{\big(16-2r_{(1)}\big)}{4}  \bigg( -\frac{k^{(1)}}{2} + \Big\lfloor \frac{k^{(1)}}{2} \Big\rfloor + \frac{1}{2} \bigg)^2   \nonumber \\
&& + \frac{\big(16-2r_{(2)}\big)}{4} \bigg( -\frac{k^{(2)}}{2} + \Big\lfloor \frac{k^{(2)}}{2} \Big\rfloor + \frac{1}{2} \bigg)^2 \, .
\eea
The left GSO projections relate the $U(1)_L$ charge of a given state to its $SO\big(16-2r_{(1)}\big) \times SO\big(16-2r_{(2)}\big)$ representation, while the right GSO projection relates the $U(1)_R$ charge to the chirality of the spacetime spinor.

%%%%%%%%%%%%%%%%%%%  EXAMPLES  %%%%%%%%%%%%%%%%%%%%%%

\section{Examples}

In this section, we will apply the techniques developed in the previous section to a pair of simple examples.  First, we study compactification to 6d on a $T^{2}$-fibration over a $T^{2}$ base. 
The computation at intermediate stages is remarkably messy; the fact that summing over twist sectors rather magically produces good $E_{6}\times E_8$ irreps is a dramatic check on our techniques.   

We then move on to an example of considerably greater physical importance, a 4d $\CN=2$ \nK\ $T^{2}$-fibration over $K3$ with non-trivial gauge and NS-NS flux.  As in the first example, while the intermediate steps are rather messy, summing over all twist sectors organizes all fields into good $SO(10)$ irreps, a remarkable check of our computation and our results.  An additional 4d example, corresponding to an Iwasawa-like compactification, is presented in an appendix \ref{sec:iwasawa}.

%%%%%%%%%%%%%%%%%%%% T^2 Example --> T^4 Example %%%%%%%%%%%%%%%%

\subsection{$d=6$, $T^2$ Base}
\label{sec:T2base}

Since the analysis outlined in Sections 2 and 3 did not depend on the dimension of the target space, we will take for our first example a TLSM whose naive 1-loop geometry is a 4d $T^{2}$-fibration over a $T^{2}$ base.  Of course, there are no interesting (orientable) circle bundles over $T^{2}$, so the end result should be either a novel non-geoemtric compactification or simply a familiar $T^{4}$ compactification at some point in its Narain moduli space.  

For simplicity, we'll consider a geometry where we fiber only one of the circles over a $T^2$ base and leave the other unfibered, and take the vector bundle $E$ to have rank 3 with $c_1(E) = 0$.
We take the charges to be
\be
{\setlength\arraycolsep{8pt}
\begin{array}{|c||c|c|c|c|c|}
\hline  & \boldsymbol{\Phi^{i=1,\ldots,3}}  & \boldsymbol{P} & \boldsymbol{\tG} & \boldsymbol{\G^{m=1,\ldots,4}} & \boldsymbol{\Theta}  \\
\hline\hline \boldsymbol{U(1)}   &  1  &  -4  &  -3  &  1  &  (3R)      \\
\hline \boldsymbol{U(1)_L}   &  \frac{1}{4}  &  0  &  -\frac{3}{4}  &  -\frac{3}{4}  &  \big(  \frac{3}{4}R \big)    \\
\hline \boldsymbol{U(1)_R}   &  \frac{1}{4}  &  0  &  \frac{1}{4}  &  \frac{1}{4}  &  \big(  \frac{3}{4} R  \big)  \\
\hline
\end{array}}
\ee
where the $S^1$ radii are $R = S= \frac{1}{\sqrt{3}}$.  These charges are non-anomalous and give $c_L = \hat{c}_R + r_L = 4 + 3$.  We will take this model to be embedded in a heterotic $E_8 \times E_8$ model, so the gauge-bundle fermions will transform in an $SU(3)$ subgroup of the first $E_8$.  Since the left-handed fermions have the same $U(1)_L$ and $U(1)_R$ charges, let $\kappa=0,\ldots,4$ and define $\gamma^0\equiv\tilde{\gamma}$ and $J^0(\phi) \equiv G(\phi)$.  In the Landau-Ginzburg orbifold phase, we see that we must orbifold by the discrete group $\G = \mathbb{Z}_4$, and so our left GSO group will be isomorphic to $\mathbb{Z}_{8} \times \mathbb{Z}_2$.  In a twisted sector $\big( k^{(1)}, k^{(2)} \big)$, the oscillators are
{\setlength\arraycolsep{0pt}
\bea
&& \phi^i_{n - \frac{k^{(1)}}{8}}  \, , \qquad \bar{\phi}{}^i_{n + \frac{k^{(1)}}{8}}  \, ,  \qquad   \gamma^\kappa_{n+\frac{3k^{(1)}}{8}}  \, ,  \qquad   \bar{\gamma}{}^\kappa_{n-\frac{3k^{(1)}}{8}}  \, , \qquad    \lambda^{I_{(1)}}_{n+\frac{k^{(1)}}{2}} \, ,  \qquad  \lambda^{I_{(2)}}_{n+\frac{k^{(2)}}{2}} \, ,   \nonumber \\
&&   X^M_{n} \, ,  \qquad  \vt_{n}   \, ,   \qquad  \bar{\vt}_{n}  \, ,  \qquad  \bar{\chi}_0 \, ,
\eea
where the handling of zero modes was explained in section \ref{sec:LG}.

We have $16$ twisted sectors $k^{(1)}=0,\ldots,7$, $k^{(2)}=0,1$ however, life isn't as awful as it seems, spacetime CPT invariance relates sector $\big( k^{(1)},k^{(2)} \big)$ to $\big( 8-k^{(1)}, 2-k^{(2)} \big)$ with $q_{L,R}\rightarrow -q_{L,R}$, so we need only keep track of $k^{(1)}=0,\ldots,4$, $k^{(2)}=0,1$.  Of course, $k^{(2)}=0$ and $k^{(2)}=1$ differ in energy by $\Delta L_0 = 1$, so we will only list $k^{(2)}=0$ when it doesn't take us above $L_0 = 0$.  For these sectors, the ground state quantum numbers and relevant oscillators are
\be
{\setlength\arraycolsep{6pt}
\begin{array}{|c||c|c|c|c|c|}
\hline \!\!\! \boldsymbol{\big(k^{(1)}, k^{(2)}\big)} \!\!\! &\boldsymbol{L_0} & \boldsymbol{q_L} & \boldsymbol{q_R} & \!\! \boldsymbol{(\hat{n}_1 , \hat{n}_2 )} \!\!   &  \mathit{Relevant~Oscillators} \\
\hline\hline  \boldsymbol{(0,1)} & 0 & -\frac{3}{2} & -1 & (0,0)  &   \mathrm{Zero~modes} \\
\hline \boldsymbol{(1,1)} & -1 & 0 &  -1 & (0,0)  & \!\! \phi^i_{-\frac{1}{8}}  ,~  \bar{\phi}{}^i_{-\frac{7}{8}}   , ~  \gamma^\kappa_{-\frac{5}{8}}  , ~ \bar{\gamma}{}^\kappa_{-\frac{3}{8}}  ,   \!\!  \\
& & & &  & \lambda^{I_{(1)}}_{-\frac{1}{2}}  , ~ \lambda^{I_{(2)}}_{-\frac{1}{2} }   , ~ X^M_{-1}  , ~ \vt_{-1}  , ~ \bar{\vt}_{-1}  , ~ \bar{\chi}_0     \\
\hline \boldsymbol{(1,0)} & 0 & 0 & -1 & (0,0) &  \lambda_0^{I_{(2)}}  , ~~ \bar{\chi}_0  \\
\hline \boldsymbol{(2,1)} & 0 & \frac{3}{2} &  -1 &  (0,0)  &  \lambda_0^{I_{(1)}} , ~~ \bar{\chi}_0  \\
\hline \boldsymbol{(3,1)} & -\frac{1}{8} & -\frac{3}{4} & \frac{1}{4} & (0,0)  &  \bar{\gamma}{}_{-\frac{1}{8}}^\kappa \, , ~~ \bar{\chi}_0   \\
\hline \boldsymbol{(4,1)} & \frac{1}{2} &  \frac{3}{4} & \frac{1}{4} &  (0,0)  &    \\
\hline
\end{array}}
\ee
where $(\hat{n}_1,\hat{n}_2)$ indicates the winding sector.

In sections \ref{sec:T2k=01} and \ref{sec:T2k=11}, we work out the spectrum explicitly for the sectors $(k^{(1)},k^{(2)})=(0,1),(1,1)$.  The spectrum, organized by twisted sector and charge under the linearly realized $\big( SO(10) \times U(1)_L \big) \times SO(16) \times U(1)_R \subset E_6 \times E_8 \times U(1)_R $ and six-dimensional Poincar\'e group,~is
\be
{\setlength\arraycolsep{8pt}
\begin{array}{|c||c|c|c|c|c|}
\hline \big( k^{(1)}, k^{(2)} \big) & \mathit{Degeneracy} & SO(10)_{q_L}  &  SO(16) & q_R  &  SO(5,1)          \\
\hline\hline \boldsymbol{(0,1)}   &  1 & \boldsymbol{16_{-\frac{3}{2}}}  &  \boldsymbol{1} & -1 & \boldsymbol{4}    \\
\hline   &  1 & \boldsymbol{16_{-\frac{3}{2}}}  &  \boldsymbol{1} & 0 & \boldsymbol{4'}    \\
\hline    &  1 & \boldsymbol{\overline{16}_{\frac{3}{2}}}  &  \boldsymbol{1} & 0 & \boldsymbol{4'}    \\
\hline   &  1 & \boldsymbol{\overline{16}_{\frac{3}{2}}}  &  \boldsymbol{1} & 1 & \boldsymbol{4}    \\
\hline & n_{27} & \boldsymbol{\overline{16}_{-\frac{1}{2}}} & \boldsymbol{1} & -1 & \boldsymbol{4}  \\
\hline & 2\, n_{27} & \boldsymbol{\overline{16}_{-\frac{1}{2}}} & \boldsymbol{1} & 0 & \boldsymbol{4'} \\
\hline & n_{27} & \boldsymbol{\overline{16}_{-\frac{1}{2}}} & \boldsymbol{1} & 1 & \boldsymbol{4} \\
\hline & n_{27} & \boldsymbol{16_{\frac{1}{2}}} & \boldsymbol{1} & -1 & \boldsymbol{4}  \\
\hline & 2\, n_{27} & \boldsymbol{16_{\frac{1}{2}}} & \boldsymbol{1} & 0 & \boldsymbol{4'} \\
\hline & n_{27} & \boldsymbol{16_{\frac{1}{2}}} & \boldsymbol{1} & 1 & \boldsymbol{4} \\
\hline\hline \boldsymbol{(1,1)} & 1 & \boldsymbol{1_0} & \boldsymbol{1} &  -1  &  \boldsymbol{6_v} \otimes \boldsymbol{4}  \\
\hline & 1 & \boldsymbol{1_0} & \boldsymbol{1} &  0  &  \boldsymbol{6_v} \otimes \boldsymbol{4'}  \\
\hline  & 3+n_1 & \boldsymbol{1_0} & \boldsymbol{1} &  -1  &  \boldsymbol{4}  \\
\hline  & 20 + 2n_1 & \boldsymbol{1_0} & \boldsymbol{1} &  0  &  \boldsymbol{4'}  \\
\hline  & 17+n_1 & \boldsymbol{1_0} & \boldsymbol{1} &  1  &  \boldsymbol{4}  \\
\hline & n_{27} & \boldsymbol{10_1} & \boldsymbol{1} & -1 & \boldsymbol{4} \\
\hline & 2\, n_{27} & \boldsymbol{10_1} & \boldsymbol{1} & 0 & \boldsymbol{4'} \\
\hline & n_{27} & \boldsymbol{10_1} & \boldsymbol{1} & 1 & \boldsymbol{4} \\
\hline & n_{27} & \boldsymbol{1_2} & \boldsymbol{1} & -1 & \boldsymbol{4} \\
\hline & 2\, n_{27} & \boldsymbol{1_2} & \boldsymbol{1} & 0 & \boldsymbol{4'} \\
\hline & n_{27} & \boldsymbol{1_2} & \boldsymbol{1} & 1 & \boldsymbol{4} \\
\hline & 1 & \boldsymbol{45_0} &  \boldsymbol{1} & 0 & \boldsymbol{4'} \\
\hline & 1 & \boldsymbol{45_0} &  \boldsymbol{1} & 1 & \boldsymbol{4} \\
\hline   &  1  &  \boldsymbol{1_0}  &  \boldsymbol{120}  &  -1  &  \boldsymbol{4}     \\
\hline   &  1  &  \boldsymbol{1_0}  &  \boldsymbol{120}  &  0  &  \boldsymbol{4'}     \\
\hline\hline  \boldsymbol{(1,0)}  &  1  &  \boldsymbol{1_0}  &  \boldsymbol{128}  &  -1  &  \boldsymbol{4}     \\
\hline   &  1  &  \boldsymbol{1_0}  &  \boldsymbol{128}  &  0  &  \boldsymbol{4'}     \\
\hline\hline \boldsymbol{(2,1)}  &  1  &  \boldsymbol{\overline{16}_{\frac{3}{2}}}  &  \boldsymbol{1}  &  -1  &  \boldsymbol{4}  \\
\hline  &  1  &  \boldsymbol{\overline{16}_{\frac{3}{2}}}  &  \boldsymbol{1}  &  0  &  \boldsymbol{4'}  \\
\hline\hline \boldsymbol{(3,1)}   &  6  &  \boldsymbol{1_0}  &  \boldsymbol{1}  &  0  &  \boldsymbol{4'}  \\
\hline   &  6  &  \boldsymbol{1_0}  &  \boldsymbol{1}  &  1  &  \boldsymbol{4}  \\
\hline
\end{array}}
\ee
Here, $n_1$ and $n_{27}$ depend on specific details of the defining polynomials $J^\kappa(\phi)$, as can be seen in the following subsections.  In particular, $n_{27}$ has the nice interpretation as the dimension of the subspace of degree 4, quasi-homogeneous polynomials in the local algebra
\be
\frac{\mathbb{C}[\phi^1,\phi^2,\phi^3]}{\langle J^0, \ldots, J^4 \rangle} \, .
\ee 
If there were only three $J^\kappa$'s, the Poincar\'e polynomial would tell us that $n_{27}$ was $6$.  However, since there are five such polynomials, this only tells us an upper bound: $n_{27}\leq 6$.  It can, in fact, be $0$ if the $J^\kappa$ are suitably chosen.  For example, if we take
\bea
\label{eqn:T2example}
G & \, =\,  & \sum_i (\phi^i)^3 + a \phi^1 \phi^2\phi^3 \, , \qquad J^1 = (\phi^1)^3 \, , \qquad J^2 = (\phi^2)^3 \, ,    \nonumber \\
J^3 &\, =\, & (\phi^3)^3 + (\phi^1)^2 \phi^2 + \phi^1 \phi^2 \phi^3 \, ,  \qquad  J^4  = (\phi^3)^3 +\phi^1 (\phi^3)^2+(\phi^2)^2 \phi^3
% NOTE: this example gives $n_{27}=6$ and $n_1=11$:
% G = \sum_i (\phi^i)^3 + a \phi^1 \phi^2 \phi^3 \, , J^1 = J^3 = \phi^1, J^2 = J^4 = \phi^2
\eea
with $a^3 \neq -27$ and $a\neq 0,1$, then $n_{27} = 0$ and $n_1 = 0$ ($a^3 \neq -27$ is necessary for $G$ to be transverse).

Meanwhile, noting the decomposition under $E_6 \rightarrow SO(10)_{U(1)}$
\bea
\boldsymbol{78} &~~ \longrightarrow~~ & \boldsymbol{45_0} \oplus \boldsymbol{16_{-\frac{3}{2}}} \oplus \boldsymbol{\overline{16}_{\frac{3}{2}}} \oplus \boldsymbol{1_0}   \nonumber \\
\boldsymbol{27} & \longrightarrow & \boldsymbol{16_{\frac{1}{2}}} \oplus \boldsymbol{10_{-1}} \oplus \boldsymbol{1_2} \nonumber
\eea
we see that the linearly realized $SO(10)\times U(1)_L$ is enhanced to $E_6$, just as $SO(16)$ is enhanced to $E_8$.  
The six-dimensional, fermionic spectrum organizes nicely into $\mathcal{N}=(1,1)$ multiplets:
\be
{\setlength\arraycolsep{8pt}
\begin{array}{|c|c|c|}
\hline 6d~\mathcal{N}=(1,1)~\mathit{Repr.}  & \mathit{Degeneracy} &  E_6 \times E_8   \\
\hline\hline \mathrm{Supergravity} & 1 & \mathbf{1}\otimes\mathbf{1}  \\
\hline \mathrm{Vector}  &  1  &  \mathbf{78}\otimes\mathbf{1}    \\
\hline \mathrm{Vector} & 1 & \mathbf{1}\otimes\mathbf{248}  \\
\hline \mathrm{Vector} & n_{27} & \mathbf{27}\otimes\mathbf{1} \\
\hline \mathrm{Vector} & n_{27} & \mathbf{\overline{27}}\otimes\mathbf{1} \\
\hline \mathrm{Vector} & 25+2n_1 & \mathbf{1} \otimes\mathbf{1}  \\
\hline
\end{array}}
\ee
Note that, since the spectrum organizes into irreps of the non-chiral $\CN=(1,1)$ superalgebra, the familiar $\CN=(0,1)$ anomaly, $n_H - n_V + 29 n_T - 273$, automatically vanishes.

If the $n_{i}$ in the table above were allowed to take general values, these models would violate a hallowed constraint of  heterotic model building:\footnote{We thank Shamit Kachru for very helpful discussions on this point.} the rank of the gauge group in a 6d compactification with $(1,1)$ \susy\ should be bounded by 24.
This implies both that the $n_{i}$ must be bounded from above, and that the $25+2n_{1}$ $E_{6}$-scalars, together with the single {\bf 78} and the  $n_{27}$ {\bf 27}'s and $\overline{\bf 27}$'s, must assemble into the adjoint of some enhanced gauge group with reduced rank.  For example, if $n_{27}=1$ and $n_{1}=0$, the {\bf 78}, {\bf 27}, $\overline{\bf 27}$ and a single {\bf 1} can combine into the {\bf 133} adjoint of $E_{7}$, with the remaining $24$ scalars joining into, say, $SU(5)$.  Simlarly, if $n_{27}=3$ we can have enhancement to $E_{8}$.  Whether such an enhancement in fact obtains depends on the interactions and thus requires a more refined analysis.  In any case, for sufficiently large $n_{1}$ and $n_{27}$, no enhancement can satisfy the rank condition.

%%%%%%%%%%%%% Change-Josh %%%%%%%%%%%%%%
Fortunately, the $n_{i}$ {\em cannot} be arbitrarily large, as can be seen already at the level of the GLSM for the base.  As discussed above, the $n_{i}$ are determined by the structure of the superpotential.  However, not every superpotential compatible with the charge assignments defines a good GLSM.  Two conditions hold particular importance:  first, the superpotential must be transverse in order to ensure that the branch structure of the GLSM is non-singular, leading to an upper bound on the $n_{i}$ --- for example, in this case we found $n_{27}\leq 6$; second, the resulting vector bundle must be semi-stable to avoid destabilization of the vacuum by worldsheet instantons, placing heavier constraints on the $J^\kappa$ (for our example, this implies that, at a minimum, the $J^\kappa$ must be linearly independent).  It would be interesting to work out the full list of allowed $n_{i}$ and check that it matches against known results for the allowed ranks of $\CN=(1,1)$ \susic\ heterotic compactifications to 6d.

%%%%%%%%%%%%%%%%%% T^2 (k1,k2)=(0,1) %%%%%%%%%%%%%%%%%%%%

\subsubsection{$\big(k^{(1)},k^{(2)} \big) = (0,1)$ for $T^2$ Base}
\label{sec:T2k=01}

In this twisted sector, the most generic states surviving the GSO projection are
\be
\sum_{\kappa_1,\ldots,\kappa_i=0}^{4} \bar{\gamma}^{[\kappa_1}_0  \cdots \bar{\gamma}^{\kappa_i]}_0 P^{[\kappa_1 \ldots \kappa_i ]}_{i+4d} \big( \phi_0 \big) |0\rangle
\ee
where $i=0,\ldots,5$, $d\geq -4\big\lfloor\frac{i}{4}\big\rfloor$ with $d\in\mathbb{Z}$, and $i+4d$ denotes the degree of the ${{5}\choose{i}}$ polynomials $P^{[\kappa_1\ldots\kappa_i]}_{i+4d}\big(\phi\big)$; we also have the same states with an insertion of $\bar{\chi}_0$.  These states have the quantum numbers $(q_L, q_R) = \big( -\frac{3}{2} + i + d , -1 + d \big)$ ($q_R = d$ when we have an insertion of $\bar{\chi}_0$).  Acting with $\mathcalQB_1$ yields
\be
\sim \sum_{\kappa_1,\ldots,\kappa_i} \bar{\gamma}_0^{[\kappa_1} \cdots \bar{\gamma}_0^{\kappa_{i-1}} J^{\kappa_i]}\big(\phi_0\big) P^{[\kappa_1\ldots\kappa_i]}_{i+4d}\big(\phi_0\big) |0\rangle \, .
\ee
We note that states with $i=0$ are annihilated by $\mathcalQB_1$.

Since the $(0,1)$ twisted sector is self-dual under spacetime CPT, we know that states must come in CPT pairs in this sector, allowing us to greatly reduce our work.  In particular, there are no states with $q_L < -\frac{3}{2}$ or $q_R < -1$, which means that there will be no states with $q_L > \frac{3}{2}$ or $q_R > 1$.  We list here the reduced list of potential states paired with their CPT duals --- the derivation of the degeneracies follows the table:
\be
\label{eqn:n27}
{\setlength\arraycolsep{8pt}
\begin{array}{|c||c|c|c|}
\hline \mathit{State} & \mathit{Degeneracy} & q_L & q_R  \\
\hline\hline |0\rangle & 1 & -\frac{3}{2} & -1  \\
\bar{\chi}_0 |0\rangle & 1 & -\frac{3}{2} & 0 \\
\hline \bar{\gamma}\bar{\gamma}P_6(\phi) |0\rangle & 1 & \frac{3}{2} & 0 \\
\bar{\gamma}\bar{\gamma} P_6(\phi) \bar{\chi}_0 |0\rangle & 1 & \frac{3}{2} & 1 \\
\hline\hline P_4(\phi) |0\rangle & n_{27} & -\frac{1}{2} & 0  \\
P_4(\phi) \bar{\chi}_0 |0\rangle & n_{27} & -\frac{1}{2} & 1\\
\hline \bar{\gamma}\bar{\gamma} P_2(\phi) |0\rangle  & n_{27} &  \frac{1}{2} & -1 \\
\bar{\gamma}\bar{\gamma} P_2(\phi)\bar{\chi}_0 |0\rangle & n_{27} & \frac{1}{2} & 0 \\
\hline\hline \bar{\gamma} P_1(\phi) |0\rangle &  n_{27} & -\frac{1}{2} & -1 \\
\bar{\gamma} P_1(\phi) \bar{\chi}_0 |0\rangle & n_{27} & -\frac{1}{2} & 0 \\
\hline \bar{\gamma} P_5(\phi) |0\rangle & n_{27} & \frac{1}{2} & 0 \\
\bar{\gamma} P_5(\phi) \bar{\chi}_0 |0\rangle & n_{27} & \frac{1}{2} & 1 \\
\hline
\end{array}}
\ee
Clearly, $|0\rangle$ contributes one element to $\mathcalQB$-cohomology, so we need only count $P_4(\phi)|0\rangle$ and $\bar{\gamma}P_1(\phi)|0\rangle$.  In fact, this is even simpler; there are ${{6}\choose{4}}=15$ potential $P_4$'s,  all of which are $\mathcalQB$-closed, and $5\times 3 = 15$ potential $\bar{\gamma}P_1$'s, none of which are $\mathcalQB$-exact.  Since $\mathcalQB_1 \big( \bar{\gamma} \cdot P_1 \big) \sim P_1 \cdot J$, we see immediately that there are the same number of $\mathcalQB$-cohomology representatives of the form $P_4(\phi)|0\rangle$ as $\bar{\gamma}P_1(\phi)|0\rangle$.  This is just the dimension of the degree 4 subspace of the local algebra
\be
\frac{\mathbb{C}[\phi^1,\phi^2,\phi^3]}{\langle J^0 , \ldots , J^4 \rangle} \, .
\ee
The exact number depends on the explicit choice of polynomials $J^{\kappa}(\phi)$, which we have presciently called $n_{27}$.  However, if we were quotienting by an ideal generated by three non-degenerate, degree 3 polynomials, the Poincar\'e polynomial would tell us that $n_{27}$ was 6.  This places an upper bound: $n_{27}\leq 6$.  The defining polynomials can be chosen such that $n_{27}=0$, as is the case with the example in equation (\ref{eqn:T2example}).

%%%%%%%%%%%%%%%%%% T^2 (k1,k2)=(1,1) %%%%%%%%%%%%%%%%%%%%

\subsubsection{$\big(k^{(1)},k^{(2)} \big) = (1,1)$ for $T^2$ Base}
\label{sec:T2k=11}

In this sector,
\bea
\mathcalQB_1 &=& -\sqrt{2}\, i  \sum_{i_1,i_2,i_3,\kappa} J^\kappa_{i_1 i_2 i_3} \bigg(  3\gamma^\kappa_{-\frac{5}{8}}  \phi^{i_1}_{\frac{7}{8}} \phi^{i_2}_{-\frac{1}{8}} \phi^{i_3}_{-\frac{1}{8}}  + \gamma^\kappa_{\frac{3}{8}}  \phi^{i_1}_{-\frac{1}{8}} \phi^{i_2}_{-\frac{1}{8}} \phi^{i_3}_{-\frac{1}{8}}   \bigg) + \ldots
\eea
where we only list the terms with strictly relevant oscillator modes.  

First, we catalog all states with $L_0 = 0$ that survive the left GSO projections, then we will restrict to $\mathcalQB_1$ cohomology; the right GSO projection will restrict to states with half-integral R-charges and will correlate $q_R$ with spacetime chirality.  For $(\hat{n}_1,\hat{n}_2) = (0,0)$,
\be
{\setlength\arraycolsep{8pt}
\begin{array}{|c||c|c|c|}
\hline State & Degeneracy & q_L & q_R        \\
\hline\hline X_{-1}^M |0\rangle & 4 & 0 &  -1 \\
\hline  \lambda^{I_{(1)}}_{-\frac{1}{2}} \lambda^{J_{(1)}}_{-\frac{1}{2}} |0\rangle & {{10}\choose{2}} &  0 &   -1  \\
\hline  \lambda^{I_{(2)}}_{-\frac{1}{2}} \lambda^{J_{(2)}}_{-\frac{1}{2}} |0\rangle & {{16}\choose{2}} &  0 &   -1  \\
\hline \vt_{-1} |0\rangle & 1 &  0 & -1 \\
\hline \bar{\vt}_{-1} |0 \rangle & 1 &  0 &  -1 \\
\hline  \phi^i_{-\frac{1}{8}} \bar{\phi}^j_{-\frac{7}{8}} |0\rangle  & 9   &   0  &   -1  \\
\hline   \gamma^{\kappa_1}_{-\frac{5}{8}}\bar{\gamma}^{\kappa_2}_{-\frac{3}{8}} |0\rangle    &  25 &  0  &  -1    \\
\hline \gamma^\kappa_{-\frac{5}{8}} P_3 \big( \phi_{-\frac{1}{8}} \big) |0\rangle  &  5\times {{5}\choose{3}} & 0 &  0   \\
\hline \lambda^{I_{(1)}}_{-\frac{1}{2}} \bar{\gamma}^\kappa_{-\frac{3}{8}} \phi^i_{-\frac{1}{8}} |0 \rangle  & 10 \times 15 &  1  &  -1  \\
\hline \lambda^{I_{(1)}}_{-\frac{1}{2}} P_4\big( \phi_{-\frac{1}{8}}\big) |0\rangle  &  10 \times {{6}\choose{4}} &  1 &  0 \\
\hline \bar{\gamma}^{\kappa_1}_{-\frac{3}{8}} \bar{\gamma}^{\kappa_2}_{-\frac{3}{8}} P_2\big( \phi_{-\frac{1}{8}} \big) |0\rangle   &  10 \times {{4}\choose{2}}  & 2 &  -1  \\
\hline \bar{\gamma}^\kappa_{-\frac{3}{8}} P_5\big( \phi_{-\frac{1}{8}}\big) |0\rangle    &  5 \times {{7}\choose{5}}  &  2 &  0 \\
\hline P_{8}\big( \phi_{-\frac{1}{8}} \big) |0\rangle     &  {{10}\choose{8}}  & 2 &  1 \\
\hline
\end{array}}
\ee
as well as the same states with an insertion of $\bar{\chi}_0$, which differ only by a shift $q_R \rightarrow q_R + 1$.  States with no dependence on $\bar{\phi}$ and $\bar{\gamma}$ are obviously annihilated by $\mathcalQB_1$.  As for the rest,
\bea
\label{eqn:QphiphibT2}
\mathcalQB_1 \Big( \phi^i_{-\frac{1}{8}} \bar{\phi}^j_{-\frac{7}{8}} |0\rangle \Big)   & \sim &    \sum_\kappa \phi^i_{-\frac{1}{8}} \p_j J^\kappa\big(\phi_{-\frac{1}{8}}\big) \gamma^\kappa_{-\frac{5}{8}} |0\rangle    \\
\label{eqn:QggbT2}
\mathcalQB_1 \Big( \gamma^{\kappa_1}_{-\frac{5}{8}} \bar{\gamma}^{\kappa_2}_{-\frac{3}{8}} |0\rangle \Big)   & \sim &    J^{\kappa_2}\big(\phi_{-\frac{1}{8}}\big) \gamma^{\kappa_1}_{-\frac{5}{8}} |0\rangle   \\
\mathcalQB_1 \Big( \lambda^{I_{(1)}}_{-\frac{1}{2}} \bar{\gamma}^\kappa_{-\frac{3}{8}} \phi^i_{-\frac{1}{8}} |0\rangle \Big)  & \sim &   \lambda^{I_{(1)}}_{-\frac{1}{2}} \phi^i_{-\frac{1}{8}} J^\kappa \big( \phi_{-\frac{1}{8}} \big) |0\rangle   \\
\mathcalQB_1\Big( \bar{\gamma}^{\kappa_1}_{-\frac{3}{8}} \bar{\gamma}^{\kappa_2}_{-\frac{3}{8}} P_2\big( \phi_{-\frac{1}{8}} \big) |0\rangle \Big)  & \sim &  \bar{\gamma}^{[\kappa_1}_{-\frac{3}{8}} J^{\kappa_2]}\big( \phi_{-\frac{1}{8}} \big) P_2\big( \phi_{-\frac{1}{8}} \big) |0\rangle  \\
\mathcalQB_1 \Big( \bar{\gamma}^\kappa_{-\frac{3}{8}} P_5\big(\phi_{-\frac{1}{8}}\big) |0\rangle \Big)  & \sim &   J^\kappa\big( \phi_{-\frac{1}{8}}\big) P_5\big(\phi_{-\frac{1}{8}}\big) |0\rangle   
\eea
Since the $J^\kappa$ are quasihomogeneous, it's clear that a linear combination of $\sum_i \phi^i_{-\frac{1}{8}} \bar{\phi}^i_{-\frac{7}{8}} |0\rangle$ and $\sum_\kappa \gamma^\kappa_{-\frac{5}{8}} \bar{\gamma}^\kappa_{-\frac{3}{8}}|0\rangle$ will be in $\mathcalQB$ cohomology.  If this were the only linear combination of (\ref{eqn:QphiphibT2}) and (\ref{eqn:QggbT2}) that vanished, then there would be $5 {{5}\choose{3}} - (9+25-1) = 17$ states of the form $\gamma^\kappa P_3(\phi) |0\rangle$ in $\mathcalQB$ cohomology however, the exact number depends on the defining polynomials, so call the totals $1+n_1$ and $17+n_1$.  The degeneracies of the rest of the states follow immediately from the analysis surrounding (\ref{eqn:n27}).  Including representations under the linearly realized $\big( SO(8) \times U(1)_L \big) \times SO(16) \subset SO(10) \times E_8$, we find that within $\mathcalQB$-cohomology we have
\be
{\setlength\arraycolsep{8pt}
\begin{array}{|c||c|c|c|c|}
\hline State & Degeneracy & SO(10)_{q_L}  &  SO(16) & q_R          \\
\hline\hline X_{-1}^M |0\rangle & 4 & \boldsymbol{1_0} &  \boldsymbol{1}  &  -1 \\
\hline  \lambda^{I_{(1)}}_{-\frac{1}{2}} \lambda^{J_{(1)}}_{-\frac{1}{2}} |0\rangle & 1 &  \boldsymbol{45_0}  &  \boldsymbol{1}  &   -1  \\
\hline  \lambda^{I_{(2)}}_{-\frac{1}{2}} \lambda^{J_{(2)}}_{-\frac{1}{2}} |0\rangle & 1  &  \boldsymbol{1_0}  &  \boldsymbol{120}  &   -1  \\
\hline \vt_{-1} |0\rangle & 1 &  \boldsymbol{1_0} & \boldsymbol{1} & -1  \\
\hline \bar{\vt}_{-1} |0 \rangle & 1 &  \boldsymbol{1_0}  &  \boldsymbol{1} &  -1 \\
\hline  \sim \big( \phi^i_{-\frac{1}{8}} \bar{\phi}^j_{-\frac{7}{8}}  +  \gamma^{\kappa_1}_{-\frac{5}{8}}\bar{\gamma}^{\kappa_2}_{-\frac{3}{8}} \big)   |0\rangle  & 1 + n_1   &   \boldsymbol{1_0}  &  \boldsymbol{1}  &   -1  \\
\hline \gamma^\kappa_{-\frac{5}{8}} P_3 \big( \phi_{-\frac{1}{8}} \big) |0\rangle  &  17+n_1 & \boldsymbol{1_0} &  \boldsymbol{1}  &  0    \\
\hline \lambda_{-\frac{1}{2}}^{I_{(1)}} \bar{\gamma}_{-\frac{3}{8}}^\kappa \phi^i_{-\frac{1}{8}} |0\rangle   &  n_{27} & \boldsymbol{10_1} & \boldsymbol{1} & -1 \\
\hline \lambda_{-\frac{1}{2}}^{I_{(1)}} P_4\big(\phi_{-\frac{1}{8}}\big) |0\rangle & n_{27} & \boldsymbol{10_1} & \boldsymbol{1} & 0  \\
\hline \bar{\gamma}^{\kappa_1}_{-\frac{3}{8}} \bar{\gamma}^{\kappa_2}_{-\frac{3}{8}} P_2\big(\phi_{-\frac{1}{8}}\big) |0\rangle  & n_{27} & \boldsymbol{1_2} & \boldsymbol{1} & -1  \\
\hline \bar{\gamma}^\kappa_{-\frac{3}{8}} P_5\big( \phi_{-\frac{1}{8}}\big) |0\rangle & n_{27} & \boldsymbol{1_2} & \boldsymbol{1} & 0 \\
\hline
\end{array}}
\ee
as well as the states multiplied by $\bar{\chi}_0$ which have $q_R \rightarrow q_R+1$.  
For the example in equation (\ref{eqn:T2example}), $n_1=0$.

%%%%%%%%%%%%%%%%%%%%% K3 BASE %%%%%%%%%%%%%%%%%%%%%%

\subsection{$d=4$, $K3$ Base}

For our next example, consider the model with charges
\be
{\setlength\arraycolsep{8pt}
\begin{array}{|c||c|c|c|c|c|}
\hline  & \boldsymbol{\Phi^{i=1,\ldots,4}}  & \boldsymbol{P} & \boldsymbol{\tG} & \boldsymbol{\G^{m=1,\ldots,5}} & \boldsymbol{\Theta}  \\
\hline\hline \boldsymbol{U(1)}   &  1  &  -5  &  -4  &  1  &  (3R+i S)      \\
\hline \boldsymbol{U(1)_L}   &  \frac{1}{5}  &  0  &  -\frac{4}{5}  &  -\frac{4}{5}  &  \big(  \frac{3}{5}R + \frac{i}{5}S \big)    \\
\hline \boldsymbol{U(1)_R}   &  \frac{1}{5}  &  0  &  \frac{1}{5}  &  \frac{1}{5}  &  \big(  \frac{3}{5} R + \frac{i}{5} S \big)  \\
\hline
\end{array}}
\ee
where the $S^1$ radii are $R = \frac{1}{\sqrt{3}}$ and $S=1$.  These charges are non-anomalous and give $c_L = \hat{c}_R + r_L = 6 + 4$.  We will take this model to be embedded in a heterotic $E_8 \times E_8$ model, so the gauge-bundle fermions will transform in an $SU(4)$ subgroup of, say, the first $E_8$.  Since the left-handed fermions have the same $U(1)_L$ and $U(1)_R$ charges, let $\kappa=0,\ldots,5$ and define $\gamma^0 \equiv \tilde{\gamma}$ and $J^0(\phi)\equiv G(\phi)$.  In the Landau-Ginzburg orbifold phase, we see that we must orbifold by the discrete group $\G = \mathbb{Z}_5$, and so our left GSO group will be isomorphic to $\mathbb{Z}_{10} \times \mathbb{Z}_2$.  In a twisted sector $\big( k^{(1)}, k^{(2)} \big)$, the oscillators are
{\setlength\arraycolsep{0pt}
\bea
&& \phi_{n - \frac{k^{(1)}}{10}}^i  \, , \qquad \bar{\phi}{}_{n + \frac{k^{(1)}}{10}}^i  \, ,  \qquad     \gamma_{n+\frac{2k^{(1)}}{5}}^\kappa  \, ,  \qquad   \bar{\gamma}{}_{n-\frac{2k^{(1)}}{5}}^\kappa  \, ,  \qquad  \lambda_{n+\frac{k^{(1)}}{2}}^{I_{(1)}} \, ,  \qquad  \lambda_{n+\frac{k^{(2)}}{2}}^{I_{(2)}} \, ,    \nonumber \\
&&  X_{n}^{M} \, ,  \qquad  \vt_{n}   \, ,   \qquad  \bar{\vt}_{n}  \, ,  \qquad  \bar{\chi}_0 \, , 
\eea}
where the handling of zero modes was explained in section \ref{sec:LG}.

We have $20$ twisted sectors $k^{(1)}=0,\ldots,9$, $k^{(2)}=0,1$ however, spacetime CPT invariance means we need only keep track of $k^{(1)}=0,\ldots,5$, $k^{(2)}=0,1$.  Of course, $k^{(2)}=0$ and $k^{(2)}=1$ differ in energy by $\Delta L_0 = 1$, so we will only list $k^{(2)}=0$ when it doesn't take us above $L_0 = 0$.  For these sectors, the ground state quantum numbers and relevant oscillators are
\be
{\setlength\arraycolsep{6pt}
\begin{array}{|c||c|c|c|c|c|}
\hline \!\!\! \boldsymbol{\big(k^{(1)}, k^{(2)}\big)} \!\!\! &\boldsymbol{L_0} & \boldsymbol{q_L} & \boldsymbol{q_R} & \!\! \boldsymbol{(\hat{n}_1 , \hat{n}_2 )} \!\!   &  \mathit{Relevant~Oscillators} \\
\hline\hline  \boldsymbol{(0,1)} & 0 & -2 & -\frac{3}{2} & (0,0)  &   \mathrm{Zero~modes} \\
\hline \boldsymbol{(1,1)} & -1 & 0 &  -\frac{3}{2} & (0,0)  & \!\! \phi_{-\frac{1}{10}}^{i}  ,~  \bar{\phi}{}_{-\frac{9}{10}}^{i}  ,  ~  \gamma_{-\frac{3}{5}}^\kappa  , ~ \bar{\gamma}{}_{-\frac{2}{5}}^\kappa  ,  ~  \lambda_{-\frac{1}{2}}^{I_{(1)}}  ,   \!\!  \\
& & & &  &  \lambda_{-\frac{1}{2}}^{I_{(2)} }   , ~ X_{-1}^{M}  , ~ \vt_{-1}  , ~ \bar{\vt}_{-1}  , ~ \bar{\chi}_0     \\
\hline \boldsymbol{(1,1)}   &   -\frac{1}{5} & -\frac{2}{5} & -\frac{19}{10} & (0,-1)  &  \phi_{-\frac{1}{10}}^{i} \, , ~~  \bar{\chi}_0    \\
\hline \boldsymbol{(1,0)} & 0 & 0 & -\frac{3}{2} & (0,0) &  \lambda_{0}^{I_{(2)}}  , ~~ \bar{\chi}_0  \\
\hline \boldsymbol{(2,1)} & 0 & 2 &  - \frac{3}{2} &  (0,0)  &  \lambda_{0}^{I_{(1)}} , ~~ \bar{\chi}_0  \\
\hline \boldsymbol{(3,1)} & -\frac{1}{5} & -\frac{4}{5} & -\frac{3}{10} & (0,0)  &   \bar{\gamma}{}_{-\frac{1}{5}}^\kappa \, , ~~ \bar{\chi}_0   \\
\hline \boldsymbol{(4,1)} & \frac{2}{5} &  \frac{6}{5} & -\frac{3}{10} &  (0,0)  &    \\
\hline \boldsymbol{(5,1)} & 1 &  -\frac{8}{5}  &  \frac{9}{10} &  (0,0)  &   \\
\hline
\end{array}}
\ee
where $(\hat{n}_1,\hat{n}_2)$ indicates the winding sector.

The computation of the massless spectrum now proceeds in exactly the same way as \cite{Kachru:1993pg} and \cite{Distler:1993mk}.  In subsections \ref{sec:K3k=01} and \ref{sec:k=11}, we'll reproduce the $\big( k^{(1)} , k^{(2)} \big) = (0,1)$ and $(1,1)$ sectors as an example, with the rest left to the reader.

The spectrum, organized by twisted sector and charge under the linearly realized $\big( SO(8) \times U(1)_L \big) \times SO(16) \times U(1)_R \subset SO(10) \times E_8 \times U(1)_R $ and four-dimensional Poincar\'e group, is: \newpage
\be
{\setlength\arraycolsep{8pt}
\begin{array}{|c||c|c|c|c|c|}
\hline \big( k^{(1)}, k^{(2)} \big) & \mathit{Degeneracy} & SO(8)_{q_L}  &  SO(16) & q_R  &  SO(3,1)          \\
\hline\hline \boldsymbol{(0,1)}  & n_{10} & \boldsymbol{8_{s,0}} & \boldsymbol{1} & -\frac{3}{2} & \boldsymbol{\bar{2}} \\
\hline    &  68 + 3n_{10} & \boldsymbol{8_{s,0}}  &  \boldsymbol{1} & -\frac{1}{2} & \boldsymbol{2}    \\
\hline   &  68 + 3n_{10} & \boldsymbol{8_{s,0}}  &  \boldsymbol{1} & \frac{1}{2} & \boldsymbol{\bar{2}}    \\
\hline & n_{10} & \boldsymbol{8_{s,0}} & \boldsymbol{1} & \frac{3}{2} & \boldsymbol{2} \\
\hline & 1 & \boldsymbol{8_{s,-2}}  &  \boldsymbol{1} & -\frac{3}{2}  &  \boldsymbol{\bar{2}}       \\
\hline & 1 & \boldsymbol{8_{s,-2}}  &  \boldsymbol{1} & -\frac{1}{2}  & \boldsymbol{2}      \\
\hline & 1 & \boldsymbol{8_{s,2}} & \boldsymbol{1}  &  \frac{1}{2}  &  \boldsymbol{\bar{2}}  \\
\hline & 1 & \boldsymbol{8_{s,2}} & \boldsymbol{1}  &  \frac{3}{2}  &  \boldsymbol{2}  \\
\hline & n_{16} & \boldsymbol{8'_{s,-1}} & \boldsymbol{1} & -\frac{3}{2} & \boldsymbol{\bar{2}}  \\
\hline & 32 + 2n_{16} & \boldsymbol{8'_{s,-1}} & \boldsymbol{1}  &  -\frac{1}{2}  &  \boldsymbol{2}  \\
\hline & 32 + n_{16} & \boldsymbol{8'_{s,-1}} & \boldsymbol{1}  &  \frac{1}{2}  &  \boldsymbol{\bar{2}}  \\
\hline & 32 + n_{16} & \boldsymbol{8'_{s,1}} & \boldsymbol{1}  &  -\frac{1}{2} &  \boldsymbol{2}  \\
\hline & 32 + 2n_{16} & \boldsymbol{8'_{s,1}} & \boldsymbol{1}  &  \frac{1}{2} &  \boldsymbol{\bar{2}}  \\
\hline & n_{16} & \boldsymbol{8'_{s,1}} & \boldsymbol{1} & \frac{3}{2} & \boldsymbol{2}  \\
\hline\hline \boldsymbol{(1,1)} & 1 & \boldsymbol{1_0} & \boldsymbol{1} &  -\frac{3}{2}  &  \boldsymbol{4_v} \otimes \boldsymbol{\bar{2}}  \\
\hline  & 1 & \boldsymbol{1_0} & \boldsymbol{1} &  -\frac{1}{2}  &  \boldsymbol{4_v} \otimes \boldsymbol{2}  \\
\hline  & 1 &  \boldsymbol{28_0}  &  \boldsymbol{1}  &   -\frac{3}{2}  &  \boldsymbol{\bar{2}} \\
\hline  & 1 &  \boldsymbol{28_0}  &  \boldsymbol{1}  &   -\frac{1}{2}  &  \boldsymbol{2} \\
\hline   & 1  &  \boldsymbol{1_0}  &  \boldsymbol{120}  &   -\frac{3}{2}  &  \boldsymbol{\bar{2}}  \\
\hline  & 1 & \boldsymbol{1_0} & \boldsymbol{120} & -\frac{1}{2}  &  \boldsymbol{2}   \\
\hline  & 13+n_1 &  \boldsymbol{1_0} & \boldsymbol{1} & -\frac{3}{2}  &  \boldsymbol{\bar{2}}   \\
\hline  &  172+2n_1 & \boldsymbol{1_0} &  \boldsymbol{1}  &  -\frac{1}{2}  & \boldsymbol{2}  \\
\hline & 159+n_1 & \boldsymbol{1_0} & \boldsymbol{1} & \frac{1}{2}  &  \boldsymbol{\bar{2}}   \\
\hline & n_{16} & \boldsymbol{8_{v,1}} & \boldsymbol{1} & -\frac{3}{2} & \boldsymbol{\bar{2}}  \\
\hline  &  32+2n_{16}  &  \boldsymbol{8_{v,1}} &  \boldsymbol{1}  &  -\frac{1}{2}   & \boldsymbol{2} \\
\hline  &  32+n_{16}  & \boldsymbol{ 8_{v,1}} &  \boldsymbol{1}  &  \frac{1}{2}   & \boldsymbol{\bar{2}} \\
\hline & n_{10} & \boldsymbol{1_2} & \boldsymbol{1} & -\frac{3}{2}  & \boldsymbol{\bar{2}} \\
\hline    &  68+3n_{10}  &  \boldsymbol{1_2}  &  \boldsymbol{1}  &  -\frac{1}{2} &  \boldsymbol{2}    \\
\hline    &  68+3n_{10}  &  \boldsymbol{1_2}  &  \boldsymbol{1}  &  \frac{1}{2} &  \boldsymbol{\bar{2}}    \\
\hline & n_{10} & \boldsymbol{1_2} & \boldsymbol{1} & \frac{3}{2}  & \boldsymbol{2} \\
\hline\hline  \boldsymbol{(1,0)}  &  1  &  \boldsymbol{1_0}  &  \boldsymbol{128}  &  -\frac{3}{2}  &  \boldsymbol{\bar{2}}     \\
\hline   &  1  &  \boldsymbol{1_0}  &  \boldsymbol{128}  &  -\frac{1}{2}  &  \boldsymbol{2}     \\
\hline\hline \boldsymbol{(2,1)}  &  1  &  \boldsymbol{8_{s,2}}  &  \boldsymbol{1}  &  -\frac{3}{2}  &  \boldsymbol{\bar{2}}  \\
\hline  &  1  &  \boldsymbol{8_{s,2}}  &  \boldsymbol{1}  &  -\frac{1}{2}  &  \boldsymbol{2}  \\
\hline\hline \boldsymbol{(3,1)}   &  6  &  \boldsymbol{1_0}  &  \boldsymbol{1}  &  -\frac{1}{2}  &  \boldsymbol{2}  \\
\hline   &  6  &  \boldsymbol{1_0}  &  \boldsymbol{1}  &  \frac{1}{2}  &  \boldsymbol{\bar{2}}  \\
\hline
\end{array}}
\ee
$n_1$, $n_{10}$, and $n_{16}$, depend on specific details of the defining polynomials $J^\kappa(\phi)$, as can be seen in the following subsections.  In particular, $n_{10}$ and $(32+n_{16})$ have the nice interpretations as the dimensions of the subspaces of degrees 10 and 5 polynomials, respectively, in the local algebra
\be
\frac{\mathbb{C}[\phi^1,\phi^2,\phi^3,\phi^4]}{\langle J^0, \ldots, J^5 \rangle} \, .
\ee 
If there were only four $J^\kappa$'s, the Poincar\'e polynomial would tell us that $(32+n_{16})$ was $40$ and $n_{10}$ was $10$.  However, since there are six such polynomials, this only yields upper bounds: $n_{10}\leq 10$, $n_{16}\leq 8$.  (The interpretation of $n_1$ is not as clean, so we leave its explanation for the subsections.)
Meanwhile, the linearly realized $SO(8)\times U(1)_L$ is enhanced to $SO(10)$, just as $SO(16)$ is enhanced to $E_8$.  We can see the former by noting the decomposition of $SO(10)\longrightarrow SO(8)_{U(1)}$ representations:
\bea
\mathbf{45} &\longrightarrow &\mathbf{28_0} \oplus \mathbf{8_{s,-2}} \oplus \mathbf{8_{s,2}} \oplus \mathbf{1_0}  \\
\mathbf{16} &\longrightarrow & \mathbf{8'_{s,-1}} \oplus \mathbf{8_{v,1}}  \\
\mathbf{\overline{16}} &\longrightarrow & \mathbf{8'_{s,1}} \oplus \mathbf{8_{v,-1}}  \\
\mathbf{10} &\longrightarrow & \mathbf{8_{s,0}} \oplus \mathbf{1_{-2}} \oplus \mathbf{1_2} 
\eea

As explained in \cite{Kachru:1993pg}, spacetime fermions with $q_R = \pm \frac{3}{2}$ fall into $\mathcal{N}=1$ vector multiplets, and spacetime fermions with $q_R = \pm \frac{1}{2}$ fall into $\mathcal{N}=1$ chiral/anti-chiral multiplets.  Recalling that $\mathcal{N}=2$ vector multiplets contain an $\mathcal{N}=1$ vector and chiral multiplet, and that $\mathcal{N}=2$ hypermultiplets contain two $\mathcal{N}=1$ chiral multiplets, we see that the fermionic spectrum actually organizes into $\mathcal{N}=2$ multiplets.  Summarizing,
\be
{\setlength\arraycolsep{8pt}
\begin{array}{|c|c|c|}
\hline d=4,~\mathcal{N}=2~\mathit{Repr.}  & \mathit{Degeneracy} &  SO(10) \times E_8   \\
\hline\hline \mathrm{Supergravity} & 1 & \mathbf{1}\otimes\mathbf{1}  \\
\hline \mathrm{Vector}  &  1  &  \mathbf{45}\otimes\mathbf{1}    \\
\hline \mathrm{Vector} & 1 & \mathbf{1}\otimes\mathbf{248}  \\
\hline \mathrm{Vector} & n_{10} & \mathbf{10} \otimes\mathbf{1} \\
\hline \frac{1}{2}\textrm{-Vector} & n_{16} & \mathbf{16}\otimes\mathbf{1} \\
\hline \frac{1}{2}\textrm{-Vector} & n_{16} & \mathbf{\overline{16}}\otimes\mathbf{1} \\
\hline \mathrm{Vector} & 12+n_1 & \mathbf{1} \otimes\mathbf{1}  \\
\hline \frac{1}{2}\textrm{-Hyper} & 68+2n_{10} & \mathbf{10} \otimes \mathbf{1}  \\
\hline \frac{1}{2}\textrm{-Hyper} & 32+n_{16} & \mathbf{16} \otimes \mathbf{1} \\
\hline \frac{1}{2}\textrm{-Hyper} & 32+n_{16} & \mathbf{\overline{16}} \otimes \mathbf{1}  \\
\hline \textrm{Hyper} & 165+n_1 & \mathbf{1}\otimes\mathbf{1} \\
\hline
\end{array}}
\ee

For a concrete example, take
\be
\label{eqn:K3example}
G = \sum_i \big(\phi^i\big)^4 + a \big( \phi^1 \phi^3\big)^2 \, ,  \quad   J^i = \big(\phi^i\big)^4 \, , \quad  J^5 = \big( \phi^2 \phi^4 \big)^2
% NOTE: this one has $n_{10}=1$, $n_{16}$=2, and $n_1=3$:
%J^5 = \big( \phi^1 \big)^3 \phi^3 \, .
\ee
In this case, we have $n_{10} =0$, $n_{16}=0$, and $n_1=3$, so there is no enhancement from $SO(10)$.

%%%%%%%%%%%%%%%%%% K3 (k1,k2)=(0,1) %%%%%%%%%%%%%%%%%%%%

\subsubsection{$\big(k^{(1)},k^{(2)} \big) = (0,1)$ for $K3$ Base}
\label{sec:K3k=01}

In this twisted sector, the most generic states surviving the GSO projection are
\be
\sum_{\kappa_1,\ldots,\kappa_i=0}^{5} \bar{\gamma}^{[\kappa_1}_0  \cdots \bar{\gamma}^{\kappa_i]}_0 P^{[\kappa_1 \ldots \kappa_i ]}_{i+5d} \big( \phi_0 \big) |0\rangle
\ee
where $i=0,\ldots,6$, $d\geq -5\big\lfloor\frac{i}{5}\big\rfloor$ with $d\in\mathbb{Z}$, and $i+5d$ denotes the degree of the ${{6}\choose{i}}$ polynomials $P^{[\kappa_1\ldots\kappa_i]}_{i+5d}\big(\phi\big)$; we also have the same states with an insertion of $\bar{\chi}_0$.  These states have the quantum numbers $(q_L, q_R) = \big( -2 + i + d , -\frac{3}{2} + d \big)$ ($q_R = -\frac{1}{2}+ d$ when we have an insertion of $\bar{\chi}_0$).  Acting with $\mathcalQB_1$ yields
\be
\sim \sum_{\kappa_1,\ldots,\kappa_i} \bar{\gamma}_0^{[\kappa_1} \cdots \bar{\gamma}_0^{\kappa_{i-1}} J^{\kappa_i]}\big(\phi_0\big) P^{[\kappa_1\ldots\kappa_i]}_{i+5d}\big(\phi_0\big) |0\rangle \, .
\ee
We note that states with $i=0$ are annihilated by $\mathcalQB_1$.

Since the $(0,1)$ twisted sector is self-dual under spacetime CPT, we know that states must come in CPT pairs in this sector, allowing us to greatly reduce our work.  In particular, there are no states with $q_L < -2$ or $q_R < -\frac{3}{2}$, which means that there will be no states with $q_L > 2$ or $q_R > \frac{3}{2}$.  We list here the reduced list of potential states paired with their CPT duals --- the derivation of the degeneracies follows the table:
\be
\label{eqn:n16}
{\setlength\arraycolsep{8pt}
\begin{array}{|c||c|c|c|}
\hline \mathit{State} & \mathit{Degeneracy} & q_L & q_R  \\
\hline\hline |0\rangle & 1 & -2 & -\frac{3}{2}  \\
\bar{\chi}_0 |0\rangle & 1 & -2 & -\frac{1}{2} \\
\hline \bar{\gamma}\bar{\gamma}P_{12}(\phi) |0\rangle & 1  & 2 & \frac{1}{2} \\
\bar{\gamma}\bar{\gamma} P_{12}(\phi) \bar{\chi}_0 |0\rangle & 1 & 2 & \frac{3}{2} \\
\hline\hline P_5(\phi) |0\rangle &  32 + n_{16} & -1 & -\frac{1}{2}  \\
P_5(\phi) \bar{\chi}_0 |0\rangle &  32+n_{16} & -1 & \frac{1}{2} \\
\hline \bar{\gamma}\bar{\gamma} P_7(\phi) |0\rangle  & 32+n_{16} &  1 & -\frac{1}{2} \\
\bar{\gamma}\bar{\gamma} P_7(\phi)\bar{\chi}_0 |0\rangle & 32+n_{16}  & 1 & \frac{1}{2} \\
\hline\hline P_{10}(\phi) |0\rangle &  n_{10} & 0 & \frac{1}{2}  \\
P_{10}(\phi) \bar{\chi}_0 |0\rangle &  n_{10} & 0 & \frac{3}{2} \\
\hline \bar{\gamma}\bar{\gamma} P_2(\phi) |0\rangle  &  n_{10} &  0 & -\frac{3}{2} \\
\bar{\gamma}\bar{\gamma} P_2(\phi)\bar{\chi}_0 |0\rangle &  n_{10} & 0 & -\frac{1}{2} \\
\hline\hline \bar{\gamma} P_1(\phi) |0\rangle &  n_{16} & -1 & -\frac{3}{2} \\
\bar{\gamma} P_1(\phi) \bar{\chi}_0 |0\rangle &  n_{16} & -1 & -\frac{1}{2} \\
\hline \bar{\gamma} P_{11}(\phi) |0\rangle &  n_{16} & 1 & \frac{1}{2}  \\
\bar{\gamma} P_{11}(\phi) \bar{\chi}_0 |0\rangle &  n_{16} & 1 & \frac{3}{2}  \\
\hline\hline \bar{\gamma} P_6(\phi) |0\rangle &  68+2n_{10} & 0 & -\frac{1}{2} \\
\bar{\gamma} P_6(\phi) \bar{\chi}_0 |0\rangle &  68+2n_{10} & 0 & \frac{1}{2} \\
\hline
\end{array}}
\ee
Clearly, $|0\rangle$ contributes one element to $\mathcalQB$-cohomology, so we need only count $P_5(\phi)|0\rangle$, $P_{10}(\phi)|0\rangle$, $\bar{\gamma} P_1(\phi)|0\rangle$, and $\bar{\gamma}P_6(\phi)|0\rangle$.  Since the number of elements in $\mathcalQB$-cohomology from $\bar{\gamma}\bar{\gamma}P_2(\phi)$ and $P_{10}(\phi)$ is the same, call it $n_{10}$, and since $\mathcalQB_1 (\bar{\gamma}\bar{\gamma}P_2(\phi)) \sim \bar{\gamma} P_2(\phi) J(\phi)$ and $\mathcalQB_1(\bar{\gamma} P_6(\phi) ) \sim P_6(\phi) J(\phi)$, we can do a simple counting to find
\be
{\setlength\arraycolsep{8pt}
\begin{array}{|c||c|c|c|c|}
\hline \mathit{State} & \mathit{\#~of~States} & \mathit{\#~of~Exact} & \mathit{\#~of~Closed} & \mathit{\#~in~Cohomology} \\
\hline\hline \bar{\gamma}\bar{\gamma} P_2(\phi)|0\rangle  & 150 & 0  &  n_{10} & n_{10}  \\
\hline \bar{\gamma}P_6(\phi)|0\rangle  & 504 & 150-n_{10} & 218+n_{10} & 68+2n_{10}  \\
\hline P_{10}(\phi)|0\rangle & 286 & 286-n_{10} & 286 & n_{10}  \\
\hline
\end{array}}
\ee
Similarly, since none of the $24~$ $\bar{\gamma}P_1(\phi)|0\rangle$'s are exact, since all of 56 $P_5(\phi)|0\rangle$'s are closed, and since $\mathcalQB_1(\bar{\gamma} P_1(\phi)) \sim P_1(\phi) J(\phi)$, we see that if there are $n_{16}~$ $\bar{\gamma}P_1(\phi)|0\rangle$'s, then there will be $32+n_{16}~$ $P_5(\phi)|0\rangle$'s.  We can state this more cleanly by recognizing that $n_{10}$ and $(32+n_{16})$ are the dimensions of the subspaces of degrees 10 and 5 polynomials, respectively, of the local algebra
\be
\frac{\mathbb{C}[\phi^1,\phi^2,\phi^3,\phi^4]}{\langle J^0 , \ldots, J^5 \rangle} \, .
\ee
If we had four non-degenerate $J^\kappa$'s, then the Poincar\'e polynomial would tell us that $n_{10}$ is 10 and $n_{16}$ is 8.  Instead, this can only place upper bounds: $n_{10}\leq 10$ and $n_{16}\leq 8$, while both can be zero if the $J^\kappa$ are chosen appropriately (the example in equation (\ref{eqn:K3example}) yields $n_{10}=n_{16}=0$).
Bundle stability conditions will place additional constraints on the $n_i$.

%%%%%%%%%%%%%%%%%% (k1,k2)=(1,1) %%%%%%%%%%%%%%%%%%%%

\subsubsection{$\big(k^{(1)},k^{(2)} \big) = (1,1)$}
\label{sec:k=11}

In this twisted sector, we have
\bea
\mathcalQB_1 &=& -\sqrt{2}\, i  \sum_{i_1,\ldots,i_4,\kappa} J^\kappa_{i_1\ldots i_4} \bigg(  4\gamma^\kappa_{-\frac{3}{5}}  \phi^{i_1}_{\frac{9}{10}} \phi^{i_2}_{-\frac{1}{10}} \phi^{i_3}_{-\frac{1}{10}} \phi^{i_4}_{-\frac{1}{10}} + \gamma^\kappa_{\frac{2}{5}}  \phi^{i_1}_{-\frac{1}{10}} \cdots \phi^{i_4}_{-\frac{1}{10}}   \bigg) + \ldots
\eea
where we only list the terms with strictly relevant oscillator modes.  

First, we catalog all states with $L_0 = 0$ that survive the left GSO projections, then we will restrict to $\mathcalQB_1$ cohomology; the right GSO projection will restrict to states with half-integral R-charges and will correlate $q_R$ with spacetime chirality.  For $(\hat{n}_1,\hat{n}_2) = (0,0)$,
\be
{\setlength\arraycolsep{8pt}
\begin{array}{|c||c|c|c|}
\hline State & Degeneracy & q_L & q_R        \\
\hline\hline X_{-1}^M |0\rangle & 2 & 0 &  -\frac{3}{2} \\
\hline  \lambda^{I_{(1)}}_{-\frac{1}{2}} \lambda^{J_{(1)}}_{-\frac{1}{2}} |0\rangle & {{8}\choose{2}} &  0 &   -\frac{3}{2}  \\
\hline  \lambda^{I_{(2)}}_{-\frac{1}{2}} \lambda^{J_{(2)}}_{-\frac{1}{2}} |0\rangle & {{16}\choose{2}} &  0 &   -\frac{3}{2}  \\
\hline \vt_{-1} |0\rangle & 1 &  0 & -\frac{3}{2} \\
\hline \bar{\vt}_{-1} |0 \rangle & 1 &  0 &  -\frac{3}{2} \\
\hline  \phi^i_{-\frac{1}{10}} \bar{\phi}^j_{-\frac{9}{10}} |0\rangle  & 16   &   0  &   -\frac{3}{2}  \\
\hline   \gamma^{\kappa_1}_{-\frac{3}{5}}\bar{\gamma}^{\kappa_2}_{-\frac{2}{5}} |0\rangle    &  36 &  0  &  -\frac{3}{2}    \\
\hline \gamma^\kappa_{-\frac{3}{5}} P_4 \big( \phi_{-\frac{1}{10}} \big) |0\rangle  &  6\times {{7}\choose{4}} & 0 &  -\frac{1}{2}   \\
\hline \lambda^{I_{(1)}}_{-\frac{1}{2}} \bar{\gamma}^\kappa_{-\frac{2}{5}} \phi^i_{-\frac{1}{10}} |0 \rangle  & 8 \times 24 &  1  &  -\frac{3}{2}  \\
\hline \lambda^{I_{(1)}}_{-\frac{1}{2}} P_5\big( \phi_{-\frac{1}{10}}\big) |0\rangle  &  8 \times {{8}\choose{5}} &  1 &  -\frac{1}{2} \\
\hline \bar{\gamma}^{\kappa_1}_{-\frac{2}{5}} \bar{\gamma}^{\kappa_2}_{-\frac{2}{5}} P_2\big( \phi_{-\frac{1}{10}} \big) |0\rangle   &  15 \times {{5}\choose{2}}  & 2 &  -\frac{3}{2}  \\
\hline \bar{\gamma}^\kappa_{-\frac{2}{5}} P_6\big( \phi_{-\frac{1}{10}}\big) |0\rangle    &  6 \times {{9}\choose{6}}  &  2 &  -\frac{1}{2} \\
\hline P_{10}\big( \phi_{-\frac{1}{10}} \big) |0\rangle     &  {{13}\choose{10}}  & 2 &  \frac{1}{2} \\
\hline
\end{array}}
\ee
as well as the same states with an insertion of $\bar{\chi}_0$, which differ only by a shift $q_R \rightarrow q_R + 1$.  States with no dependence on $\bar{\phi}$ and $\bar{\gamma}$ are obviously annihilated by $\mathcalQB_1$.  As for the rest,
\bea
\label{eqn:Qphiphib}
\mathcalQB_1 \Big( \phi^i_{-\frac{1}{10}} \bar{\phi}^j_{-\frac{9}{10}} |0\rangle \Big)   & \sim &    \sum_\kappa \phi^i_{-\frac{1}{10}} \p_j J^\kappa\big(\phi_{-\frac{1}{10}}\big) \gamma^\kappa_{-\frac{3}{5}} |0\rangle    \\
\label{eqn:Qggb}
\mathcalQB_1 \Big( \gamma^{\kappa_1}_{-\frac{3}{5}} \bar{\gamma}^{\kappa_2}_{-\frac{2}{5}} |0\rangle \Big)   & \sim &    J^{\kappa_2}\big(\phi_{-\frac{1}{10}}\big) \gamma^{\kappa_1}_{-\frac{3}{5}} |0\rangle   \\
\label{eqn:QLgbphi}
\mathcalQB_1 \Big( \lambda^{I_{(1)}}_{-\frac{1}{2}} \bar{\gamma}^\kappa_{-\frac{2}{5}} \phi^i_{-\frac{1}{10}} |0\rangle \Big)  & \sim &   \lambda^{I_{(1)}}_{-\frac{1}{2}} \phi^i_{-\frac{1}{10}} J^\kappa \big( \phi_{-\frac{1}{10}} \big) |0\rangle   \\
\label{eqn:Qgbgbphi}
\mathcalQB_1\Big( \bar{\gamma}^{\kappa_1}_{-\frac{2}{5}} \bar{\gamma}^{\kappa_2}_{-\frac{2}{5}} P_2\big( \phi_{-\frac{1}{10}} \big) |0\rangle \Big)  & \sim &  \bar{\gamma}^{[\kappa_1}_{-\frac{2}{5}} J^{\kappa_2]}\big( \phi_{-\frac{1}{10}} \big) P_2\big( \phi_{-\frac{1}{10}} \big) |0\rangle  \\
\label{eqn:Qgbphi}
\mathcalQB_1 \Big( \bar{\gamma}^\kappa_{-\frac{2}{5}} P_6\big(\phi_{-\frac{1}{10}}\big) |0\rangle \Big)  & \sim &   J^\kappa\big( \phi_{-\frac{1}{10}}\big) P_6\big(\phi_{-\frac{1}{10}}\big) |0\rangle   
\eea
Since the $J^\kappa$ are quasihomogeneous, it's clear that a linear combination of $\sum_i \phi^i_{-\frac{1}{10}} \bar{\phi}^i_{-\frac{9}{10}} |0\rangle$ and $\sum_\kappa \gamma^\kappa_{-\frac{3}{5}} \bar{\gamma}^\kappa_{-\frac{2}{5}}|0\rangle$ will be in $\mathcalQB$ cohomology.  Depending on the choices of the $J^\kappa(\phi)$, there may be more, say $1+n_1$.  Then there will be $6 {{7}\choose{4}} - (16+36-1-n_1) = 159+n_1$ states of the form $\gamma^\kappa P_4(\phi) |0\rangle$ in $\mathcalQB$-cohomology.   The analysis surrounding (\ref{eqn:n16}) explains the contribution to $\mathcalQB$-cohomology of the remaining states.

Summarizing the results, and including representations under the linearly realized $\big( SO(8) \times U(1)_L \big) \times SO(16) \subset SO(10) \times E_8$, we find that within $\mathcalQB$ cohomology we have
\be
{\setlength\arraycolsep{8pt}
\begin{array}{|c||c|c|c|c|}
\hline State & Degeneracy & SO(8)_{q_L}  &  SO(16) & q_R          \\
\hline\hline X_{-1}^M |0\rangle & 2 & \boldsymbol{1_0} &  \boldsymbol{1}  &  -\frac{3}{2} \\
\hline  \lambda^{I_{(1)}}_{-\frac{1}{2}} \lambda^{J_{(1)}}_{-\frac{1}{2}} |0\rangle & 1 &  \boldsymbol{28_0}  &  \boldsymbol{1}  &   -\frac{3}{2}  \\
\hline  \lambda^{I_{(2)}}_{-\frac{1}{2}} \lambda^{J_{(2)}}_{-\frac{1}{2}} |0\rangle & 1  &  \boldsymbol{1_0}  &  \boldsymbol{120}  &   -\frac{3}{2}  \\
\hline \vt_{-1} |0\rangle & 1 &  \boldsymbol{1_0} & \boldsymbol{1} & -\frac{3}{2}  \\
\hline \bar{\vt}_{-1} |0 \rangle & 1 &  \boldsymbol{1_0}  &  \boldsymbol{1} &  -\frac{3}{2} \\
\hline  \sim \big(  \phi^i_{-\frac{1}{10}} \bar{\phi}^j_{-\frac{9}{10}}  +  \gamma^{\kappa_1}_{-\frac{3}{5}}\bar{\gamma}^{\kappa_2}_{-\frac{2}{5}} \big)   |0\rangle  & 1 +n_1  &   \boldsymbol{1_0}  &  \boldsymbol{1}  &   -\frac{3}{2}  \\
\hline \gamma^\kappa_{-\frac{3}{5}} P_4 \big( \phi_{-\frac{1}{10}} \big) |0\rangle  &  159+n_1 & \boldsymbol{1_0} &  \boldsymbol{1}  &  -\frac{1}{2}    \\
\hline \lambda^{I_{(1)}}_{-\frac{1}{2}} \bar{\gamma}^\kappa_{-\frac{2}{5}} \phi^i_{-\frac{1}{10}}|0\rangle & n_{16} & \boldsymbol{8_{v,1}} & \boldsymbol{1} & -\frac{3}{2}   \\
\hline \lambda^{I_{(1)}}_{-\frac{1}{2}} P_5\big( \phi_{-\frac{1}{10}}\big) |0\rangle  &  32 + n_{16}  &  \boldsymbol{8_{v,1}} &  \boldsymbol{1}  &  -\frac{1}{2}  \\
\hline \bar{\gamma}^{\kappa_1}_{-\frac{2}{5}} \bar{\gamma}^{\kappa_2}_{-\frac{2}{5}} P_2\big(\phi_{-\frac{1}{10}}\big)|0\rangle  &  n_{10} & \boldsymbol{1_2} & \boldsymbol{1} & -\frac{3}{2} \\
\hline \bar{\gamma}^\kappa_{-\frac{2}{5}} P_6\big( \phi_{-\frac{1}{10}}\big) |0\rangle    &  68 +2n_{10} &  \boldsymbol{1_2}  &  \boldsymbol{1}  &  -\frac{1}{2} \\
\hline P_{10}\big(\phi_{-\frac{1}{10}}\big) |0\rangle &  n_{10} & \boldsymbol{1_2} & \boldsymbol{1} & \frac{1}{2}   \\
\hline
\end{array}}
\ee
as well as the states multiplied by $\bar{\chi}_0$ which have $q_R \rightarrow q_R+1$.  For the example in equation (\ref{eqn:K3example}), $n_1=3$.

The $(\hat{n}_1, \hat{n}_2) = (0,-1)$ sector contributes
\be
{\setlength\arraycolsep{8pt}
\begin{array}{|c||c|c|c|c|}
\hline State & Degeneracy & SO(8)_{q_L}  &  SO(16) & q_R          \\
\hline\hline P_2\big(\phi_{-\frac{1}{10}}\big) |0\rangle   &   10  &  \boldsymbol{1_0}  &  \boldsymbol{1}  &  -\frac{3}{2}  \\
\hline P_2 \big( \phi_{-\frac{1}{10}}\big) \bar{\chi}_0 |0\rangle & 10 & \boldsymbol{1_0} & \boldsymbol{1} & -\frac{1}{2} \\
\hline
\end{array}}
\ee

%%%%%%%%%%%%%%%%%%%%  CONCLUSION  %%%%%%%%%%%%%%%%%%%

\section{Conclusion and Outlook}

In this paper, we have used the TLSM and generalizations of standard LG-orbifold techniques to compute the exact
spectrum of massless fermions in
a class of heterotic flux vacua based on rigid $T^{2}$-fibrations over \Ka\ bases decorated with gauge and NS-NS fluxes satisfying the modified Bianchi identity.  In accordance with expectations, the spectra we found had no chiral fermions, \ie\ their generation numbers were all zero.  This outcome is easily traced to the free fermions in the torsion multiplet.  

This suggests a natural extension of our results.  As discussed in \cite{Adams:2009av}, the original TLSM is a special case of a larger class of semi-linear models describing heterotic flux vacua in which the gauge anomaly of a GLSM with anomalous gauge group $G$ is cancelled by coupling the anomalous vector to a chiral gauged-WZW model for the coset, $G \backslash \CG / H$, where $\CG/H$ is a complex coset defined by a right-action of $H$ on $\CG$.  When $G$ is abelian we recover the models studied in the present paper; more generally, we get new models.  Importantly, when $G$ is non-abelian, the fermions in the WZW model (which generalizes the torsion multiplet) are no longer free.  The resulting spectrum is thus no longer forced to be non-chiral.  It would be extremely interesting to study the spectra of such models.

Another natural move would be to construct another variant of the TLSM, for example with $T^{2}$ fibres no longer rigid, or perhaps with additional interactions or with a double fibration structure or non-generaic superpotential, in which the curvature of the $T^{2}$-bundles are {\em not} anti-self-dual $(1,1)$-forms, with the corresponding fermions again not free.  If such models can in fact be built, it is likely that the techniques in this paper would again obtain.

\section*{Acknowledgments}

We would like to thank 
K.~Becker,
M.~Becker,
R.~Eager,
M.~Ernebjerg, 
H.~Jockers,
A.~Lawrence,
R.~Martinez,
J.~\mbox{McGreevy},
D. Morrison,
 M. Ro\v cek,
 S.~Sethi,
E.~Silverstein,
A. Strominger,
W.~Taylor,
L.-S.~Tseng,
S.-T.~Yau,
and especially S.~Kachru,
for many fun and enlightening conversations, with particular thanks to S.~Kachru for comments on a draft of this paper.
A.A.~would also like to thank the Kavli Institute for Theoretical Physics at UCSB,
%``Quantum Criticality and the AdS/CFT Correspondence'' Miniprogram 
the Aspen Center for Physics,
%``String Duals of Finite Temperature and Low-Dimensional Systems''
and TEDGlobal 2009 
for hospitality during the completion of this paper.
The work of A.A.~is supported in part by the U.S.~Department of Energy (D.O.E.) under cooperative research agreement DE-FG0205ER41360.
The work of J.L.~is supported in part by the National Science Foundation under Grant No. PHY05-51164 and Grant No. PHY07-57035.

%%%%%%%%%%%%%%%%%% APPENDICES %%%%%%%%%%%%%%

\appendix

\section{Large Radius Limits, or Lack Thereof}
\label{app:no-large-radius}

In heterotic string theory, it is evident from the worldsheet nonlinear sigma model that there is an invariance under $(G,B,\a') \rightarrow \lambda (G,B,\a')$, for constant $\lambda$ ($A$ and $\Phi$ are unchanged).  This translates into an obvious scaling behavior of the equations of motions under $(G,H)\rightarrow\lambda(G,H)$:
\be
E_\Psi\big(\lambda^{\vec{\Phi}}\vec{\Phi}\big) = \sum_{n\geq 0} \alpha'^n E^{(n)}_\Psi \big(\lambda^{\vec{\Phi}}\vec{\Phi}\big) = \lambda^{\Psi-1} \sum_{n\geq 0} \big(\tfrac{\a'}{\lambda}\big)^n E^{(n)}_\Psi \big(\vec{\Phi}\big) \, ,
\ee
where $\vec{\Phi}$ refers collectively to the various supergravity fields and $\lambda^\Psi$ refers to the scaling of the field $\Psi$.  Suppose now that $\vec{\Phi}$ solves the equations of motion to all orders in $\alpha'$,
\be
E_{\vec{\Psi}}\big(\vec{\Phi}\big) = \vec{0} \, .
\ee
\emph{If we arrive at the solution $\vec{\Phi}$ by solving the equations perturbatively in $\alpha'$}, then we have
\be
\vec{\Phi} = \sum_{n\geq 0} \big(\tfrac{\alpha'}{\ell^2}\big)^n \vec{\Phi}^{(n)} \, ,
\ee
where $\ell$ is some length scale associated with the solution (by ``solve perturbatively,'' we mean expand $E\big(\vec{\Phi}\big)$ as a power series in $\alpha'$ whose coefficients involve only powers of $\Phi^{(n)}$ as well as $E^{(n)}(\Phi^{(0)})$ and its derivatives, then independently set the coefficient of each power of $\alpha'$ to zero).  We can then immediately write down a family solutions
\be
\vec{\Phi}_\lambda \equiv \lambda^{\vec{\Phi}} \sum_{n\geq 0} \big(\tfrac{\alpha'}{\lambda\ell^2}\big)^n \vec{\Phi}^{(n)} \, .
\ee
Since the original solution $\vec{\Phi}$ has a convergent expansion by assumption, $\vec{\Phi}_\lambda$ will surely exist for $\lambda \geq 1$.  In particular, this implies that if we have a perturbative solution $(G,H,F)$, then for the family of solutions parameterized by $\lambda$, the volume of the manifold $V_\lambda$ tends to $ \lambda^3 V^{(0)}$ as $\lambda\rightarrow\infty$, yielding a large radius limit --- we will call these families ``large-radius scaling solutions.''

Consider the implications for perturbative, supersymmetric, heterotic solutions.  These imply that the zeroeth-order terms in the expansion of our solutions should satisfy
\be
H^{(0)} \propto i(\p-\bar{\p}) \omega^{(0)} \, , \qquad i\p\bar{\p}\omega^{(0)} = 0 \, ,  \qquad \ldots .
\ee
For such a perturbative solution to exist, then, the topology of the manifold must admit solutions to the zeroeth-order equations (this argument was presented in the original paper on heterotic with torsion \cite{Strominger:1986uh}).  A classification of real 6-folds with $SU(3)$ structure illuminates the topological restrictions (see, {\it e.g.}, \cite{LopesCardoso:2002hd}), showing that the manifold must be topologically Calabi-Yau.  Thus, \emph{perturbative supersymmetric solutions, which admit large-radius scaling limits, must live on topologically Calabi-Yau manifolds}.

The contrapositive tells us that \emph{supersymmetric heterotic solutions involving topologically non-K\"ahler manifolds cannot be realized as perturbative solutions, thus they will not admit large-radius scaling solutions}.  The scaling argument is certainly the cleanest way to be comfortable with supergravity analyses of CY${}_3$ compactifications; since it does not apply to topologically non-K\"ahler compactifications, one should justify a supergravity analysis of non-K\"ahler compactifications by checking that curvature invariants are small.  The Bianchi identity is the sticking point since it is the equation that is \emph{not} being solved in a perturbative fashion:
\be
i\p\bar{\p}\omega^{(0)} \propto \frac{\a'}{4} \Big[ \tr\big( R^{(0)}\wedge R^{(0)} \big) - \tr\big( F^{(0)}\wedge F^{(0)}\big)\Big]\, ,
\ee
suggesting\footnote{In a perturbative solution, we would instead require
\be
i\p\bar{\p}\omega^{(1)} \sim \frac{\ell^2}{4} \Big[ \tr\big( R^{(0)}\wedge R^{(0)} \big) - \tr\big( F^{(0)}\wedge F^{(0)}\big)\Big]\, .
\ee} that non-K\"ahler manifolds (arising in supersymmetric compactifications of heterotic) will typically be stuck with some string-scale radii.  In the presently known examples, it is certainly true that this is the case.  Since the Bianchi identity must be satisfied as an equation on forms and not just in cohomology, one can wedge with \emph{any} 2-form $\alpha_2$ and integrate:
\be
\int_K dH \wedge \alpha_2 = \frac{\alpha'}{4} \int_K \Big[ \tr\big(R\wedge R\big) - \tr\big(F \wedge F\big) \Big] \wedge \alpha_2 \, .
\ee
In particular, for torus-fibered manifolds, $K$, we have vertical one-forms $\rho$ and $\bar{\rho}$, so we can choose $\alpha_2 = \rho\wedge\bar{\rho}$~as one test of this equality.  This allows the equation to be integrated along the $T^2$ fiber and results in an equation on the real 4-fold base, $S$ (see (\ref{eqn:anomaly-t2-bundle}) in our case).  The right-hand site describes topological invariants (dimensionless numbers) multiplied by $\alpha'$, while the left-hand side relates instead to the Hermitian form on $K$, as well as to topological data of the fibering --- it is thus a dimensionless number multiplied by a dimensionful scale that is related to some scale of the compactification.  This immediately relates this particular dimensionful scale to $\alpha'$ and, in our case, actually bounds it to not be parametrically larger than $\a'$.  A more thorough discussion of the volume of the compactifications in this paper can be found in \cite{Cyrier:2006pp}.

\section{Conventions}
\label{app:conventions}

The following is a lightning review of the salient features of $(0,2)$ gauged linear sigma models; for more complete discussions see \cite{Witten:1993yc,Distler:1995mi}.  Our conventions and notation follow \cite{Hori:2003xxx}, with all factors of $\a'$ suppressed throughout the paper.
We take the \ZT\ superspace coordinates to be $(y^+, y^-,\thp,\thpb)$,
where $y^\pm = (y^0\pm y^1)$.  We begin with the gauge multiplet.

The right-moving gauge covariant superderivatives ${\cal{D}}_+,
{\bbar{\cal{D}}}_+$, satisfy the algebra
\be
\CD_+^2 = \CDB_+^2 =0, \qquad  -\tfrac{i}{4}\{\ \!\CD_+,
\CDB_+ \}\ =\nabla_{+} =\p_+ +iQv_+ ,
\ee
where $Q$ is the charge of the field on which they act.  These imply that in a suitable basis we can identify
\be  
\CD_{+} =\frac{\partial}{\partial \theta^+} - 2i {\bar{\theta}}^+ \nabla_{+}, \qquad \CDB_{+} = -\frac{\partial}{\partial {\bar{\theta}}^+} + 2i\thp  \nabla_{+},\qquad \CD_{-} = \p_{-} + \tfrac{i}{2} QV_{-}, \nonumber
\ee
\be
Q_+ = \frac{\p}{\p\thp} + 2i\thpb \nabla_+ \, ,      \qquad    \QB_+ = -\frac{\p}{\p\thpb} - 2i\thp\nabla_+
\ee
where $V$ and $V_{-}$ are real vector superfields which transform under a gauge transformation with (uncharged) chiral gauge parameter $\CDB_{+}\Lambda=0$ as $\delta_{\Lambda} V_{-}$=$\p_-(\Lambda + \bbar{\Lambda})$ and $\delta_{\Lambda} V$=$\frac{i}{2} (\Lambda-\bbar{\Lambda})$; $\nabla_\pm$ are the usual gauge covariant derivatives.  This allows us to fix to Wess-Zumino gauge in which 
\be
V = \theta^+ {\bar{\theta}}^+ 2v_{+}  \qquad
V_{-} = 2v_{-} - 2i \theta^+ {\bar{\lambda}}_- -2i {\bar{\theta}}^+ \lambda_- +2 \theta^+ {\bar{\theta}}^+ D. \nonumber
\ee
Note that $V_{-}$ contains a complex left-moving gaugino. Finally, the natural field strength is a fermionic chiral superfield, 
\be 
\Ups= 2[{\bbar{\mathcal{D}}}_+ ,\CD_{-}] =\CDB_+(2\p_-V + i V_{-}) =
-2\{\lambda_- -i\thp(D+2iv_{+-}) - 2i\thp\bar{\thp}\p_+\lambda_- \}, 
\ee 
for which the natural action is
\be 
(4\pi) S_{\Ups} = - \frac{1}{8e^2}\int\! d^2y\, d\thp d\thpb ~\bbar{\Ups}\Ups =\frac{1}{e^2}
\int d^2y\,
\left\{ 2v_{+-}^2  + 2i {\bar{\lambda}}_- \p_+ \lambda_-  +\frac{1}{2} D^2  \right\},
\ee 
where $d^2y = dy^0 dy^1 = \frac{1}{2}dy^+ dy^-$ and we use conventions where $\int d\thp \thp = \int \thpb d\thpb = 1$.  
%Note that
%\bes
%\int d^2y d\thp [\ldots] = \int d^2y D_+ [\ldots] \Big|_{\thp=\thpb=0}~,   \qquad    \int d^2y d\thpb [\ldots] = \int d^2y \DB_+ [\ldots] \Big|_{\thp=\thpb=0}~.
%\ees
%Also, integration by parts in superspace says that
%\bes
%\int d^2y\, d^2\theta (D_+A)B  = (-1)^{a+1} \int d^2y\, d^2\theta A(D_+B)~, 
%\ees
%where $a$ is the grading of $A$ ($0$ for bosonic and $1$ for fermionic).

Matter multiplets are similarly straightforward.  A bosonic superfield satisfying $\CDB_{+}  \Phi =0$ is called a \emph{chiral} supermultiplet and contains a complex scalar and a right-moving complex fermion $\Phi = \phi +\rt\theta^+\psi_+ - 2i\thp\thpb \nabla_{+}\phi$, and under gauge transformations $\Phi\to e^{-iQ(\Lambda+\bbar{\Lambda})/2}\Phi$.  The gauge invariant Lagrangian is given by
\bea
S_{\Phi} &=&  - \frac{i}{4\pi} \int \, d^2y\, d^2\theta {\bbar{\Phi}} \CD_- \Phi     \\
&=&  \frac{1}{4\pi}\int d^2y\,  \Big\{ - \vert \nabla_{\alpha} \phi\vert^2 +
    2i{\bar{\psi}}_{+}\nabla_- \psi_{+} -i Q {\sqrt 2}
    {\bar{\phi}} \lambda_- \psi_{+} 
    +i Q {\sqrt 2} \phi
    {\bar{\psi}}_{+} {\bar{\lambda}}_-  + Q D \vert \phi \vert^2\Big\} , \nonumber
\eea
where the metric is given by $\eta^{+-} = -2$.

We can write (gauge) chiral superfields in terms of ordinary chiral superfields by:
\bes
\Phi = e^{QV}\Phi_0   \qquad   \textrm{and}   \qquad   \bbar{\Phi} = e^{QV}\bbar{\Phi}_0 .
\ees
We can also use this to write gauge-covriant super-derivatives in terms of ordinary super-derivatives, depending on what kind of field they act on:
\beas
\CD_+ \Phi &=& e^{-QV}D_+(e^{QV}\Phi)~,   \qquad   \CDB_+ \Phi = e^{QV}\DB_+(e^{-QV}\Phi)~,   \\
\CD_+ \bbar{\Phi} &=& e^{QV}D_+(e^{-QV}\bbar{\Phi})~,   \qquad   \CDB_+ \bbar{\Phi} = e^{-QV}\DB_+(e^{QV}\bbar{\Phi})~.
\eeas
Similarly,
\be
\CD_- \Phi = (\p_- + \tfrac{i}{2}QV_-)\Phi~,   \qquad   \CD_-\bbar{\Phi} = (\p_- - \tfrac{i}{2}QV_-)\bbar{\Phi}~.
\ee

Left-handed fermions transform in their own supermultiplet, the {\em fermi} supermultiplet, which satisfies the chiral constraint
\be\label{eq:almostchiral}
\bbar{\cal{D}}_+ \Gamma = \sqrt{2} E
\ee
and has component expansion $\Gamma = \gamma_- -\rt\thp G
- 2i\thp \thpb \nabla_+ \gamma_{-} - \sqrt{2} \thpb E,$
where ${\bbar{\cal{D}}}_+ E=0$ is a bosonic chiral superfield with the same gauge charge as $\Gamma$. 
The action for $\Gamma$ is given by
 \bea \label{LF} 
(4\pi) S_{\Gamma} &=& - \frac{1}{2}\int\! d^2y\, d^2\theta ~\bbar{\Gamma} \Gamma \\
&=& \int d^2y\, \left\{ 2i{\bar{\gamma}}_{-}  \nabla_+ \gamma_{-} + \vert G \vert^2 - \vert E \vert^2  - \sum_i \left( \bar{\gamma}_- \frac{\partial E}{\partial \phi^i} \psi^i_{+} + \bar{\psi}{}^i_{+} \frac{\partial \bbar{E}}{\partial \bar{\phi}{}^i} \gamma_- \right) \right\}. \nonumber
\eea

In general, we can add superpotential terms to our Lagrangian. Since these are integrals over a single supercoordinate, the superpotential can be written as a sum of fermi superfields $\Gamma^m$ times holomorphic functions $F^m$ of the chiral superfields,
\bea \label{superpotential}
(4\pi) S_{\cal W} &=& \frac{1}{\sqrt{2}}\int\! d^2y\,d\thp~ \sum_m \Gamma^m
F^m \vert_{{\bar\theta}^+=0} + {\rm h.c.},\\ &=& -\int d^2y\, \sum_m \left\{G^m F^m(\phi) + \gamma_{-}^m\sum_i \psi_{+}^i  \frac{\partial F^m}{\partial \phi^i} \right\}+ {\rm  h.c.}.
\nonumber
\eea
Since $\Gamma^m$ is not an honest chiral superfield but satisfies
(\ref{eq:almostchiral}), we need to impose the condition
\be
E \cdot F=0
\ee
to ensure that the superpotential is chiral.
Finally, since $\Ups$ is a chiral fermion, we can also add an FI term of the form
\be 
\label{defineFI} 
(4\pi) S_{\rm FI} = \frac{it_{\textit FI}}{4}\int\! d^2y\, d\thp~ \Ups \vert_{\thpb =0} + {\rm h.c.} = \int d^2y \left( -r_{\textit FI}D + 2\theta_{\textit FI} v_{+-} \right)
\ee
where $t_{\textit FI}=r_{\textit FI}+i\theta_{\textit FI}$  is the complexified FI parameter.

The torsion multiplet is discussed in section \ref{sec:torsion}.  For all the fields in our paper, the component expressions of the supersymmetry transformations are
\be
{\setlength\arraycolsep{8pt}
\begin{array}{lclcl} 
\delta \phi = \sqrt{2} \epsilon\psi & \qquad &  \delta \gamma = -\sqrt{2}\big( \epsilon G + \bar{\epsilon} E \big)         \\
\delta \bar{\phi} = - \sqrt{2}\bar{\epsilon}\bar{\psi}   & &  \delta \bar{\gamma} = -\sqrt{2} \big( \bar{\epsilon} \bar{G} + \epsilon \bar{E} \big)        \\
\delta \psi = -2i\sqrt{2}\bar{\epsilon}\nabla_+\phi   &  &   \delta G = \sqrt{2} \bar{\epsilon} \big( 2i\nabla_+ \gamma - \sum_i \psi^i \p_i E \big)  \\
\delta \bar{\psi} =  2i\sqrt{2}\epsilon\nabla_+ \bar{\phi}  & &   \delta \bar{G} = \sqrt{2}\epsilon \big( 2i\nabla_+\bar{\gamma} + \sum_i \bar{\psi}{}^i \bar{\p}_i \bar{E} \big)  
\end{array}}
\ee
\be
{\setlength\arraycolsep{8pt}
\begin{array}{lcl} 
\delta \vt = \sqrt{2} \epsilon\chi & \qquad &  \delta v_- = -i\epsilon \bar{\lambda} - i\bar{\epsilon}\lambda   \\
\delta \bar{\vt} = - \sqrt{2}\bar{\epsilon}\bar{\chi}   & &  \delta\lambda = \epsilon (2 v_{+-} - iD) \\
\delta \chi = -2i\sqrt{2}\bar{\epsilon}\nabla_+\vt   &  &  \delta\bar{\lambda} = \bar{\epsilon} (2 v_{+-} + iD)  \\
\delta \bar{\chi} =  2i\sqrt{2}\epsilon\nabla_+ \bar{\vt}  &  &  \delta D = - 2\epsilon\p_+ \bar{\lambda} + 2\bar{\epsilon}\p_+ \lambda   \\
& & \delta v_+ = 0
\end{array}}
\ee
where $\delta = \epsilon Q_+ - \bar{\epsilon}\QB_+ + \delta_{\Lambda_{\mathrm{WZ}}}$, and the gauge transformation that maintains Wess-Zumino gauge is $\Lambda_{\mathrm{WZ}} = 4i\thp \bar{\epsilon} v_+$.

%%%%%%%%%%%%%%%%%%%%%%%%%  NON-WESS-ZUMINO GAUGE  %%%%%%%%%%%%%%%%%%%%
%

%

\ifeq
\section{Non-WZ Gauge Conventions}

The right-moving gauge covariant superderivatives ${\cal{D}}_+,
{\bbar{\cal{D}}}_+$, satisfy the algebra
\be
\CD_+^2 = \CDB_+^2 =0, \qquad  -\tfrac{i}{4}\{\ \!\CD_+,
\CDB_+ \}\ = \p_+ - \tfrac{i}{4} \big( Q [D_+, \DB_+ ] V \big) =\p_+ +iQv_+ + \ldots ,
\ee
where $Q$ is the charge of the field on which they act.  As before,
\be  
D_{+} =\frac{\partial}{\partial \theta^+} - 2i {\bar{\theta}}^+ \p_{+}, \qquad \DB_{+} = -\frac{\partial}{\partial {\bar{\theta}}^+} + 2i\thp  \p_{+},\qquad \CD_{-} = \p_{-} + \tfrac{i}{2} QV_{-}, \nonumber
\ee
\be
Q_+ = \frac{\p}{\p\thp} + 2i\thpb \p_+ \, ,      \qquad    \QB_+ = -\frac{\p}{\p\thpb} - 2i\thp\p_+
\ee
where $V$ and $V_{-}$ are real vector superfields which transform under a gauge transformation with (uncharged) chiral gauge parameter $\DB_{+}\Lambda=0$ as $\delta_{\Lambda} V_{-}$=$\p_-(\Lambda + \bbar{\Lambda})$ and $\delta_{\Lambda} V$=$\frac{i}{2} (\Lambda-\bbar{\Lambda})$; $\nabla_\pm$ are the usual gauge covariant derivatives.  Out of Wess-Zumino gauge, 
\be
V = \alpha + \sqrt{2} \thp f_+ - \sqrt{2} \thpb \bar{f}_+ + \theta^+ {\bar{\theta}}^+ 2v_{+}  \, ,  \qquad
V_{-} = 2v_{-} - 2i \theta^+ {\bar{\lambda}}_- -2i {\bar{\theta}}^+ \lambda_- +2 \theta^+ {\bar{\theta}}^+ D\, . \nonumber
\ee
Note that $V_{-}$ contains a complex left-moving gaugino. Finally, the natural field strength is a fermionic chiral superfield, 
\bea 
\Ups &=& 2[{\bbar{\mathcal{D}}}_+ ,\CD_{-}] =\DB_+(2\p_-V + i V_{-})     \nonumber \\
&=& -2\big\{ (\lambda_- - \sqrt{2}\p_- \bar{f}_+) - i\thp\big( (D+2 \p_+\p_- \alpha) +2iv_{+-}\big) - 2i\thp\bar{\thp}\p_+(\lambda_- - \sqrt{2}\p_- \bar{f}_+) \big\}  \nonumber
\eea 
In fact, if in the actions we'll consider, $D$ always appears in the combination $D'\equiv D+2\p_+\p_-\alpha$ and $\lambda_-$ in the combination $\lambda'_- \equiv \lambda_- - \sqrt{2}\p_- \bar{f}_+$.  We can switch to the primed quantities without incurring a Jacobian factor since we leave $\alpha$ and $f_+$ unchanged.  The natural choice of gauge invariant action is
\be 
(4\pi) S_{\Ups} = - \frac{1}{8e^2}\int\! d^2y\, d\thp d\thpb ~\bbar{\Ups}\Ups =\frac{1}{e^2}
\int d^2y\,
\left\{ 2v_{+-}^2  + 2i {\bar{\lambda}'}_- \p_+ \lambda'_-  +\frac{1}{2} D'^2  \right\},
\ee 
where $d^2y = dy^0 dy^1 = \frac{1}{2}dy^+ dy^-$ and we use conventions where $\int d\thp \thp = \int \thpb d\thpb = 1$.  Neglecting boundary terms,
\bes
\int d^2y d\thp [\ldots] = \int d^2y D_+ [\ldots] \Big|_{\thp=\thpb=0}~,   \qquad    \int d^2y d\thpb [\ldots] = \int d^2y \DB_+ [\ldots] \Big|_{\thp=\thpb=0}~.
\ees
Also, integration by parts in superspace says that
\bes
\int d^2y\, d^2\theta (D_+A)B  = (-1)^{a+1} \int d^2y\, d^2\theta A(D_+B)~, 
\ees
where $a$ is the grading of $A$ ($0$ for bosonic and $1$ for fermionic).

Matter multiplets are similarly straightforward.  A bosonic superfield satisfying $\CDB_{+}  \Phi =0$ is called a \emph{chiral} supermultiplet and contains a complex scalar and a right-moving complex fermion.  $\Phi \equiv e^{QV} \Phi_0$, where $\Phi_0 = \phi +\rt\theta^+\psi_+ - 2i\thp\thpb \p_{+}\phi$ satisfies $\DB_+ \Phi_0 = 0$ and under gauge transformations $\Phi\to e^{-iQ(\Lambda+\bbar{\Lambda})/2}\Phi$.  The gauge invariant Lagrangian is given by
\bea
S_{\Phi} &=&  - \frac{i}{4\pi} \int \, d^2y\, d^2\theta {\bbar{\Phi}} \CD_- \Phi     \\
&=&  \frac{1}{4\pi}\int d^2y\,  e^{2Q\alpha} \Big\{ - \vert \tilde{\nabla}_{\mu} \phi\vert^2 +
    2i{\bar{\psi}'}_{+}\tilde{\nabla}_- \psi'_{+} -i Q {\sqrt 2}
    {\bar{\phi}} \lambda'_- \psi'_{+} 
    +i Q {\sqrt 2} \phi
    {\bar{\psi}'}_{+} {\bar{\lambda}'}_-  + Q D' \vert \phi \vert^2\Big\} , \nonumber
\eea
where the metric is given by $\eta^{+-} = -2$ and we have defined
\be
\tilde{\nabla} \equiv e^{-Q\alpha} \nabla e^{Q\alpha} \, ,    \qquad   \psi'_+ \equiv \psi_+ + 2 Q \phi f_+
\ee
which, again, causes no introduction of Jacobian factor.

We can write (gauge) chiral superfields in terms of ordinary chiral superfields by:
\bes
\Phi = e^{QV}\Phi_0   \qquad   \textrm{and}   \qquad   \bbar{\Phi} = e^{QV}\bbar{\Phi}_0 .
\ees
We can also use this to write gauge-covriant super-derivatives in terms of ordinary super-derivatives, depending on what kind of field they act on:
\beas
\CD_+ \Phi &=& e^{-QV}D_+(e^{QV}\Phi)~,   \qquad   \CDB_+ \Phi = e^{QV}\DB_+(e^{-QV}\Phi)~,   \\
\CD_+ \bbar{\Phi} &=& e^{QV}D_+(e^{-QV}\bbar{\Phi})~,   \qquad   \CDB_+ \bbar{\Phi} = e^{-QV}\DB_+(e^{QV}\bbar{\Phi})~.
\eeas
Similarly,
\be
\CD_- \Phi = (\p_- + \tfrac{i}{2}QV_-)\Phi~,   \qquad   \CD_-\bbar{\Phi} = (\p_- - \tfrac{i}{2}QV_-)\bbar{\Phi}~.
\ee

Left-moving fermions transform in their own supermultiplet, the {\em fermi} supermultiplet, which satisfies the chiral constraint
\be\label{eq:almostchiral}
\bbar{\cal{D}}_+ \Gamma = \sqrt{2} E
\ee
and has component expansion $\Gamma_0 = \gamma_- -\rt\thp G - 2i\thp \thpb \p_+ \gamma_{-} - \sqrt{2} \thpb E(\Phi_0),$ where $E(\Phi)$ is a bosonic chiral superfield with the same gauge charge as $\Gamma$.   The action for $\Gamma$ is given by
 \bea \label{LF} 
(4\pi) S_{\Gamma} &=& - \frac{1}{2}\int\! d^2y\, d^2\theta ~\bbar{\Gamma} \Gamma \\
&=& \int d^2y\,  e^{2q\alpha} \Big\{ 2i{\bar{\gamma}}_{-}  \tilde{\nabla}_+ \gamma_{-} + \vert G' \vert^2 - \vert E \vert^2  - \bar{\gamma}_- \psi'_{+i} \partial_i E + \gamma_- \bar{\psi}'_{+i} \bar{\p}_i \bbar{E} \Big\} \nonumber
\eea
where again
\be
G' \equiv G - 2qf_+ \gamma_-
\ee
introduces no Jacobian factor.

In general, we can add superpotential terms to our Lagrangian. Since these are integrals over a single supercoordinate, the superpotential can be written as a sum of fermi superfields $\Gamma_m$ times holomorphic functions $J^m$ of the chiral superfields,
\bea 
\label{superpotential}
(4\pi) S_{\cal W} &=& \frac{1}{\sqrt{2}}\int\! d^2y\,d\thp~ \Gamma_m
J^m \vert_{{\bar\theta}^+=0} + {\rm h.c.},   \nonumber \\
&=& -\int d^2y\, \Big\{G'_m J^m(\phi_i) + \gamma_{-m} \psi'_{+i}  \p_i J^m \Big\}+ {\rm  h.c.}\, .
\eea
Since $\Gamma_m$ is not an honest chiral superfield but satisfies
(\ref{eq:almostchiral}), we need to impose the condition
\be
E \cdot J=0
\ee
to ensure that the superpotential is chiral.  Since $\Ups$ is chiral, we can also add an FI term of the form
\be 
\label{defineFI} 
(4\pi) S_{\rm FI} = \frac{it}{4}\int\! d^2y\, d\thp~ \Ups \vert_{\thpb =0} + {\rm h.c.} = \int d^2y \left( -rD' + 2\theta v_{+-} \right)
\ee
where $t=r+i\theta$  is the complexified FI parameter.  Another term that we'll be interested in is of the generic form of a gauge anomaly, similar to the FI term but with the FI parameter $t$ replaced by a bosonic, chiral superfield $\Lambda \equiv a + \sqrt{2}\thp \xi_+ - 2i\thp\thpb \p_+ a$:
\be
\frac{1}{2} \int d^2y d\thp \, \Lambda \Ups  +  {\rm h.c.}  =  \int d^2y \, \Big\{ \sqrt{2}\bar{\xi}_+ \bar{\lambda}'_- - \sqrt{2}\xi_+ \lambda'_- - 2Im(a)D' - 4Re(a)v_{+-} \Big\}
\ee
This is the term that gives us problems when we want to go to Wess-Zumino gauge, because doing so requires modifying the SUSY transformations by the addition of a gauge transformation to remain in WZ gauge.  In our models, the measure is variant under a gauge transformation, which means that the WZ gauge transform of the measure will cancel with that of the classical action.  Thus, our action in WZ gauge will not be manifestly supersymmetric.

For the torsion multiplet, we have
\bea
(4\pi)S_{tor} &=& \int \! d^2y \, \Big\{  - \big| \nabla_\mu \vt' \big|^2  +  2i\bar{\chi}' \p_- \chi' + 2 \big(M^a\bar{\vt}' + \MB{}^a \vt' \big) v_{+-a}      \nonumber \\
&&\qquad  \qquad + 2i\sqrt{2} M^{(a}\MB{}^{b)} \big( f_a \lambda'_b + \bar{f}_a\bar{\lambda}'_b \big)  +  2 M^{(a}\MB{}^{b)} \alpha_a D'_b \Big\} 
\eea
where
\be
\vt' \equiv \vt - i M^a \alpha_a  \, ,   \qquad   \chi' \equiv \chi - 2iM^a f_a \, .
\ee

\subsection{Non-WZ Component Expressions}

\be
{\setlength\arraycolsep{8pt}
\begin{array}{lcl}
\CD_+ \Phi \big|_{\thp=\thpb=0} = \sqrt{2}e^{Q\alpha} \psi'   &  \qquad  & \CD_- \Phi \big|_{\thp=\thpb=0} = e^{Q\alpha} \tilde{\nabla}_- \phi \\
 \CDB_+ \CD_- \Phi = \tfrac{1}{2} Q \Ups \Phi  & &   \CD_+ \CD_- \Phi  =  \CD_- \CD_+ \Phi - \tfrac{1}{2} Q \overline{\Ups} \Phi    \\
 \CD_+ \Gamma \big|_{\thp=\thpb=0}  =  -\sqrt{2}e^{q\alpha} G'   & &     \CDB_+ \Gamma  =  \sqrt{2} E(\Phi) \\
\CD_- \Theta \big|_{\thp=\thpb=0}  =  \nabla_- \vt'    & &    D_+ \big( \Theta - 2iMV\big) \big|_{\thp=\thpb=0} = \sqrt{2}\chi'   \\
D_+ \CD_-\Theta \big|_{\thp=\thpb=0}  =  \sqrt{2} \p_- \chi'  -  i M\bar{\lambda}'    &  &   \\
\{ \CD_+ , \CDB_+ \}  =  \big( 4i\p_+ - Q[D_+,\DB_+]V \big)      &  &  D_+ \Ups \big|_{\thp=\thpb=0}  =  2i \big( D' + 2iv_{+-} \big)     \\
D_+V \big|_{\thp=\thpb=0}  =  \sqrt{2}f    & &    \DB_+ D_+ V\big|_{\thp=\thpb=0}  =  - 2 v_+ + 2i\p_+ \alpha    \\
\tfrac{1}{4}[D_+,\DB_+] V = v_+ - i\sqrt{2}\, \p_+\big( \thp f  + \thpb \bar{f} \big)  - 2\thp\thpb \p_+^2 \alpha   \!\!\!\!\!\!\!\!\!\!\!\!\!\!\!\!\!\!\!\!\!\!\!\!\!\! &  &
\end{array}}
\ee

\subsection{Non-WZ Component Field Transformations}

It is useful to record the variations of the component fields under a SUSY transformation $\delta = \epsilon Q_+ - \bar{\epsilon} \QB_+$ are, for the gauge fields
\be
{\setlength\arraycolsep{8pt}
\begin{array}{lcl} 
\delta_\epsilon \alpha = \sqrt{2} (\epsilon f - \bar{\epsilon}\bar{f}) & \qquad &  \delta_\epsilon v_- = -i\epsilon \bar{\lambda} - i\bar{\epsilon}\lambda   \\
\delta_\epsilon f = \sqrt{2} \bar{\epsilon}( v_+ - i\p_+\alpha )  & &  \delta_\epsilon\lambda = \epsilon (2 \p_+ v_{-} - iD) \\
\delta_\epsilon \bar{f} = \sqrt{2}\epsilon( v_+ + i \p_+\alpha )    &  &  \delta_\epsilon\bar{\lambda} = \bar{\epsilon} (2 \p_+ v_{-} + iD)  \\
\delta_\epsilon v_+ = -i \sqrt{2}( \epsilon \p_+ f + \bar{\epsilon} \p_+ \bar{f} )  &  &  \delta_\epsilon D = - 2\epsilon\p_+ \bar{\lambda} + 2\bar{\epsilon}\p_+ \lambda  
\end{array}}
\ee
and for the matter fields
\be
{\setlength\arraycolsep{8pt}
\begin{array}{lclcl} 
\delta_\epsilon \phi = \sqrt{2} \epsilon\psi & \qquad &  \delta_\epsilon \gamma = -\sqrt{2}\big( \epsilon G + \bar{\epsilon} E \big)         \\
\delta_\epsilon \bar{\phi} = - \sqrt{2}\bar{\epsilon}\bar{\psi}   & &  \delta_\epsilon \bar{\gamma} = -\sqrt{2} \big( \bar{\epsilon} \bar{G} + \epsilon \bar{E} \big)        \\
\delta_\epsilon \psi = -2i\sqrt{2}\bar{\epsilon}\p_+\phi   &  &   \delta_\epsilon G = \sqrt{2} \bar{\epsilon} \big( 2i\p_+ \gamma - \psi_i \p_i E \big)  \\
\delta_\epsilon \bar{\psi} =  2i\sqrt{2}\epsilon\p_+ \bar{\phi}  & &   \delta_\epsilon \bar{G} = \sqrt{2}\epsilon \big( 2i\p_+\bar{\gamma} + \bar{\psi}_i \bar{\p}_i \bar{E} \big)  
\end{array}}
\ee
Recall the field redefinitions that were useful above:
\be
{\setlength\arraycolsep{8pt}
\begin{array}{lclcl}
\lambda' \equiv \lambda - \sqrt{2} \p_- \bar{f}  \, ,  &  ~~~ & D' \equiv D+2\p_+\p_-\alpha \, ,   &  ~~~  &  \psi' \equiv \psi + 2Q\phi f  \, ,   \\
G' \equiv G - 2qf\gamma  \, , & ~~~ &  \vt' \equiv \vt - i M \alpha \, ,   & ~~~ &  \chi' \equiv \chi - 2iM f \, .
\end{array}}
\ee
It's useful to recast the component SUSY variations in terms of the primed variables
\be
{\setlength\arraycolsep{8pt}
\begin{array}{lcl} 
\delta_\epsilon \alpha = \sqrt{2} (\epsilon f - \bar{\epsilon}\bar{f}) & \qquad &  \delta_\epsilon v_- = -i\epsilon (\bar{\lambda}' + \sqrt{2}\p_- f) - i\bar{\epsilon}(\lambda' + \sqrt{2}\p_- \bar{f})   \\
\delta_\epsilon f = \sqrt{2} \bar{\epsilon}( v_+ - i\p_+\alpha )  & &  \delta_\epsilon\lambda' = \epsilon (2 v_{+-} - iD') \\
\delta_\epsilon \bar{f} = \sqrt{2}\epsilon( v_+ + i \p_+\alpha )    &  &  \delta_\epsilon \bar{\lambda}' = \bar{\epsilon} (2 v_{+-} + iD')  \\
\delta_\epsilon v_+ = -i \sqrt{2}( \epsilon \p_+ f + \bar{\epsilon} \p_+ \bar{f} )  &  &  \delta_\epsilon D' = - 2\epsilon\p_+ \bar{\lambda}' + 2\bar{\epsilon}\p_+ \lambda'  
\end{array}}
\ee
\be
{\setlength\arraycolsep{8pt}
\begin{array}{lclcl} 
\delta_\epsilon \phi = \sqrt{2} \epsilon(\psi' - 2 Q \phi f) & \qquad &  \delta_\epsilon \gamma = -\sqrt{2}\big( \epsilon (G' + 2qf\gamma) + \bar{\epsilon} E \big)         \\
\delta_\epsilon \bar{\phi} = - \sqrt{2}\bar{\epsilon}(\bar{\psi}' - 2Q\bar{\phi}\bar{f})   & &  \delta_\epsilon \bar{\gamma} = -\sqrt{2} \big( \bar{\epsilon} (\bar{G}' - 2q\bar{f}\bar{\gamma}) + \epsilon \bar{E} \big)        \\
\delta_\epsilon \psi' = -2i\sqrt{2} (\bar{\epsilon}\tilde{\nabla}_+\phi + i \epsilon Q \psi' f )   &  &   \delta_\epsilon G' = \sqrt{2} \Big( \bar{\epsilon} \big( 2i\tilde{\nabla}_+ \gamma - \psi'_i \p_i E \big) - \epsilon\, 2qfG' \Big)  \\
\delta_\epsilon \bar{\psi}' =  2i\sqrt{2}( \epsilon\tilde{\nabla}_+ \bar{\phi} + i \bar{\epsilon} Q \bar{\psi}'\bar{f})  & &   \delta_\epsilon \bar{G}' = \sqrt{2}\Big( \epsilon \big( 2i\tilde{\nabla}_+\bar{\gamma} + \bar{\psi}'_i \bar{\p}_i \bar{E} \big)  + \bar{\epsilon}\, 2q\bar{f}\bar{G}' \Big)
\end{array}}
\ee
or
\be
{\setlength\arraycolsep{8pt}
\begin{array}{l} 
\delta_\epsilon (e^{Q\alpha} \phi ) = \sqrt{2} e^{Q\alpha} \big[ \epsilon \psi' - Q \phi ( \epsilon f + \bar{\epsilon}\bar{f} ) \big]          \\
\delta_\epsilon (e^{Q\alpha}\psi') = \sqrt{2} e^{Q\alpha}\big[-2i\bar{\epsilon}\tilde{\nabla}_+\phi -  Q \psi' (\epsilon f + \bar{\epsilon} \bar{f} ) \big]    \\
\delta_\epsilon (e^{q\alpha} \gamma) = \sqrt{2}e^{q\alpha} \big[ - \epsilon G'  - \bar{\epsilon} E - q\gamma(\epsilon f + \bar{\epsilon}\bar{f}) \big]   \\
\delta_\epsilon (e^{q\alpha}G') = \sqrt{2} e^{q\alpha} \big[ \bar{\epsilon} ( 2i\tilde{\nabla}_+ \gamma - \psi'_i \p_i E ) - qG' (\epsilon f + \bar{\epsilon}\bar{f}) \big]\end{array}}
\ee
and finally for the torsion multiplet
\be
\begin{array}{l}
\delta_\epsilon \vt' = \sqrt{2}\big( \epsilon \chi' + i M ( \epsilon f + \bar{\epsilon} \bar{f}) \big)  \\
\delta_\epsilon \bar{\vt}' = \sqrt{2} \big( - \bar{\epsilon} \bar{\chi}' + i \MB (\epsilon f + \bar{\epsilon}\bar{f} ) \big)    \\
\delta_\epsilon \chi' = - 2i\sqrt{2}\bar{\epsilon} \, \nabla_+ \vt'    \\
\delta_\epsilon \bar{\chi}' = 2i\sqrt{2} \epsilon \nabla_+ \bar{\vt}'
\end{array}
\ee
We see, then, that
\be
\begin{array}{l}
\delta_\epsilon \big( \nabla_+ \vt' \big)  =  \sqrt{2} \epsilon \p_+ \chi'   \\
\delta_\epsilon \big( \nabla_- \vt' \big)  =  \sqrt{2} \epsilon \p_- \chi'  -  iM \big( \epsilon \bar{\lambda}' + \bar{\epsilon} \lambda' \big)  \\
\delta_\epsilon \big( e^{2Q\alpha} \phi \tilde{\nabla}_- \bar{\phi} \big) = e^{2Q\alpha} \Big( \sqrt{2}\epsilon \psi' \tilde{\nabla}_- \bar{\phi} - \sqrt{2}\bar{\epsilon}\phi \tilde{\nabla}_- \bar{\psi}' - Q |\phi|^2 ( \epsilon \bar{\lambda}' + \bar{\epsilon}\lambda' ) \Big)
\end{array}
\ee

Next, let's consider the component transformations under the chiral gauge transformation $\Lambda = a + \sqrt{2}\thp \xi  - 2i\thp\thpb \p_+ a$:
\be
{\setlength\arraycolsep{8pt}
\begin{array}{lcl} 
\delta_\Lambda \alpha = -Im(a)     & \qquad &  \delta_\Lambda v_- = \p_- Re(a)   \\
\delta_\Lambda f = \tfrac{i}{2}\xi  & &  \delta_\Lambda \lambda = -\tfrac{i}{\sqrt{2}} \p_- \xi  \\
\delta_\Lambda v_+ = \p_+ Re(a)  &  &  \delta_\Lambda D = 2\p_+\p_- Im(a)  
\end{array}}
\ee
and for the matter fields
\be
{\setlength\arraycolsep{8pt}
\begin{array}{lclcl} 
\delta_\Lambda \phi = -iQa\phi & \qquad &  \delta_\Lambda \gamma = -iqa\gamma         \\
\delta_\Lambda \psi = -iQa\psi - iQ\xi\phi   &  &   \delta_\Lambda G = -iqaG + i q\xi\gamma \\
\delta_\Lambda \vt = - Ma   & &   \\
\delta_\Lambda \chi = - M \xi  & &
\end{array}}
\ee
Thus, for our primed fields we have
\be
{\setlength\arraycolsep{8pt}
\begin{array}{lcl} 
\delta_\Lambda \alpha = -Im(a)     & \qquad &  \delta_\Lambda v_- = \p_- Re(a)   \\
\delta_\Lambda f = \tfrac{i}{2}\xi  & &  \delta_\Lambda \lambda' = 0  \\
\delta_\Lambda v_+ = \p_+ Re(a)  &  &  \delta_\Lambda D' = 0  
\end{array}}
\ee
\be
{\setlength\arraycolsep{8pt}
\begin{array}{lclcl} 
\delta_\Lambda \phi = -iQa\phi & \qquad &  \delta_\Lambda \gamma = -iqa\gamma         \\
\delta_\Lambda \psi' = -iQa \psi'   &  &   \delta_\Lambda G' = -iqaG' \\
\delta_\Lambda \vt' = - M Re(a)   & &   \\
\delta_\Lambda \chi' = 0  & &
\end{array}}
\ee

\subsection{Component EOMs}

\be
\begin{array}{rcl}
\boxed{\alpha_a}  &  \qquad  &  \sum_i Q_i^a e^{2Q_i \cdot \alpha} \Big( \phi_i \tilde{\nabla}^2 \bar{\phi}_i + \bar{\phi}'_i \tilde{\nabla}^2 \phi_i + 2i\bar{\psi}'_i \tilde{\nabla}_- \psi'_i + 2i\psi'_i \tilde{\nabla}_- \bar{\psi}'_i + 2Q_i^b D'_b |\phi_i|^2  \\
&& \qquad\qquad\qquad  - 2i\sqrt{2} Q_i^b \lambda'_b \bar{\phi}_i \psi'_i  -  2i\sqrt{2} Q_i^b \bar{\lambda}'_b \phi_i \bar{\psi}'_i \Big)      \\
&&  \quad +  \sum_m q_m^a e^{2q_m\cdot\alpha} \Big( 2i \bar{\gamma}_m \tilde{\nabla}_+ \gamma_m + 2i \gamma_m \tilde{\nabla}_+ \bar{\gamma}_m  +  2|G'_m|^2 - 2|E_m|^2   \\
&& \qquad\qquad\qquad  + 2\gamma_m \bar{\psi}'_i \bar{\p}_i \bar{E}_m - 2\bar{\gamma}_m \psi'_i \p_i E_m \Big)     \\
&& \quad +  2M^{(a}\MB{}^{b)}D'_b = 0   \\
\boxed{D'_a}  &  &  \frac{1}{e^2} D'_a + \sum_i Q_i^a e^{2Q_i\cdot\alpha}|\phi_i|^2 - r_a + 2M^{(a}\MB{}^{b)} \alpha_b  = 0   \\
\boxed{f_a }  &  &  M^{(a}\MB{}^{b)} \lambda'_b  =  0    \\
\boxed{v_{+a} }  & &  \frac{2}{e^2}  \p_- v_{+-a} + i\sum_i  Q_i^a e^{2Q_i\cdot\alpha} ( \phi_i \tilde{\nabla}_- \bar{\phi}_i - \bar{\phi}_i \tilde{\nabla}_- \phi_i )  -  \sum_m q_m^a e^{2q_m\cdot\alpha} \bar{\gamma}_m \gamma_m    \\
&&  \quad +  2 \nabla_- ( \MB{}^a\vt' + M^a \bar{\vt}' )  -  2 M^{(a}\MB{}^{b)} v_{-b}  =  0  \\
\boxed{v_{-a} }  & &   \frac{2}{e^2} \p_+ v_{+-a} - i\sum_i Q_i^a e^{2Q_i\cdot\alpha}( \phi_i \tilde{\nabla}_+ \bar{\phi}_i - \bar{\phi}_i \tilde{\nabla}_+ \phi_i  + i \bar{\psi}'_i \psi'_i)  -  2 M^{(a} \MB{}^{b)} v_{+b} = 0    \\
\boxed{\lambda'_a }  & &  \frac{2i}{e^2} \p_+ \bar{\lambda}'_a - \sum_i i\sqrt{2}Q_i^a e^{2Q_i\cdot\alpha} \bar{\phi}_i \psi'_i   -  2i\sqrt{2} M^{(a}\MB{}^{b)} f_b  =  0    \\
\boxed{\phi_i}  &&  e^{2Q_i\cdot\alpha} \Big( \tilde{\nabla}^2 \bar{\phi}_i + i\sqrt{2}Q_i^a \bar{\psi}'_i \bar{\lambda}'_a + Q_i^a D'_a \bar{\phi}_i \Big)  -  \sum_m G'_m \p_i F_m     \\
&& \quad - \sum_{j,m} \gamma_m \psi'_j \p_i\p_j F_m +  \sum_m e^{2q_m\cdot\alpha}\Big( - \p_i E_m \bar{E}_m - \bar{\gamma}_m \psi'_j \p_i\p_j E_m \Big)  =  0     \\
\boxed{\psi'_i} &&  e^{2Q_i\cdot\alpha} \Big( 2 \tilde{\nabla}_- \bar{\psi}'_i + \sqrt{2} Q_i^a \bar{\phi}_i \lambda'_a \Big) - \sum_m i \Big( \gamma_m \p_i F_m  +   e^{2q_m\cdot\alpha} \bar{\gamma}_m \p_i E_m \Big) = 0     \\
\boxed{\gamma_m} &&  e^{2q_m\cdot\alpha} \Big( 2\tilde{\nabla}_+ \bar{\gamma}_m  - i\sum_i \bar{\psi}'_i \bar{\p}_i \bar{E}_m  \Big) + i\sum_i \psi'_i \p_i F_m = 0     \\
\boxed{G'_m}  &&  e^{2q_m\cdot\alpha} \bar{G}' - F_m = 0  \\
\boxed{\vt'} &&  \p_+ \nabla_-  \bar{\vt}' - \MB{}^a v_{+-a}  =  \p_- \nabla_+ \bar{\vt}' = 0  \\
\boxed{\chi'} &&  \p_- \bar{\chi}'  =  0
\end{array}
\ee

Plugging matter EOMs into the $\alpha$ EOM as well as $\p_+ (v_+\textrm{-EOM}) - \p_- (v_- \textrm{-EOM})$ yields
\be
M^{(a}\MB{}^{b)}D'_b  =  M^{(a}\MB{}^{b)} v_{+-b} = 0 \, .
\ee
In particular, we can replace the $\alpha$ EOM by $M^{(a}\MB{}^{b)}D'_b = 0$ or, as we'll see, $M^a D'_a=0$.  If we denote $x^a$ by a column vector $x$, we can write these conditions as
\be
(MM^\dagger +  M^* M^T ) x = 0
\ee
Thus,
\be
x^\dagger (MM^\dagger + M^* M^T )x = \big| M^\dagger x \big|^2  +  \big| M^T x \big|^2 = 0 
\ee
and so we see that this condition actually implies that
\be
M^\dag x = M^T x = 0 \, .
\ee
Adding in the $f$ EOM, we see that
\be
M^a \lambda'_a = \MB{}^a \lambda'_a = M^a v_{+-a} = \MB{}^a v_{+-}a = M^a D'_a = \MB{}^a D'_a = 0 \, ,
\ee
or, more succinctly, the classical equations of motion imply that
\be
M^a \Ups_a = \MB{}^a \Ups_a = 0\, .
\ee

\fi

\section{Torsional $T^2$ Bundles}
\label{app:fy}

The underlying manifold satisfying all of the supersymmetry constraints unrelated to the gauge bundle was studied by Goldstein and Prokushkin in \cite{Goldstein:2002pg}.  Their solution involved constructing the complex 3-fold as a $T^2$ bundle over a $T^4$ or $K3$ base.  Fu and Yau \cite{Fu:2006vj} used this underlying manifold and constructed a gauge bundle satisfying the remaining supersymmetry constraints as well as the modified Bianchi identity.  We start by explaining the underlying manifold.

%%%%%%%%%%%%%%%%%%%%%%%%%%%%
%Unlike the Calabi-Yau case, in non-K\"ahler compactifications one cannot embed the Strominger-connection in the gauge connection because the curvature form will have $(0,2)$ and $(2,0)$ components, so the choice of gauge bundle satisfying Strominger's system becomes much more complicated.  Fu and Yau undertook the difficult task of constructing just such a gauge bundle and were able to prove the existence of solutions to Strominger's system \cite{Fu:2006vj}.\footnote{In their original paper \cite{Fu:2005sm}, Fu and Yau proved the existence of a solution to the system of equations considered in section \ref{Strominger} but with opposite sign for (\ref{eqn:bianchi}).  In \cite{Fu:2006vj}, they have solved the system of equations from section \ref{Strominger}, which are the solutions relevant to heterotic compactifications.  The sign difference dates to a sign error in \cite{Strominger:1986uh}.  Fu and Yau have also considered a wider class of gauge bundles in the more recent paper.}  We briefly review these constructions here.
%%%%%%%%%%%%%%%%%%%%%%%%%%%%%

Let $S$ be a complex Hermitian 2-fold and choose\footnote{Actually, to preserve supersymmetry it is only required that $\omega_P + i\omega_Q$ have no $(0,2)$-component.}
\be
\frac{\omega_P}{2\pi},\frac{\omega_Q}{2\pi}\in H^2(S;\bb{Z}) \cap \Lambda^{1,1}T^*_S .
\ee
where $\omega_P$ and $\omega_Q$ are anti self-dual. Being elements of integer cohomology, there are two $\mathbb{C}^*$-bundles over $S$, call them $P$ and $Q$, whose curvature 2-forms are $\omega_P$ and $\omega_Q$, respectively.  We can then restrict to unit-circle bundles $S_P^1$ and $S_Q^1$ of $P$ and $Q$ respectively, and take the product of the two circles over each point in $S$ to form a $T^2$ bundle over $S$ which we will refer to as $K$ ($T^2\rightarrow K \stackrel{\pi}{\rightarrow}  S$).

Given this setup, Goldstein and Prokushkin showed that if $S$ admits a non-vanishing, holomorphic $(2,0)$-form, then $K$ admits a non-vanishing, holomorphic $(3,0)$-form.  Furthermore, they showed that if $\omega_P$ or $\omega_Q$ are nontrivial in cohomology on $S$, then $K$ admits \emph{no} K\"ahler metric.  They constructed the non-vanishing holomorphic $(3,0)$-form and a Hermitian metric on $K$ from data on $S$.

The curvature 2-form $\omega_P$ determines a non-unique connection $\nabla$ on $S^1_P$ (and similarly for $\omega_Q$ on $S^1_Q$).  A connection determines a split of $T_K$ into a vertical and horizontal subbundle --- the horizontal subbundle is composed of the elements of $T_K$ that are annihilated by the connection $1$-form, the vertical subbundle is then, roughly speaking, the elements of $T_K$ tangent to the fibers.  Over an open subset $U\subset S$, we have a local trivialization of $K$ and we can use unit-norm sections, $\xi\in\Gamma(U;S^1_P)$ and $\zeta\in\Gamma(U;S^1_Q)$, to define local coordinates for $z\in U\times T^2$ by
\be
\label{eqn:coords}
z = (p,e^{i\theta_P}\xi(p),e^{i\theta_Q}\zeta(p)) , 
\ee
where $p = \pi(z)\in U$.  The sections $\xi$ and $\zeta$ also define connection 1-forms via
\be
\nabla \xi = i\alpha_P\otimes \xi      \qquad     \textrm{and}     \qquad     \nabla \zeta = i\alpha_Q\otimes\zeta,
\ee
where $\omega_P = d\alpha_P$ and $\omega_Q = d\alpha_Q$ on $U$, and the $\alpha_i$ are necessarily real to preserve the unit-norms of $\xi$ and $\zeta$.

The complex structure is given on the fibres by $\partial_{\theta_P}\rightarrow\partial_{\theta_Q}$ and $\partial_{\theta_Q}\rightarrow -\partial_{\theta_P}$ while on the horizontal distribution it is induced by projection onto $S$.\footnote{Actually, this just gives an almost complex structure, but Goldstein and Prokushkin proved that it is integrable \cite{Goldstein:2002pg}.}  Given a Hermitian $2$-form $\omega_S$ on $S$, the 2-form
\be
\label{eqn:hermitian form}
\omega_u = \pi^*\left(e^{u} \omega_M\right) + (d\theta_P+\pi^*\alpha_P)\wedge(d\theta_Q+\pi^*\alpha_Q),
\ee
where $u$ is some smooth function on $S$, is a Hermitian $2$-form on $K$ with respect to this complex structure.  The connection $1$-form
\be
\rho \equiv (d\theta_P+\pi^*\alpha_P) + i(d\theta_Q+\pi^*\alpha_Q)
\ee
annihilates elements of the horizontal distribution of $T_K$ while reducing to $d\theta_P+id\theta_Q$ along the fibres.  These data define the complex Hermitian $3$-fold $(K,\omega_u)$.  Explicitly,
\bea
ds^{2}_{K} &=& \pi^*\left(e^{u}ds^{2}_{S}\right) + (d\theta_P+\pi^*\alpha_P)^{2} + (d\theta_Q+\pi^*\alpha_Q)^{2} \nonumber \\
J_{K} &=& \pi^*\left(e^{u}J_{S}\right) + \tfrac{1}{2} \rho\wedge\bar{\rho}  \nonumber \\
\Omega_{K} &=& \pi^*\left(\Omega_{S}\right)\wedge\rho  \nonumber \\
H &=& \sum_{i=P,Q} (d\theta_{i}+\pi^*\alpha_{i})\wedge\pi^*\omega_{i},  \nonumber 
\eea
where $\Omega_{S}$ is the nowhere-vanishing, holomorphic $(2,0)$-form on $S$ ($K3$ or $T^4$).  It is straightforward check that all the supersymmetry constraints are satisfied by this Ansatz, however for a valid heterotic compactifications a gauge bundle still needed to be constructed to satisfy the Bianchi identity.

%%%%%%%%%%%%%%%%%%%%%%%%%%%%%%%%%%%%
%\bea
%ds^{2}_{X} &=& e^{2\phi}ds^{2}_{K3} + (d\theta_{1}+\a_{1})^{2} + (d\theta_{2}+\a_{2})^{2} \non \\
%\vartheta &=&  (d\theta_{1}+\a_{1}) + i (d\theta_{2}+\a_{2})  \non \\
%J_{X} &=& e^{2\phi}J_{K3} + \half \vartheta\wedge\bar{\vartheta}  \non \\
%\Omega_{X} &=& \Omega_{K3}\wedge\vartheta  \non \\
%H &=& \sum_{i=1}^{2} (d\theta_{i}+\a_{i})\wedge\w_{i},  \non 
%\eea
%where $\w_{i}\in H^{2}(K3,\IZ)$ are the curvatures for the two $S^{1}$-bundles, $\a_{i}$ are local one-forms with $\w_{i}=d\a_{i}$, $\vartheta$ is a globally-defined vertical (0,1)-form on the $T^{2}$-fibration, $\Omega_{K3}$ is the holomorphic 2-form on K3, and $\phi$, the dilaton, is a general function on $K3$.  it is easy to check that all the conditions above except one are trivially satisfied by this Ansatz; the Bianchi identity, however, becomes a complicated non-linear PDE for the dilaton, $\phi$.  Fu and Yau proved, under mild assumptions, the existence of a solution to this PDE, and thus, for this Ansatz, to the full superstring equations of motion at one loop in $\a'$.
%%%%%%%%%%%%%%%%%%%%%%%%%%%%%%%%%%%%

Fu and Yau proved the existence of gauge bundles over these manifolds with Hermitian-Yang-Mills connections satisfying the Bianchi identity (\ref{eqn:bianchi}).  They took the Hermitian form (\ref{eqn:hermitian form}) and converted the Bianchi identity into a differential equation for the function $u$.  Under the assumption
\be
\label{eqn:u assumptions}
\left(\int_{K3} e^{-4u}\frac{\omega_{K3}^2}{2} \right)^{1/4} \ll 1 = \int_{K3} \frac{\omega_{K3}^2}{2} ,
\ee
they showed that there exists a solution $u$ to the Bianchi identity for \emph{any} compatible choice of gauge bundle $\mathcal{V}_{K}$ and curvatures $\omega_P$ and $\omega_Q$ such that the gauge bundle $\mathcal{V}_{K}$ over $K$ is the pullback of a stable, degree 0 bundle $\mathcal{V}_{K3}$ over $K3$, $\mathcal{V}_{K} = \pi^*\mathcal{V}_{K3}$ \cite{Fu:2006vj}.\footnote{In fact, \cite{Fu:2006vj} arrived at this result using the Hermitian connection in the Bianchi identity (\ref{eqn:bianchi}).  It has since been pointed out \cite{Becker:2009df} that if one uses the ``$+H$'' connection, which seems more natural from supergravity \cite{Bergshoeff:1989de}, the differential equation that one must solve for $u$ is a Laplacian equation instead of Monge-Amp\`{e}re.}

Note that by a ``compatible'' choice of gauge bundle and $\omega_i$'s we mean the following: choose the gauge bundle $\mathcal{V}_{K}$ and the curvature forms to satisfy the integrated Bianchi identity
\be\label{eq:topcond}
\chi(S) - \tr F^2 = \int_S \sum_i \omega_i^2 .
\ee
In particular, note that the right-hand side and $\tr F^2$ are manifestly non-negative, since $*_S F=-F$ and $F$ is anti-Hermitian. Hence, the only possible solution for a $T^4$ base is to take the gauge bundle \emph{and} the $T^2$ bundle to be trivial, leaving us with a Calabi-Yau solution $T^2\times T^4$ \cite{Becker:2006et,Fu:2006vj}.  This is in agreement with arguments from string duality ruling out the Iwasawa manifold as a solution to the heterotic supersymmetry constraints \cite{Gauntlett:2003cy}.

\section{The Iwasawa Example: $d=4$, $T^4$ Base}
\label{sec:iwasawa}

In \cite{Becker:2006et}, it was argued that supersymmetric solutions do not allow for a $T^4$ base with non-trivial $T^2$ fibration.  In our linear sigma model, we find this statement to be slightly incomplete, as we will elucidate after giving the TLSM example.  The example is essentially two copies of the $T^2$ base of section \ref{sec:T2base},
\be
{\setlength\arraycolsep{8pt}
\begin{array}{|c||c|c|c|c|c|c|c|c|c|}
\hline  & \boldsymbol{\Phi_{i}}  & \boldsymbol{\Phi'_{i}} & \boldsymbol{P} & \boldsymbol{P'} & \boldsymbol{\tG} & \boldsymbol{\tG'} & \boldsymbol{\G_{m}} & \boldsymbol{\G'_{m}} &  \boldsymbol{\Theta}  \\
\hline\hline \boldsymbol{U(1)_1}   &  1  & 0 &  -4  & 0 &  -3  & 0 &  1  & 0 &  (3R)      \\
\hline \boldsymbol{U(1)_2}   &  0  & 1 &  0 & -4 &  0  & -3 &  0  & 1 &  (3iS)      \\
\hline \boldsymbol{U(1)_{L_1}}   &  \frac{1}{4}  & 0 &  0  & 0 &  -\frac{3}{4}  & 0 &  -\frac{3}{4}  & 0 & \big(  \frac{3}{4}R \big)    \\
\hline \boldsymbol{U(1)_{L_2}}   &  0 & \frac{1}{4}  &  0  & 0 & 0 &  -\frac{3}{4}   & 0 &  -\frac{3}{4}  &   \big(  \frac{3}{4}iS \big)    \\
\hline \boldsymbol{U(1)_R}   &  \frac{1}{4}  & \frac{1}{4} &  0  & 0 &  \frac{1}{4}  & \frac{1}{4} &  \frac{1}{4}  & \frac{1}{4} &  \big(  \frac{3}{4} R  + \frac{3}{4} iS \big)  \\
\hline
\end{array}}
\ee
where $R=S=\frac{1}{\sqrt{3}}$.  These satisfy all anomaly conditions and give $c_L = \hat{c}_R + r^{(1)}_L + r^{(2)}_L = 6 + 3 + 3$.

The computation of the massless fermions corresponding the the Ramond ground states is almost identical to section \ref{sec:T2base}, so we'll cut to the chase and simply list the ground state quantum numbers of each twisted sector containing zero-energy states, followed by the massless spectrum.  The rest we leave to the reader (we omit winding numbers as there are no winding states contributing to the massless fermion spectrum)
\be
{\setlength\arraycolsep{6pt}
\begin{array}{|c||c|c|c|c|}
\hline \!\!\! \boldsymbol{\big(k^{(1)}, k^{(2)}\big)} \!\!\! &\boldsymbol{L_0} & \boldsymbol{q_{L_1}} & \boldsymbol{q_{L_2}} & \boldsymbol{q_R}  \\
\hline\hline  \boldsymbol{(0,1)} & 0 & -\frac{3}{2} & 0 & -\frac{3}{2}  \\
\hline \boldsymbol{(0,7)} & 0 & -\frac{3}{2} & 0 &  -\frac{1}{2}   \\
\hline \boldsymbol{(1,0)} & 0 & 0 & -\frac{3}{2} &  -\frac{3}{2}  \\
\hline \boldsymbol{(1,1)} & -1 & 0 & 0 & -\frac{3}{2}   \\
\hline \boldsymbol{(1,2)} & 0 & 0 & \frac{3}{2} & -\frac{3}{2}   \\
\hline \boldsymbol{(1,3)} & -\frac{1}{8} & 0 & -\frac{3}{4} & -\frac{1}{4}     \\
\hline \boldsymbol{(1,5)} & -\frac{1}{8} & 0 & \frac{3}{4} & -\frac{7}{4}   \\
\hline \boldsymbol{(1,6)} & 0 & 0 & -\frac{3}{2} & -\frac{1}{2}   \\
\hline \boldsymbol{(1,7)} & -1 & 0 & 0 & -\frac{1}{2}   \\
\hline \boldsymbol{(2,1)} & 0 & \frac{3}{2} & 0 & -\frac{3}{2}   \\
\hline \boldsymbol{(2,7)} & 0 & \frac{3}{2} & 0 & -\frac{1}{2}   \\
\hline \boldsymbol{(3,1)} & -\frac{1}{8} & -\frac{3}{4} & 0 & -\frac{1}{4}   \\
\hline \boldsymbol{(3,7)} & -\frac{1}{8} & -\frac{3}{4} & 0 & \frac{3}{4}   \\
\hline
\end{array}}
\ee
Notably, since we can only compute the spectrum of Ramond ground states, corresponding to massless spacetime fermions, we cannot identify the massless bosonic spectrum without appeal to spacetime \susy.  In this putatively non-supersymmetric example, we can thus only extract partial information about the full massless spectrum.  However, the spectrum in fact contains two gravitini in the (1,1)-twisted sector.  Thus, while we expected spacetime supersymmetry to be broken based on arguments form the supergravity approximation, the fermionic spectrum is still neatly organized into $d=4$, $\mathcal{N}=4$ supermultiplets, so we state the results this way, again with the caveat that we do not expect this model to yield spacetime supersymmetry:
\be
{\setlength\arraycolsep{8pt}
\begin{array}{|c||c|c|}
\hline d=4,~\mathcal{N}=4~\mathit{Repr.}  &  \mathit{Degeneracy} & E_6 \times E_6   \\
\hline\hline \textrm{Supergravity} & 1 & \boldsymbol{1}\otimes\boldsymbol{1}  \\
\hline \textrm{Vector} & 1 & \boldsymbol{78} \otimes \boldsymbol{1}  \\
\hline \textrm{Vector} & 1 & \boldsymbol{1} \otimes \boldsymbol{78}  \\
\hline \textrm{Vector} & n_{27} & \big(\boldsymbol{27}\oplus\boldsymbol{\overline{27}} \big) \otimes \boldsymbol{1}  \\
\hline \textrm{Vector} & n'_{27} & \boldsymbol{1} \otimes \big( \boldsymbol{27}\oplus \boldsymbol{\overline{27}} \big)  \\
\hline \textrm{Vector} & 45 + 2n_1 + 2n'_1 & \boldsymbol{1}\otimes\boldsymbol{1} \\
\hline
\end{array}}
\ee
where $n_{27}$, $n'_{27}$, $n_1$, and $n'_1$, depend on the explicit choice of superpotential, as in earlier sections.   It would be interesting to resolve this apparent contradiction.

\bibliographystyle{hunsrt}
\bibliography{spectrum}

\end{document}